\documentclass[final,3p]{elsarticle}

\usepackage{xcolor} \definecolor{darkgreen}{rgb}{0,0.5,0}

\usepackage{amsmath,amssymb}

\graphicspath{{Figures/}}
\usepackage{psfrag} 
\usepackage{graphicx} 
\usepackage{overpic}
\usepackage{array}

\usepackage{colortbl}

\newcommand{\matr}[1]{\mathbf{#1}} 
\renewcommand{\vec}[1]{\mathbf{#1}} 

\newcommand{\norm}[1]{\vert\vert #1 \vert\vert}

\newcolumntype{L}[1]{>{\raggedright\let\newline\\\arraybackslash\hspace{0pt}}m{#1}}
\newcolumntype{C}[1]{>{\centering\let\newline\\\arraybackslash\hspace{0pt}}m{#1}}
\newcolumntype{R}[1]{>{\raggedleft\let\newline\\\arraybackslash\hspace{0pt}}m{#1}}

\def\mysingleq#1{`#1'}

\usepackage{filecontents}

\DeclareGraphicsExtensions{.pdf,.png,.jpg}
\begin{document}

\begin{frontmatter} 
  \title{Sparsity enabled cluster reduced-order models for control} 
  \author[uw]{Eurika Kaiser\corref{cor}} \ead{eurika@uw.edu}
  \cortext[cor]{Corresponding author} 
  \author[pu]{Marek Morzy\'nski} 
  \author[cerfacs]{Guillaume Daviller} 
  \author[uwam]{J. Nathan Kutz}
  \author[uwne]{Bingni W. Brunton}
  \author[uw,uwam]{Steven L.~Brunton}

\address[uw]{Department of Mechanical Engineering, 
	University of Washington, Seattle, WA 98195, United States}
\address[pu]{Chair of Virtual Engineering, Pozna\'n University of Technology,
	60-965 Pozna\'n, Poland}  
\address[cerfacs]{CERFACS, F-31057 Toulouse CEDEX 01, France} 
\address[uwam]{Department of Applied Mathematics, 
	University of Washington, Seattle, WA 98195, United States}
\address[uwne]{Department of Biology and Institute of Neuroengineering,  
	University of Washington, Seattle, WA 98195, United States}
    
    \begin{abstract}
        Characterizing and controlling nonlinear, multi-scale phenomena play important roles in science and engineering.
        Cluster-based reduced-order modeling (CROM) was introduced to exploit the underlying low-dimensional dynamics of complex systems.  CROM builds a data-driven discretization of the Perron-Frobenius operator, resulting in a probabilistic model for ensembles of trajectories.  
        A key advantage of CROM is that it embeds nonlinear dynamics in a linear framework,
        and uncertainty can be managed with data assimilation.
        CROM is typically computed on high-dimensional data;  however, access to and computations on this full-state data limit the {\em online} implementation of CROM for prediction and control.
        Here, we address this key challenge by identifying a small subset of critical measurements to learn an efficient CROM, referred to as sparsity-enabled CROM. 
        In particular, we leverage compressive measurements to faithfully embed the cluster geometry and preserve the probabilistic dynamics. 
        Further, we show how to identify fewer optimized sensor locations tailored to a specific problem that outperform random measurements.  
        Both of these sparsity-enabled sensing strategies significantly reduce the burden of data acquisition and processing for low-latency in-time estimation and control.  
        %
        %
        We illustrate this unsupervised learning approach on three different high-dimensional nonlinear dynamical systems from fluids  with increasing complexity, with one application in flow control.
        Sparsity-enabled CROM is a critical facilitator for real-time implementation on high-dimensional systems where full-state information may be inaccessible. 
    \end{abstract}

\end{frontmatter}


\section{Introduction}
Nonlinear, multi-scale phenomena are ubiquitous in many fields in science and engineering; examples include 
the spread of infectious diseases, 
global planetary processes such as the Earth's climate system,
neural brain activity,
autonomous behavior of robotic systems,
sustainable energy production, and greener transport systems.
The high-dimensionality of these systems poses a challenge to understand and realistically model these phenomena. Moreover, low-latency real-time prediction and control is still a difficult endeavor, despite continually increasing computing power and memory storage. 
The long history of model reduction exhibits numerous examples of compact representations of such high-dimensional systems, such as POD-Galerkin models~\cite{Holmes2012book,benner2015survey}, that successfully capture the principal mechanisms. 
An alternative representation of nonlinear systems is based on infinite-dimensional linear operators on functions of the state space, such as the Koopman and Perron-Frobenius operators.
The critical motivation for these operator-based approaches is the ability to apply powerful linear estimation and control techniques to nonlinear systems.   
Cluster-based reduced-order modeling (CROM)~\cite{Kaiser2014jfm} was recently introduced to approximate the Perron-Frobenius operator in an unsupervised manner from high-dimensional data yielding a low-dimensional, linear model in probability space.  
The present work combines CROM with sparsity-promoting techniques, particularly the sparse sensor placement optimization for classification (SSPOC) architecture~\cite{Brunton2016}, as a critical enabler for real-time prediction and control.
The {\em sparsity enabled CROM} identifies the probabilistic dynamics from few optimized measurements or compressed data facilitating its application for online prediction, estimation, and control and faster computations for high-dimensional systems.

Reduced-order models (ROMs) aim to simplify a high-dimensional system by reducing the degrees of freedom, keeping only those that are important to model the phenomenon of interest. 
The intrinsic coordinates, in which the system exhibits such a low-rank structure, are often computed by proper orthogonal decomposition (POD)~\cite{Holmes2012book}, and low dimensional dynamics are obtained via Galerkin projection.
%
%
%
ROMs for parameterized systems are enabled by efficient evaluation of the nonlinear terms using sparse sampling techniques such as gappy POD~\cite{Everson1995josa}.  
The state of the art algorithm for principled sparse sampling of ROMs is the discrete empirical interpolation method (DEIM)~\cite{Chaturantabut2010siam-jsc}, with variants including the addition of a genetic algorithm~\cite{Sargsyan:2015} and the use of pivot locations from the QR factorization~\cite{Drmac2016siam-jsc}.
More generally, sparsity-promoting techniques play an increasingly important role for model identification~\cite{Brunton2015jcd,Brunton2016pnas}, mode selection~\cite{Jovanovic2014pof}, and sensor placement~\cite{Yildirim2009om,Bright2013pof,Brunton2014siam,Sargsyan:2015,Brunton2016} as well as for classification~\cite{Kim2011proc,Akhlaghi2015ol,Bai2017,Manohar2016arxiv} and reconstruction~\cite{Willcox2006cf,Carlberg2013jcp}.

Nonlinearities arising in standard ROMs remain challenging.  
For estimation and control purposes, a linear representation is highly advantageous, spurring considerable work on operator-theoretic embeddings of nonlinear dynamics; these embeddings are not to be confused with local linearization. 
Techniques for linear representation of dynamics include operator methods of Koopman~\cite{Koopman1931pnas,Mezic2005nd}, Perron-Frobenius~\cite{Perron1907ma,Ulam1964book} and Fokker-Planck~\cite{Ryter1987phys}. 
These infinite dimensional operators act on functions of the state space, providing a global linear description of the system. 
The practical computation of finite-dimensional approximations of the Koopman operator include 
dynamic mode decomposition (DMD)~\cite{Schmid2010jfm, Rowley2009jfm} and its variants~\cite{Kutz2016book,Williams2015jns}.  
The Perron-Frobenius operator is the adjoint of the Koopman operator, and it is associated with a probabilistic description of the dynamics. 
The continuous-time Liouville equation~\cite{Liouville1838} associated with Perron-Frobenius governs the evolution of the probability density function (p.d.f.) in the state space (i.e., how an ensemble of trajectories evolves).
%
%
Data-driven approximations of the Perron-Frobenius operator include identification of almost-invariant sets~\cite{Froyland1998831,Froyland2003jsc,Dellnitz2004,Froyland2013jna} via the Ulam-Galerkin method~\cite{Ulam1964book,Bollt2013book}, which reduces Perron-Frobenius to a stochastic matrix.  
In practice, Ulam's method involves a high-dimensional discretization of the state space using a box partition, which suffers from the {\it curse of dimensionality}.
If time-series data is available, the transition probabilities between those boxes can then be determined directly, but the computational burden of computing the partition and transition matrix is significant. 

Cluster-based reduced-order modeling is a particular realization of Ulam's method where a low-dimensional discretization is obtained in an unsupervised manner from data using a clustering algorithm. 
This data-driven discretization enables an efficient partitioning, while avoiding superfluous covering of regions where data is not available. 
The simplest CROM uses the k-means clustering algorithm~\cite{Li1976jat} to learn an intrinsic partition or structure directly from data by grouping similar observations~\cite{Bishop2007book}. 
CROM generally relies on the knowledge of full-state measurements, which may be inaccessible in practice, and limits its use for real-time estimation and control.  

In this work, we leverage sparsity-promoting techniques to construct an efficient CROM from few measurements, referred to as sparsity-enabled CROM, which is a critical enabler for its \emph{online} application.
We first show that a sufficient, but small number of random measurements embed the cluster geometry and preserve the probabilistic dynamics. 
Further, we demonstrate the ability to learn a minimal set of optimized sensors, using the sparse sensor placement optimization for classification (SSPOC) architecture~\cite{Brunton2016}, that are tailored to the specific CROM and provide performance on par with the high-dimensional CROM.  
Sparsity-enabled CROM allows one to identify low-dimensional probabilistic dynamics of high-dimensional systems in an unsupervised manner from sparsely-sampled data. 
Our method facilitates faster computations and is a critical enabler for real-time applications such as prediction and control.
These sparsity enabled innovations are demonstrated on three high-dimensional fluid systems of increasing complexity, and in all cases optimized sensors outperform randomly chosen sensors.  
We also show that the sparsity enabled CROM can be used for closed-loop control, resulting in control performance that is similar to that of full-state CROM.  


The remainder of the article is structured as follows:
The present work is centered around the CROM framework, compressed sensing, 
and the SSPOC architecture, which are discussed in Sec.~\ref{Sec:Background}.
The main contributions of this work are
(a) combining CROM with compressed sensing techniques 
to enable its estimation from few incoherent measurements,
and (b) combining SSPOC with CROM to identify few optimized sensor locations in an unsupervised manner,  
both presented in Sec.~\ref{Sec:Methods}.
The approaches are illustrated for three high-dimensional systems from fluids in Sec.~\ref{Sec:Results}, 
the periodic double gyre flow, a well studied model for ocean mixing, 
a separating flow over a smoothly contoured ramp, where identified sensors are used for control, 
and the spatially developing mixing layer undergoing vortex pairing, 
where sensors are learned on heavily subsampled data.
The main results are summarized and an outlook is provided in Sec.~\ref{Sec:Conclusions}.

\section{Background}
\label{Sec:Background}

This work develops reduced-order models of high-dimensional nonlinear dynamical systems using sparse measurements in a linear operator framework.   
The Perron-Frobenius operator is an infinite-dimensional linear operator for the evolution of densities in the state space of a nonlinear dynamical system.  
Although trading nonlinearity for a linear representation is desirable, a host of additional challenges arise due to the infinite-dimensional nature of the Perron-Frobenius operator.  
Thus, a finite-rank approximation of the Perron-Frobenius operator has been recently proposed~\cite{Kaiser2014jfm}, based on a data-driven discretization of phase space.  This method, Cluster-based Reduced-Order Modeling (CROM), is discussed in Sec.~\ref{Sec:CROM}.

To facilitate sparse measurements and efficient computations for real-time control, techniques from compressed sensing are employed to determine a CROM for high-dimensional systems using few measurements.  
In compressed sensing, a high-dimensional signal can be recovered from few measurements if the signal is sparse in a transform basis. 
The geometry-preserving property of compressed sensing
makes it ideal for estimating a CROM from few measurements,
which 
requires that points close in high-dimensional state space remain close in measurement space.
This is discussed in Sec.~\ref{Sec:CS}.

The Sparse Sensor Placement Optimization for Classification (SSPOC) framework~\cite{Brunton2016} leverages techniques from compressed sensing and exploits the low-rank structure occurring in many systems for optimized sensor placement, providing tailored sensor locations for a particular problem.
This work combines CROM with SSPOC to yield (1) a sparse CROM from measurements, and  
(2) an unsupervised sensor placement framework for cluster classification.
The SSPOC approach is reviewed in Sec.~\ref{Sec:SSPOC}.

In the following, we will consider a high-dimensional state $\mathbf{u}\in\mathbb{R}^N$ with $N\gg 1$, which may be obtained by discretizing a partial differential equation (PDE), that is governed by a nonlinear dynamical system
\begin{equation}\label{Eqn:DS}
\frac{d}{dt}\mathbf{u} = \mathbf{f}(\mathbf{u}).
\end{equation}
It is assumed that the governing equations exhibit low-rank structure that can be computed from the singular value decomposition. The model reduction framework represents the dynamics in a POD basis of rank $N_f$ given by the columns of the matrix $\matr{\Psi} \in \mathbb{R}^{N\times N_f}$:
\begin{equation}\label{Eqn:POD}
{\bf u}(t)  = \matr{\Psi} \vec{a}(t)
\end{equation}    
so that the dynamics are now captured by the evolution of the coefficients ${\bf a}(t) \in \mathbb{R}^{N_f}$~\cite{Holmes2012book,benner2015survey}.

%


\subsection{Cluster-based reduced-order modeling (CROM) and control}
\label{Sec:CROM}
\begin{figure}[tb]
	\centering
	\psfrag{Data}{\raisebox{0.35cm}{Data}}
	\psfrag{Kinematics}{\raisebox{0.35cm}{Kinematics}}
	\psfrag{Dynamics}{\raisebox{0.35cm}{Dynamics}}
	\includegraphics[width=\textwidth]{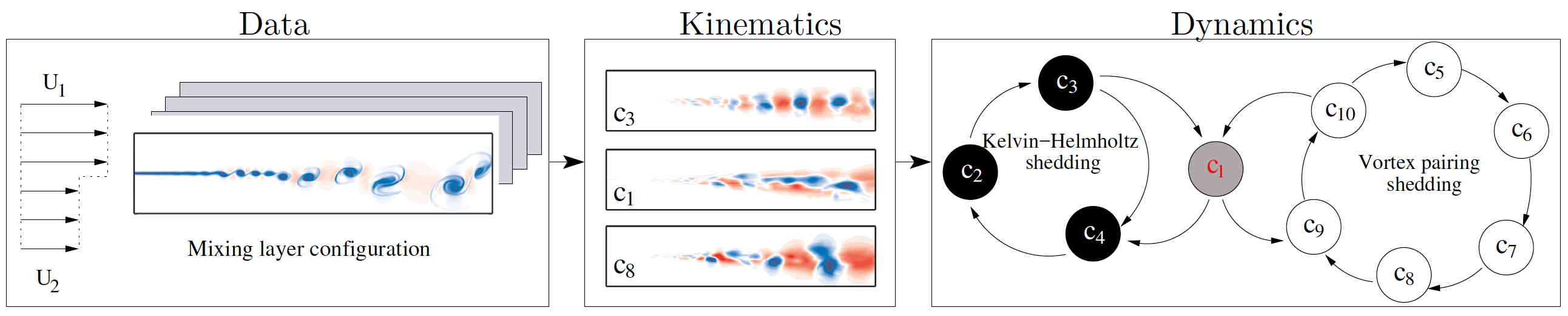}
	\caption{Cluster-based reduced-order model of a mixing layer.
		High-dimensional time-series data is partitioned into few clusters; three exemplary cluster centroids are depicted.
		CROM yields a Markov model for the probabilistic dynamics on the set of clusters, which are here represented as a graph.
		See text for details.
	}
	\label{Fig:CROM_ML}
\end{figure}

The cluster-based reduced-order modeling (CROM)~\cite{Kaiser2014jfm} framework has been recently introduced to model the coarse-grained probabilistic dynamics of high-dimensional nonlinear systems, such as fluid flows.  CROM identifies models in an unsupervised manner directly from data  (see Fig.~\ref{Fig:CROM_ML} for an example). 
The resulting low-dimensional model 
yields insights into properties of the attractor by analysis of the underlying interaction dynamics between clusters.  
Thus, a coarse-grained probability vector on the spate space is evolved, taking into account uncertainties with a well-defined prediction horizon, and capturing nonlinear mechanisms in a linear framework.   
The approach is closely related to the common Ulam-Galerkin approximation scheme~\cite{Perron1907ma,Lasota1994book}, which reduces the infinite dimensional Perron-Frobenius operator to a stochastic matrix.

In many systems, such as fluids, we are often interested in controlling statistical flow properties, 
such as the average drag on a car or average lift on an airfoil. 
Moreover, these quantities are often determined from a single time-series of data from an experiment or simulation, as opposed to an ensemble of data.   
The basis for their computation is Birkhoff's ergodic theorem~\cite{Birkhoff1931pnas}, 
which states the equivalence between time averages and space averages in ergodic systems.
Hence, 
interest in the long-term behavior leads naturally to invariant (or ergodic) probability measures on the attractor, i.e.\ these measures stay the same after transformation of the attractor. 
Moreover, a probabilistic description of complex dynamical systems can be more insightful than that of an individual trajectory, particularly in the study of transport and mixing processes~\cite{Cvitanovic2012book,Dellnitz2004,Froyland2013jna}.
The evolution of the probability density function (p.d.f.) of the state variables, i.e.\ an ensemble of trajectories,
is governed by the linear Liouville equation.
A prominent example is Hopf's derivation of a Liouville equation for the Navier-Stokes equation~\cite{Hopf1952jrma}. 
An overview of methods for the numerical approximation of functional differential equations, such as arising from Hopf's formalism of the Liouville equation, is provided in~\cite{Venturi2016arxiv}.
A prominent ROM strategy is based on the Mori-Zwanzig projection operator formalism~\cite{Mori1965ptp,Zwanzig1973jsp,Chorin2009book} for a Liouville equation~\cite{Stinis20140446,Gouasmi2016arxiv}.
The Liouville equation for a Galerkin system constitutes a simpler version and the reader is referred to~\cite{Noack2012jfm} for a detailed discussion.
Associated with the continuous-time Liouville equation is the above mentioned discrete-time Perron-Frobenius operator, 
which maps the p.d.f. forward in time; the Liouville equation may be thought of as the infinitesimal generator for the one-parameter family of Perron-Frobenius operators.  
CROM is closely related to the Ulam-Galerkin method~\cite{Ulam1964book,Li1976jat}, but with the critical distinction of a data-driven discretization of phase space that results in a much lower-dimensional model.  
Thus, CROM is closely aligned with closure schemes, 
in which a stable fixed point represents the ergodic measure for the unsteady attractor in velocity space.

\begin{figure}[t]
	\centering
	\includegraphics[scale = 1]{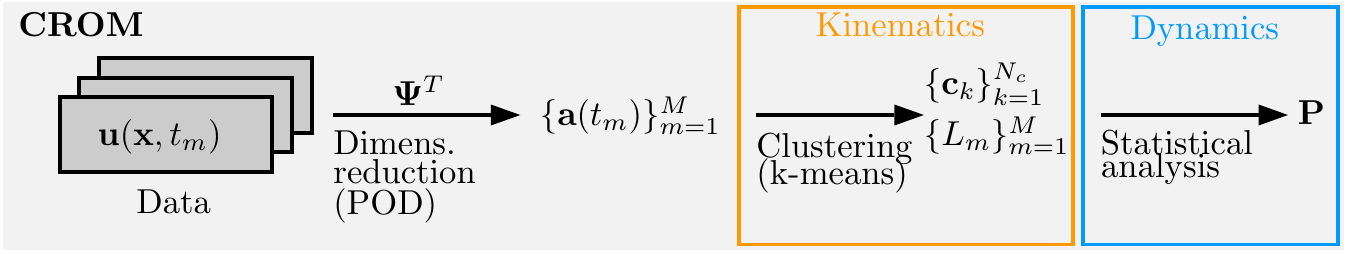}
	\caption{Schematic of the cluster-based reduced-order modeling (CROM) strategy.}
	\label{Fig:CROM_Schematic}
\end{figure}

In this work, we assume velocity fields as input data, 
which are denoted by $\{{\bf u}({\bf x}, t_m)\}_{m=1}^M$ in the following,  
where ${\bf u}({\bf x}, t_m)$ is the $m$th realization at discrete time $t_m$ over a fixed domain $\Omega$
with spatial coordinate ${\bf x}$. A constant time step $\Delta t$ is assumed.
A schematic of CROM is provided in Fig.~\ref{Fig:CROM_Schematic} and discussed below.
CROM assumes time-resolved data and relies on two steps:
First, the data is partitioned into groups of kinematically similar observations using an unsupervised clustering algorithm, such as k-means~\cite{Steinhaus1956}, to obtain a coarse-grained state space.
K-means aims to find a natural grouping or hidden structure in data by maximizing the similarity of observations in the same group, also referred to as {\it cluster}, and minimizing it for observations belonging to different groups.
The clustering algorithm assumes a pre-defined number of clusters $N_c$ 
and yields a set of centroids $\{{\bf c}_k\}_{k=1}^{N_c}$, 
where ${\bf c}_k$ represents the mean of all observations in cluster $\mathcal{C}_k$, $k=1,\ldots,N_c$, 
and a set of labels $\{L_m\}_{m=1}^{M}$ with $L_m\in\{1,\ldots,N_c\}$, which affiliates each observation ${\bf u}({\bf x},t_m)$ with a distinct cluster $\mathcal{C}_k$.
Moreover, the data space is partitioned into $N_c$ centroidal Voronoi cells,
which are defined as particular Voronoi cells for which the generating points of the
Voronoi tessellation are equal to the mass centroids of the Voronoi regions~\cite{Du_Faber_Gunzburger_SIAM_Review_1999}. 
K-means clustering has been applied in a variety of applications related to model reduction, e.g.\ for dimensionality reduction~\cite{Du2003}, trust-region reduced-order modeling~\cite{Amsallem2009aiaa,Amsallem2012nme}, 
and similarly to CROM, for the prediction of coarse-grained observables~\cite{Giannakis2011joc}, to name a few.
Second, the transitions between those clusters are modeled as a Markov process. The resulting transition probability matrix, which describes how the probability distribution evolves on the discretized state space, can be represented as a graph (Fig.~\ref{Fig:CROM_ML}).
The maximum likelihood estimator is used to determine the transition probabilities ${\bf P}=(P_{jk})$ of the Markov process, where $P_{jk} = \mathrm{Prob}\{ \vec{u}(\vec{x},t_{m+1})\in \mathcal{C}_j \vert \vec{u}(\vec{x},t_{m})\in \mathcal{C}_k  \}$ denotes the probability that a transition of the trajectory occurs from cluster $\mathcal{C}_k$ to cluster $\mathcal{C}_j$ over one time step $\Delta t$.
If the data is high-dimensional, a reduction using, e.g., proper orthogonal decomposition (POD)~\cite{Holmes2012book}, 
might be necessary to increase the feasibility of the procedure.
The clustering is then applied to the POD time coefficients $\{ {\bf a}(t_m)\}_{m=1}^M$ and the procedure continues as described above.

The representation of nonlinear dynamics in an approximate linear framework is of significant current interest, largely because of the potential to enable advanced nonlinear prediction, estimation and control using standard tools from linear systems theory~\cite{Mezic2005nd,Budivsic2012chaos,Mezic2013arfm,Brunton2016plosone,Kutz2016book}.  
CROM is a practical data-driven approach for representing high-dimensional nonlinear systems in a probabilistic linear framework.  
However, the standard CROM analysis still relies on access to high-dimensional measurement data, which may be expensive to collect.  
Moreover, computations based on this high-dimensional data introduce unacceptable latency, limiting the bandwidth for real-time feedback control.  
The goal of real-time estimation and control motivates the use of compressed sensing and sparse measurements from the following sections.  

\subsection{Compressed sensing}
\label{Sec:CS}
Compressed sensing is revolutionizing our understanding of signal compression and reconstruction~\cite{Candes2006ieee,Candes2006c,Donoho2006ieee}. 
This growing body of work relies on the fact that most natural high-dimensional signals ${\bf u}$, such as discretized solutions to PDEs, are highly compressible.  
Thus in an appropriate basis (such as a tailored POD basis, or a Fourier or wavelet basis), the high-dimensional signal may be written as a sparse vector $\vec{a}$ as in \eqref{Eqn:POD} with many zero-valued coefficients.  
For motivating examples of compressed sensing, such as image reconstruction, a generic wavelet or Fourier basis is sufficient. 
%
If the vector ${\bf a}$ has $K$ nonzero elements, we say that it is $K$-sparse.  
Instead of measuring the high-dimensional signal ${\bf u}$ directly, compressed sensing provides rigorous conditions under which it is possible to collect surprisingly few measurements with respect to the Nyquist sampling frequency and \emph{infer} the few non-zero coefficients of ${\bf a}$, and hence ${\bf u}$.  
This observation has led to a number of studies investigating the properties of random sparse measurements and the construction of sensing matrices with favorable reconstruction properties.  
In particular, consider a measurement matrix ${\bf \Phi} \in \mathbb{R}^{N_s\times N}$, with $K<N_s\ll N$.  
Then measurements ${\bf y}$ are given by:
\begin{equation}
{\bf y} = {\bf\Phi}{\bf u} = {\bf\Phi}{\bf\Psi}{\bf a} = {\bf\Theta}{\bf a}.
\end{equation}
The main result of compressed sensing is that the sparse coefficients of ${\bf a}$ may be determined with high-probability, given that the measurements are chosen so that the matrix ${\bf \Theta}$ satisfies the Restricted Isometry Property (RIP).  
In particular, there must be sufficiently many measurements, typically on the order ${N_s=\mathcal{O}(K\log(N/K))}$, and these measurements must be \emph{incoherent} with respect to the sparsifying basis ${\bf \Psi}$, so that the rows of ${\bf\Phi}$ are not too correlated with any column of ${\bf \Psi}$.  
An important set of results have shown that random measurements, where the entries of ${\bf \Phi}$ are Gaussian or Bernoulli random variables, are incoherent with a given basis ${\bf \Psi}$ with high probability.  

Without compressed sensing, searching for the sparsest vector ${\bf a}$ consistent with the measurements ${\bf y}$ amounts to an intractable brute-force search through the combinatorially many sparse vectors.  
Mathematically, this can be formulated as an optimization problem
\begin{equation}
{\bf a} = \arg \min\limits_{{\bf a}'} \vert\vert {\bf a}'\vert\vert_0\quad \text{subject to}\quad {\bf y} = {\bf\Theta}{\bf a}' \;.
\label{Eqn:CS:L0}
\end{equation}
However, the $\ell_0$ pseudo-norm, which measures the sparsity of ${\bf a}$, makes this optimization non-convex, so that it does not scale well to large problems.  
With the advent of compressed sensing, it is now possible to solve for the sparsest consistent ${\bf a}$ with high probability by relaxing the $\ell_0$ term to an $\ell_1$ norm~\cite{Candes2006c,Candes2006ieee,Donoho2006book}:
\begin{equation}
{\bf a} = \arg \min\limits_{{\bf a}'} \vert\vert {\bf a}'\vert\vert_1\quad \text{subject to}\quad {\bf y} = {\bf\Theta}{\bf a}'.
\label{Eqn:CS:L1}
\end{equation}
Solutions to \eqref{Eqn:CS:L1} can be found through convex optimization methods (e.g. using the \texttt{cvx} toolbox \cite{cvx,gb08}) or greedy algorithms such as orthogonal matching pursuit \cite{Tropp2007ieee,Tropp2006sp}.  
The number of sensors can be further reduced for classification, considered in the present study, as the bijectivity property can be relaxed. 

In many engineering applications, Gaussian random measurements from compressed sensing are not practical.  
Instead, point measurements are more physical, as they correspond to individual sensors.  
Fortunately, point sensors are optimally incoherent with respect to the Fourier basis, and many engineering signals, such as fluid velocity fields and other solutions of PDEs, are sparse in the Fourier domain.  
There has been considerable recent work combining sparsity with \emph{dynamical} systems~\cite{Bright2013pof,Ozolicnvs2013pnas,Schaeffer2013pnas,mackey2014compressive,Brunton2014siam,Tu2014ef,Brunton2015jcd,Gueniat2015pof,Kramer2015arxiv,Bai2017,Brunton2016pnas}, a perspective that is continued here.  
Throughout this work, we will leverage the fact from compressed sensing that random measurements tend to preserve the geometry of sparse vectors in the measurement space.

\subsection{Sparse sensor placement optimization for classification (SSPOC)}
\label{Sec:SSPOC}
\begin{figure}[tb]
	\centering
	\includegraphics[width=\textwidth]{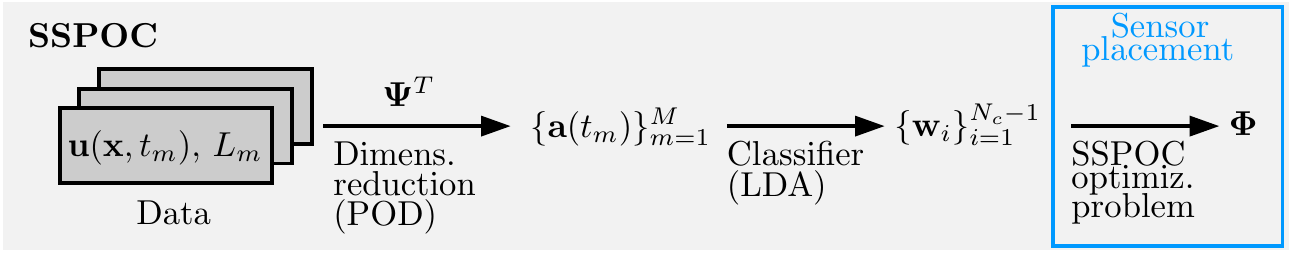}
	\caption{Schematic of the Sparse Sensor Placement Optimization for Classification (SSPOC) strategy. 
	}
	\label{Fig:SSPOC_Schematic}
\end{figure}

The sparse sensor placement optimization for classification (SSPOC) framework \cite{Brunton2016} combines dimensionality reduction and discrimination techniques with compressed sensing to learn sparse sensor locations that enable classification of a high-dimensional system from few measurements. 
\begin{figure}[tb]
	\centering
	\includegraphics[width=\textwidth]{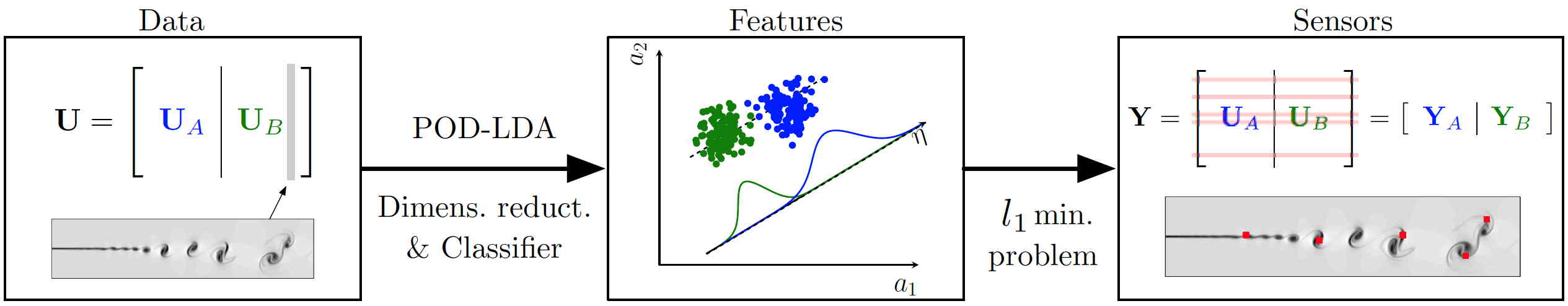}
	\caption{Sparse sensor placement optimization for classification (SSPOC) for high-dimensional systems.}
	\label{Fig:SSPOC_Schematic_LDA-PCA}
\end{figure}
SSPOC exploits the fact that many high-dimensional systems evolve on a low-dimensional attractor, and can thus be represented in a low-rank basis.  
Moreover, classification is simpler than full-state reconstruction, and can be accomplished with fewer measurements.  
It is common to combine low-rank representations such as POD with linear discriminant analysis (LDA) to learn low-dimensional classifiers.  
In addition, using POD as a pre-processing step to LDA can regularize ill-conditioned problems. 
While SSPOC is a general procedure, here we make use of the POD-LDA approach as suggested in \cite{Brunton2016} for simplicity.  
A schematic of the procedure is shown in Fig.~\ref{Fig:SSPOC_Schematic}.

We consider high-dimensional data, such as the velocity snapshot ensemble $\{{\bf u}({\bf x},t_m)\}_{m=1}^M$  introduced in Sec.~\ref{Sec:CROM}.
Moreover, this data may be associated with different classes such as different bifurcation regimes \cite{Brunton2014siam}, different control cases \cite{Bai2017}, or distinct clusters representing a coarse-grained discretization of state space, as in the present study.
The classification of each observation $\{L_m\}_{m=1}^M$
must be known in advance.
It is further assumed that the data can be represented by a low-rank feature basis 
${\bf \Psi} = [\boldsymbol{\psi}_1, \ldots,\boldsymbol{\psi}_{N_f}]\in\mathbb{R}^{N\times N_f}$, 
where $N$ is the dimension of the data and $N_f$ is the rank of the basis. 
The LDA classifier is trained using labeled data in the feature space and identifies the directions given by ${\bf w} = [{\bf w}_1, \ldots, {\bf w}_{N_c-1}]\in\mathbb{R}^{{N_f}\times N_c-1}$, in which the classes are best separated.  
SSPOC aims to find a sparse vector ${\bf s}$
that best reconstructs the discriminating directions ${\bf w}$ by solving the optimization problem
\begin{equation}
{\bf s} = \arg \min\limits_{{\bf s}'} \left\{ \vert\vert {\bf s}'\vert\vert_1 + \lambda\vert\vert{\bf s}'{\bf 1}\vert\vert_1 \right\} 
\quad \text{subject to}\quad \vert\vert{\bf\Psi}^T{\bf s}'-{\bf w} \vert\vert_F \leq \varepsilon \;,
\label{Eqn:SSPOC_nonbinary}
\end{equation}
where ${\bf 1}$ represents a column vector of $N_c-1$ ones and $\varepsilon$ is a small error tolerance (set to $\varepsilon = 10^{-10}$ in all examples).
The non-zero entries in the solution $\vec{s}\in\mathbb{R}^{N}$ are the spatial locations of the learned sparse sensors; these sensors are selected for rows in $\matr{\Psi}$ that best reconstruct $\vec{w}$. 
The coupling weight $\lambda$ tunes the number of learned sensors at the cost of decreasing the classification accuracy.  
Thus, increasing $\lambda$ amounts to strengthening the coupling between columns of $\vec{s}$, so that the same entries or measurements are re-used to reconstruct several decision vectors.

Having identified the optimal sensor locations, a sensing matrix $\matr{\Phi}\in\mathbb{R}^{N_s\times N}$ is constructed by selecting the $N_s$ rows of the $N\times N$ identity matrix corresponding to the $N_s$ nonzero rows in ${\bf s}$. 
The classification task can then be performed on low-dimensional measurements 
$\vec{y} = \matr{\Phi} {\vec{u}}$.  
Although it is possible to use the original LDA classifier on the new measurements in ${\bf y}$, it is generally advisable to train a new classifier directly in the sensor space, resulting in new discriminating directions $\hat{\bf w}\in\mathbb{R}^{{N_s}\times N_c-1}$.  
Then, a new measurement is assigned a class corresponding to the cluster $k$ whose projected centroid $\boldsymbol{\xi}_k = \hat{\vec{w}}^T \vec{c}_k$, $k=1,\ldots,N_c$, is closest to $\boldsymbol{\eta}$ (the nearest-centroid method, NCM). 

Depending on the dimensionality of the data, this sensor placement approach can be quite costly and may require considerable computational resources. 
Thus, in some cases, it is advantageous to randomly subsample the data before learning the sensor locations, as explored in~\cite{Brunton2016}.  
For many tasks, it has been shown that sensor locations learned on $10\%$ of the data perform similarly to those trained on the full state.   
Measurements ${\vec{y}}$ are then obtained by the projection ${\vec{y}} = \tilde{\matr{\Phi}} \hat{\matr{\Phi}} {\vec{u}}$ where $\hat{\matr{\Phi}}$ is the sub-sampling matrix consisting of random rows of the $N\times N$ identity matrix and $\tilde{\matr{\Phi}}$ is the sensing matrix learned in that subspace.  

The SSPOC procedure has been previously demonstrated in a number of applied contexts to streamline the sensors required for an accurate classification based on a pre-trained supervised classification scheme~\cite{Brunton2016,Bai2017}.  
The present work generalizes this algorithm to work without known labels of observations using \emph{unsupervised} clustering, such as k-means.  
More importantly, this work makes a critical generalization of SSPOC to apply to \emph{dynamical} systems, where sparse sensor selection can dramatically improve real-time estimation and control performance.  

\section{Methodology}
\label{Sec:Methods}

The major contribution of this work is in extending the CROM framework (see Sec.~\ref{Sec:CROM}) to include compressive measurements; in particular, we use the SSPOC architecture (see Sec.~\ref{Sec:SSPOC}) for sensor placement optimization.  
This combination enables the three main results of this work:

\begin{enumerate}
	\item It is possible to compute CROM from compressive measurements yielding the same probabilistic transition dynamics as CROM based on high-fidelity data.
	We refer to this as sparsity-enabled CROM.
	\item 
	We apply SSPOC with CROM to find a few, optimized point measurements tailored to the specific CROM problem. 
	This allows one to implement CROM, estimated from high-dimensional data, in real-time applications such as estimation and control. For control, we find that the optimized sensors perform similarly to full-state measurements.  
	\item We generalize SSPOC 
	to be applicable to unlabeled data and to learn sensor placement for dynamical systems, such as CROM.
	SSPOC finds sensors that perform the classification task with high accuracy, even though the data considered here is, by definition, not well separated among the partitions.
\end{enumerate}

Sparsity-enabled CROM estimated from few measurements, exploiting the geometry-preserving properties of  compressed sensing methods, is discussed in Sec.~\ref{Sec:sCROM}. 
The combination of SSPOC with CROM to learn a small number of optimized sensors is presented in Sec.~\ref{Sec:SSPOCCROM}.  

\subsection{Sparsity-enabled CROM}
\label{Sec:sCROM}
\begin{figure}[t]
	\centering
	\includegraphics[width=\textwidth]{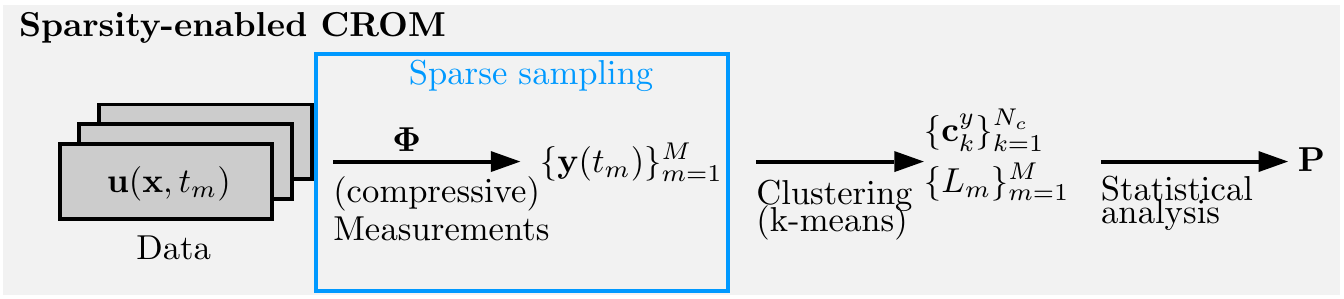}
	\caption{Schematic of the sparsity-enabled CROM strategy.}
	\label{Fig:sparseCROM_Schematic}
\end{figure}

The analysis, modeling and control of high-dimensional systems often involve algorithms that are computationally expensive, making real-time applications intractable.
In this section, we combine CROM with ideas from compressed sensing to enable a computationally efficient cluster and model identification from few incoherent measurements. 
A schematic of the sparsity-enabled CROM strategy is shown in Fig.~\ref{Fig:sparseCROM_Schematic}.

Let us consider full-state measurements $\vec{u}\in\mathbb{R}^N$ of the high-dimensional dynamical system \eqref{Eqn:DS}.
It is possible to collect compressed data $\vec{y} = \matr{\Phi}\vec{u}\in\mathbb{R}^{N_s}$,  where $\matr{\Phi}$ is the sensing matrix.
We seek a transition probability matrix from those measurements $\vec{y}$ that exhibits the same topological structure as its counterpart estimated from full-state measurements $\vec{u}$.
The transition probabilities depend solely on the cluster affiliation provided by the clustering of the time history of $\vec{u}$ or $\vec{y}$.
Thus, the k-means clustering step, yielding the state-space discretization into clusters, is crucial to preserve the probabilistic dynamics.
K-means clustering aims to partition $M$ observations into $N_c$ clusters, such that the distances between observations in the same cluster are minimized and those between observations belonging to different clusters are maximized.
Specifically, it minimizes the sum of the squared distances
\begin{equation}
\{ \vec{c}_1,\ldots,\vec{c}_{N_c} \} = \arg\min\limits_{\vec{c}'_1,\ldots,\vec{c}'_{N_c}} \sum\limits_{i=1}^{N_c}\sum\limits_{\vec{u}\in\mathcal{C}'_i}
\norm{\vec{u}-\vec{c}'_i}^2\, ,
\end{equation}
where $N_c$ is the number of clusters and $\mathcal{C}_i$ denotes the Voronoi cell associated with cluster centroid $\vec{c}_i$, $i=1,\ldots,N_c$. 
Note that we consider in the present work only the Euclidean distance metric.
This means that not only must measurements $\vec{y}$ disambiguate different high-dimensional states $\vec{u}$,
but the measurement matrix $\matr{\Phi}$ must also ensure that two states $\vec{u}_1$ and $\vec{u}_2$, which are close in state space, must also be close in sensor space.
Distances between data points must be preserved under the action of the sensing matrix $\matr{\Phi}$:
\begin{equation}
\norm{\vec{u}_1-\vec{u}_2}  \approx \norm{\vec{y}_1-\vec{y}_2}= \norm{\matr{\Phi}\vec{u}_1-\matr{\Phi}\vec{u}_2} = \norm{\matr{\Phi}(\vec{u}_1-\vec{u}_2)} \; .
\end{equation}
These geometry-preserving properties establish that the dynamics estimated from measurements are equal to those in full state space.
For this to be true, in the compressed sensing framework, the following conditions must be fulfilled:
(i) $\vec{u}$ must be sparse in transform basis  $\matr{\Psi}$,
(ii) sufficiently many measurements, typically $N_s = \mathcal{O}(K \log(N/K))$, must be collected, 
and (iii) the sensing matrix $\matr{\Phi}$ must be incoherent with respect to $\matr{\Psi}$.
Thus, if the sensing matrix $\matr{\Phi} $ satisfies the RIP,
the pair-wise distances between any two $K$-sparse vectors, i.e.\ here specifically $\vec{a}=\matr{\Psi}^T\vec{u}$, are preserved, and the high-dimensional state $\vec{u}$ can be reconstructed from $\vec{y}$ \cite{Davenport2012book}.
It can also be concluded from this property that high-fidelity cluster centroids $\{\vec{c}_i\}_{i=1}^{N_c}$ can be reconstructed from those centroids $\{\hat{\vec{c}}_i\}_{i=1}^{N_c}$ learned from measurements $\vec{y}$.

Identifying dynamics from compressive data of high-dimensional systems
has the additional advantage of making the pre-processing dimensionality-reduction step expendable.
The representation of the data $\vec{u}$ in a transform basis such as POD
has been found to increase the computational efficiency if the state is high-dimensional \cite{Kaiser2014jfm}.
Specifically, POD becomes computationally advantageous for $I N_c > (M+1)/2$, where $I$ is the number of iterations in the k-means algorithm, 
when comparing the number of distance integrals of k-means with correlation integrals of POD. 
Compressive measurements become advantageous if $(M+1)/2>N_s$, not taking into account any additional calculations for POD.

Sparsity-enabled CROM allows one to identify the probabilistic dynamics of high-dimensional, nonlinear systems from few measurements
facilitating more efficient computations and making data preprocessing steps for data compression and feature extraction superfluous.
The critical enabler is the compressed sensing paradigm which directs the design of sensing matrices, such as Gaussian random matrices, that preserve geometric properties of sparse vectors. This allows one to apply k-means clustering directly to compressive measurements.
In the following example, we demonstrate sparsity-enabled CROM for a high-dimensional system from fluids using Gaussian random and random point measurements. 
While Gaussian random measurements still rely on access to full-state data, random point measurements are more physical, corresponding to individual sensors.  
However, these are not tailored to the problem, but are instead chosen randomly, suggesting that significant improvements can be achieved by optimizing their locations.

\paragraph{Example: Sparse CROM estimated from compressive measurements of the mixing layer}
We illustrate the sparsity-enabled CROM estimated from few linear, incoherent measurements of the high-dimensional spatially developing fluid mixing layer (see Fig.~\ref{Fig:CROM_ML}). 
For details on this dataset we refer to \cite{Kaiser2014jfm} and Sec.~\ref{Sec:MixingLayer} where it is studied in detail for optimized sensor placement of point measurements.  
In particular, we are interested in comparing the cluster affiliation and the probabilistic dynamics based on few incoherent measurements with those of the full state.

We consider the time history of $M=2000$ velocity fields $\vec{u}(\vec{x},t_m)$, $m=1,\ldots,M$, which is compressed using POD. Note that the dimension of each velocity field is $N\approx 3.7\cdot 10^6$ (considering the streamwise and transverse velocity component).
Following the CROM strategy described in Sec.~\ref{Sec:CROM} and outlined in Fig.~\ref{Fig:CROM_Schematic}, 
the labels $\{L(t_m)\}_{m=1}^M$, affiliating each velocity field with a cluster and the cluster transition probability matrix (CTM), here denoted by $\matr{Q}$, are determined. 
This is the reference to which the results based on random measurements will be compared, and is in the following referred to as full-state CROM.
We consider two different sensing matrices to obtain incoherent measurements $\vec{y} = \matr{\Phi}\vec{u}$:
a Gaussian random matrix,  which can be generated using, e.g., the \texttt{randn} command in Matlab, 
and random point measurements generated from a random selection of rows of the $N\times N$ identity matrix.
In particular, we consider three cases:
\begin{itemize}\setlength\itemsep{0em}
	\item[(A)] Gaussian sensing matrix and keeping the clustering fixed,
	\item[(B)] Gaussian sensing matrix and re-clustering of measurements $\vec{y}$, and
	\item[(C)] random point measurements and re-clustering of measurements $\vec{y}$.
\end{itemize}
In case A,  centroids, cluster affiliation and CTM are re-computed from measurements using the reference labels $\{L(t_m)\}_{m=1}^M$. 
This is advantageous if CROM is learned offline on high-dimensional data for use in a real-time application with few measurements. 
In contrast, cases B and C follow the sparsity-enabled CROM strategy as shown in Fig.~\ref{Fig:sparseCROM_Schematic}, where CROM is directly learned from the measurements.

The probabilistic cluster dynamics described by the CTMs, namely $\matr{P}$ from measurements $\vec{y}$ and reference $\matr{Q}$ from the full state,
are compared via the Jensen-Shannon divergence (JSD) \cite{Lin1991ieee}
\begin{equation}\label{Eqn:JSD}
JSD(\matr{P},\matr{Q}) =  \frac{1}{2} D_{KL}(\matr{P},\matr{M}) + \frac{1}{2} D_{KL}(\matr{Q},\matr{M})\quad \text{with}\quad \matr{M} = \frac{1}{2}(\matr{P}+\matr{Q})
\end{equation}
where $D_{KL}$ denotes the Kullback-Leibler divergence \cite{Kullback1951ams,Kullback1959book,Noack2012jfm} defined by
\begin{equation}
D_{KL}(\matr{P},\matr{Q}) = \sum\limits_{i=1}^{N_c} \sum\limits_{j=1}^{N_c}\, P_{ij} \log{\frac{P_{ij}}{Q_{ij}}}\, .
\end{equation}

\begin{figure}[tb]
	\centering
	\includegraphics[width=\textwidth]{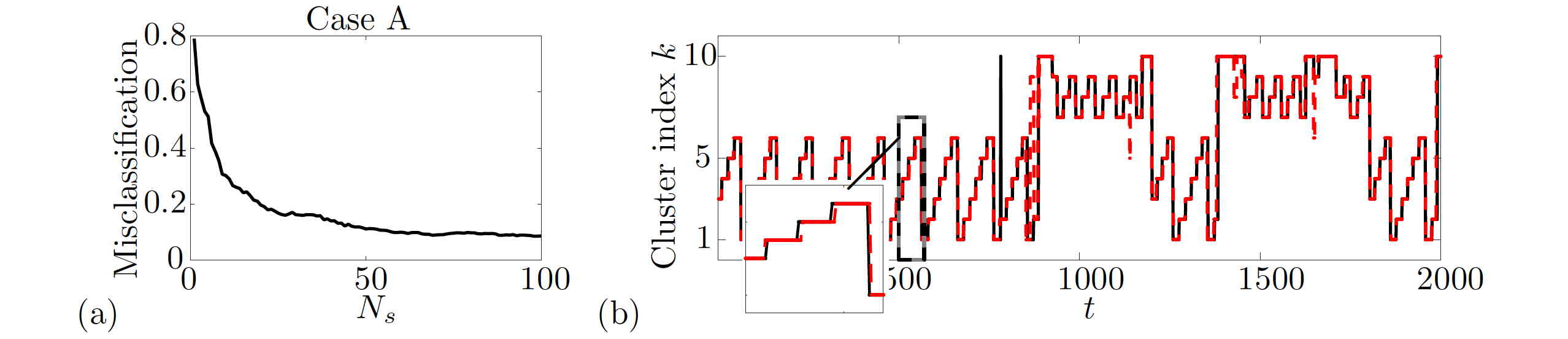}
	\caption{Classification error (a) using a Gaussian sensing matrix and (b) example time history of the cluster affiliation of full-state features (black line) and recomputed from $100$ measurements (red dashed lines).
		Misclassification mainly occurs at the cluster borders as visible in the zoomed window.}	
	\label{Fig:sCROM:randn:keep:class_error}	
\end{figure}

\begin{figure}[tb]
	\centering
	\includegraphics[width=\textwidth]{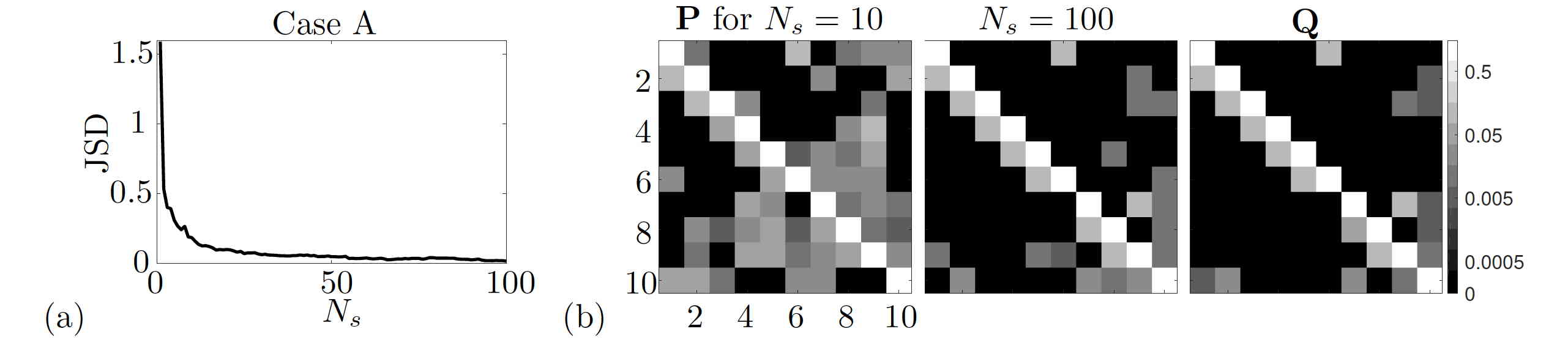}
	\caption{Sparse CROM from incoherent measurements:
		(a) Jensen-Shannon divergence comparing CTMs $\matr{P}$ estimated from measurements to the reference  $\matr{Q}$. (b) Select transition matrices $\matr{P}$ are plotted for $N_s = 50$, $N_s=100$ and reference  $\matr{Q}$ (from left to right).
		Transition probabilities are displayed for better visualization in logarithmic scale ranging from zero probability ($\blacksquare$) to  probability of $1$ ($\square$).}	
	\label{Fig:sCROM:randn:keep:CTM}
\end{figure}

The classification error as a function of the number of measurements for case A is 
shown in Fig.~\ref{Fig:sCROM:randn:keep:class_error}.
The following steps are performed to compute the cluster affiliation and CTM from measurements:
(1) centroids are re-computed from measurements using labels  $\{L(t_m)\}_{m=1}^M$,
(2) the cluster affiliation is updated based on the nearest-centroid method using the cluster centroids from (1),
(3) the CTM is re-computed, $\matr{P}$, based on the cluster affiliation from (2).
The error decays rapidly if more measurements are used. 
An example time history of the cluster affiliation closely matches that of the reference.
The JSD decays analogously and the CTM converges to the reference CTM $\matr{Q}$
with increasing number of measurements (see Fig.~\ref{Fig:sCROM:randn:keep:CTM}).

\begin{figure}[tb]
	\centering
	\includegraphics[width=\textwidth]{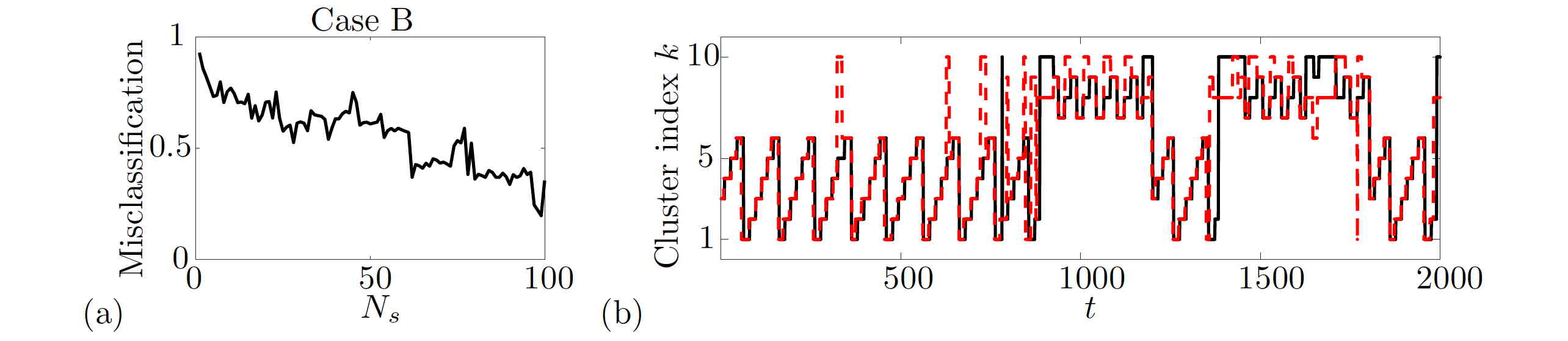}
	\caption{Analogous to Fig.~\ref{Fig:sCROM:randn:keep:class_error} but measurements are reclustered, 
		showing (a) classification error and (b) time history of cluster affiliation for $100$ measurements (red dashed lines) and reference (black line).
		Cluster indices are renumbered to match reference as good as possible.}	
	\label{Fig:sCROM:randn:new:class_error}		
\end{figure}

\begin{figure}[tb]
	\centering
	\includegraphics[width=\textwidth]{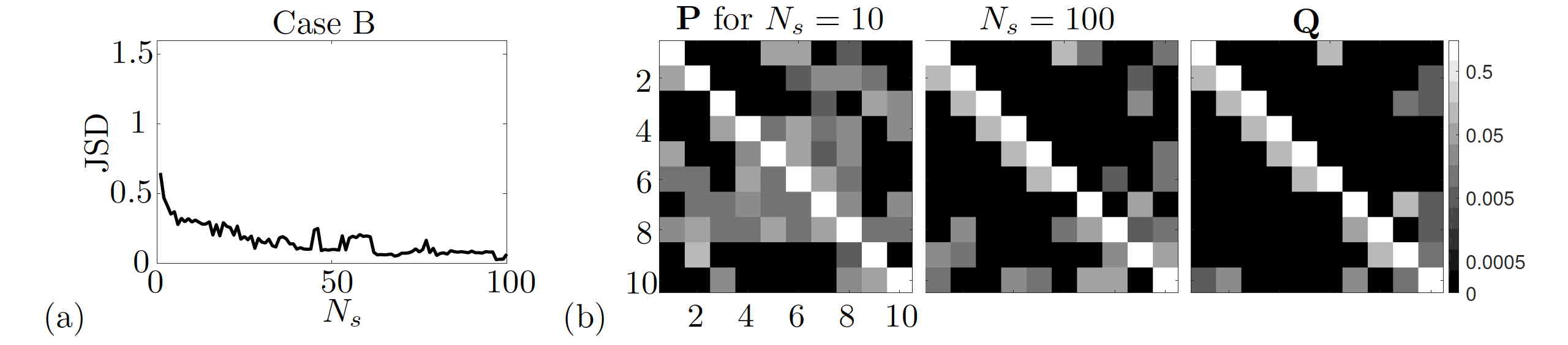}
	\caption{Sparse CROM from incoherent measurements analogous to Fig.~\ref{Fig:sCROM:randn:keep:CTM} but based on the clustering in Fig.~\ref{Fig:sCROM:randn:new:class_error}:
		(a) Jensen-Shannon divergence and (b) select transition matrices for $N_s = 10$, $N_s=100$ and reference $\matr{Q}$ (from left to right).}	
	\label{Fig:sCROM:randn:new:CTM}
\end{figure}

In the compressed sensing framework, 
few, but sufficiently many, incoherent measurements preserve geometric properties such as the cluster geometry. 
Thus, clustering algorithms such as k-means shall, in principle, yield the same results when applied directly to the measurements.
As a consequence, the transition probabilities must also be equal to those computed using the POD coefficients.
To facilitate the comparison of the results based on measurements with the full-state reference, we choose the same initial set of centroids in the iteration process of k-means.
However, small differences in the pairwise distances between observations will inevitably lead to different final clustering results; the location of the final set of centroids will be different compared to the reference.
Nevertheless, if sufficiently many measurements are taken, the cluster partition should converge.
Further, the numbering of the clusters may change, thus the clusters computed from $\vec{y}$ are renumbered to match the full-state clusters as close as possible.
The classification error and an example time history of the cluster affiliation is shown in Fig.~\ref{Fig:sCROM:randn:new:class_error}.
Despite being generally higher, the classification error shows the expected decay with increasing number of measurements.
Similarly, the CTM $\matr{P}$ converges to the true CTM $\matr{Q}$ 
while the JSD decreases as shown in Fig.~\ref{Fig:sCROM:randn:new:CTM}.

\begin{figure}[tb]
	\centering
	\includegraphics[width=\textwidth]{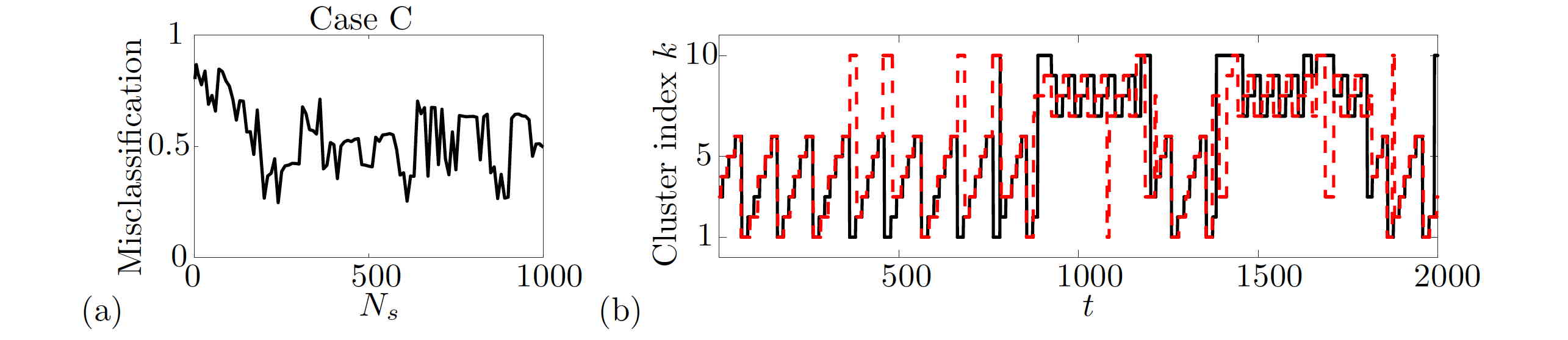}
	\caption{Analogous to Fig.~\ref{Fig:sCROM:randn:keep:class_error} but measurements are reclustered, 
		showing (a) classification error and (b) time history of cluster affiliation for $1000$ measurements (red dashed lines) and reference (black line).
		Cluster indices are renumbered to match reference as good as possible.}	
	\label{Fig:sCROM:point:new:class_error}		
\end{figure}

\begin{figure}[tb]
	\centering
	\includegraphics[width=\textwidth]{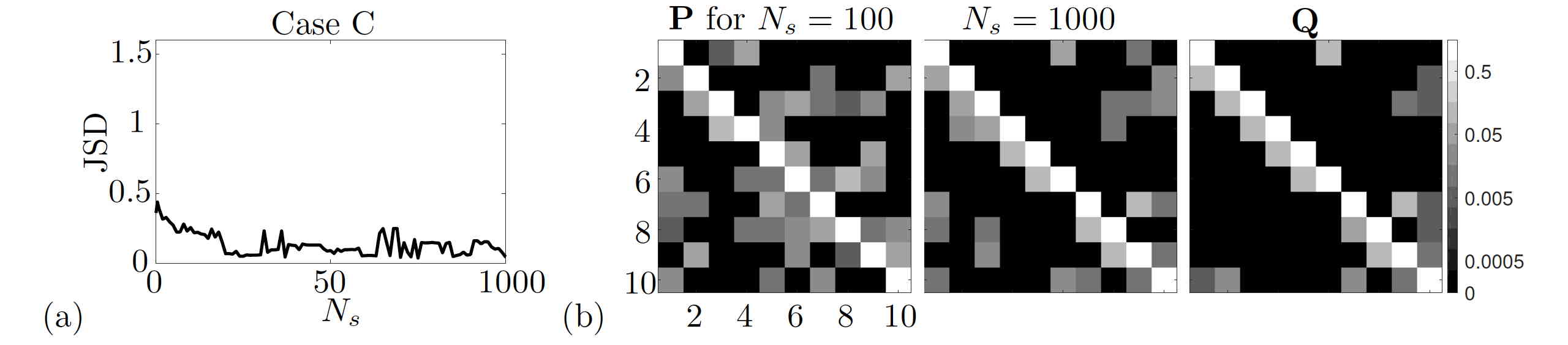}
	\caption{Sparse CROM from incoherent measurements analogous to Fig.~\ref{Fig:sCROM:randn:keep:CTM} but based on the clustering in Fig.~\ref{Fig:sCROM:randn:new:class_error}:
		(a) Jensen-Shannon divergence and (b) select transition matrices for $N_s = 100$, $N_s=100$ and reference $\matr{Q}$ (from left to right).}	
	\label{Fig:sCROM:point:new:CTM}
\end{figure}

In the following, more realistic measurements are considered corresponding to point measurements.
The classification error for case C and an example time history of the cluster affiliation based on $N_s=1000$ random point measurements at each time instant are shown in Fig.~\ref{Fig:sCROM:point:new:class_error}	.
Note that for this example up to $1000$ measurements are considered.
The classification error does not decrease as rapidly as in the previous examples, as
(1) single point measurements contain less information than Gaussian random measurements, which are obtained from taking the dot product between the Gaussian random matrix and the full state,
(2) the position change of the centroids also affects the renumbering procedure of the clusters introducing an error in the cluster affiliation. 
The fluctuation in the misclassification increases due to the strong dependency of the measurements on the selection of sensor locations in the sensing matrix.
Despite these weaknesses, the CTM converges to the full-state CTM if sufficiently many measurements are collected
(see Fig.~\ref{Fig:sCROM:point:new:CTM}).

In conclusion, we have shown that CROM from few, incoherent measurements preserves the cluster geometry and 
and topological structure of the transition probability matrix.
Thus, the same probabilistic dynamics are identified if sufficiently many measurements are collected. 
More generally, the computational cost of k-means clustering can be reduced if compressive measurements are employed.

\subsection{Sparse sensor placement optimization for CROM}
\label{Sec:SSPOCCROM}
Sparsity-enabled CROM makes possible more efficient computations using fewer measurements. 
While Gaussian random measurements are very suitable from a compressed sensing viewpoint, these are not suitable for realistic applications. 
In contrast, random point measurements can be interpreted as physically realizable individual sensors.
However, their random selection does not guarantee good performance.  
Moreover, sufficiently many measurements have to be collected to preserve the cluster geometry.

\begin{figure}[htb]
	\centering
	\begin{overpic}[width=\textwidth]{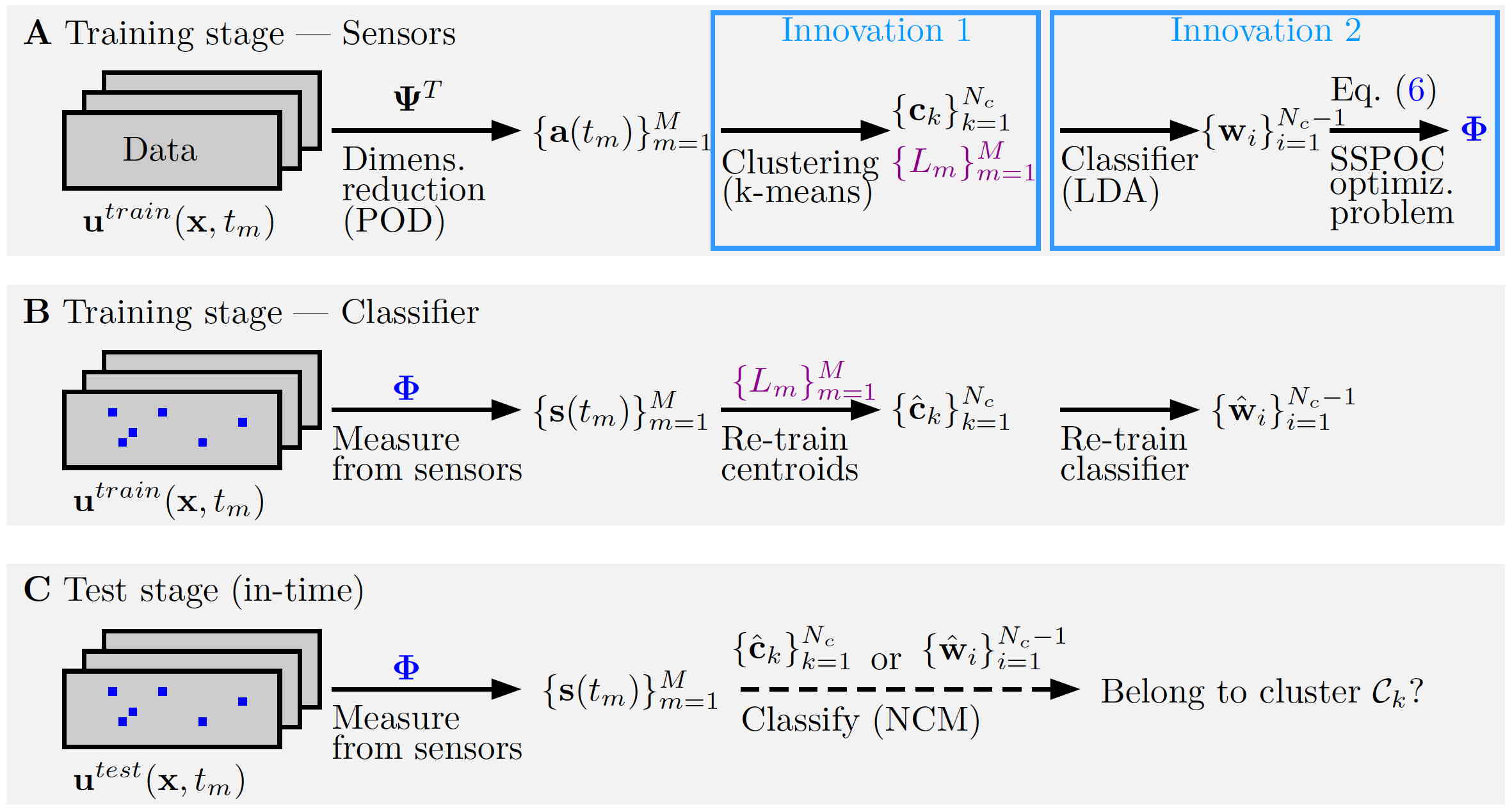}
		\put(88,47){Eq.~\eqref{Eqn:SSPOC_nonbinary}}
	\end{overpic}	
	\caption{Schematic of the sparse sensor placement optimization for CROM 
		showing the different training stages A and B as well as the subsequent (possibly real-time) application of the sensors in C. See text for details.}
	\label{Fig:SSPOCandCROM}
\end{figure}

Optimized sensor locations tailored to the specific CROM can yield improvements in accuracy, while decreasing the number of sensors.
SSPOC has been demonstrated to find few optimized sensors for accurate classification based on a pre-trained supervised classification scheme~\cite{Brunton2016,Bai2017}.
While CROM learns an intrinsic data partitioning, SSPOC exploits a known partition to find a minimal number of sensors that are most informative for discriminating those classes. 
Thus, combining CROM with SSPOC is particularly suitable and allows one to unify their respective merits.
A schematic of the sparse sensor placement strategy for CROM facilitated by SSPOC is outlined in Fig.~\ref{Fig:SSPOCandCROM}.

Both SSPOC and CROM start with a dimensionality reduction procedure such as POD.
Diverging from the standard procedures for SSPOC (compare Fig.~\ref{Fig:SSPOC_Schematic}) 
and CROM (compare Fig.~\ref{Fig:CROM_Schematic}), 
two key innovations are implemented: 
(1) K-means clustering is integrated into SSPOC as an intermediate step. 
This enables SSPOC to learn sensors in an unsupervised manner, where classes of the data are unknown and must be first discovered using an unsupervised clustering algorithm, such as k-means.
(2) The partitioning of CROM in conjunction with a supervised classifier, such as LDA, allows one to solve the $l_1$ optimization problem of SSPOC~\eqref{Eqn:SSPOC_nonbinary} to learn few optimized sensors that are key for discriminating the clusters. 

The scheme for learning a CROM and subsequent sensor optimization follows three stages as shown in Fig.~\ref{Fig:SSPOCandCROM}.
In training stage A, which is performed offline, time-series data from a high-dimensional systems is analyzed. 
The standard CROM procedure can then be applied (see Fig.~\ref{Fig:CROM_Schematic}) to the full-state data.
The cluster affiliation given by the labels $\{ L(t_m) \}_{m=1}^M$,
resulting from the learned state-space partitioning using k-means,
is provided to SSPOC yielding few optimized sensor locations. 
While CROM is trained on all features, depending on the dimensionality of the data and its sparsity in that basis, it can be suitable to reduce the number of features considered in the optimization problem, shrinking the computational costs.
Finding a good set of sensors involves two steps: 
First, sensors are determined for varying $\lambda$.
Often the total number of sensor locations found, denoted by $N_s$, reaches a plateau for a particular $\lambda$ value. 
There is generally a trade-off between the number of sensors and achieved accuracy which has to be taken into account when choosing $\lambda$. 
Second, the number of sensors can be further tuned by keeping $\lambda$ fixed, e.g. achieving the largest gain, and instead varying the number of features. 
Alternatively, the number of sensors can be adapted by sweeping through the error tolerance $\varepsilon$ in the optimization (see \eqref{Eqn:SSPOC_nonbinary}).
The $N_s$ sensor locations correspond to rows in $\vec{s}$ which have at least one non-zero entry. 
In practice, these can be found by applying the threshold $\vert s_{ij}\vert \geq \frac{\vert\vert {\bf s}\vert\vert_F}{2N_c N_f}$ \cite{Brunton2016} in order to construct the sensing matrix $\matr{\Phi}$.  
For very high-dimensional problems, such as the mixing layer flow, it can be necessary to first randomly subsample the data to make the optimization problem tractable.  

In training stage B, the classifier for discriminating clusters is re-trained in the sensor space.
Using the sensing matrix $\matr{\Phi}$ created in stage A, 
single point measurements $\{\vec{y}(t_m)\}_{m=1}^M$ are collected from the training data.
The LDA and cluster centroid classifiers are then re-trained from the measurements yielding the discriminating projection vectors $\{\hat{\vec{w}}_i\}_{i=1}^{N_c-1}$
and centroids $\{\hat{\vec{c}}_k\}_{k=1}^{N_c}$. 
Retraining classifiers in the sensor space is recommended, as this generally increases classification accuracy. 
This is done in all examples.

The last stage C marks the online phase, where in-time measurements are collected from the sensor locations and the prevailing cluster is determined. 
Thus, a few point measurements of the high-dimensional state are measured and subsequently classified into a cluster in conjunction with the classifiers. 
This is a critical enabler for low-latency in-time estimation and response.
While the classification based on the closest cluster centroid seems more natural for the considered problem, it can be advantageous to employ the LDA classifier in the sensor space, 
as sensor locations are optimized with regard to how well these reconstruct the discriminating projection vectors.
However, if sensors are learned on subsampled data, we have found that the nearest-centroid method applied to the cluster centroids in the sensor space can achieve higher accuracy.
In all examples, we compare the performance of the learned SSPOC sensors with random sensors and sensors corresponding to the pivot locations from the QR factorization~\cite{Drmac2016siam-jsc} of the transform basis $\matr{\Psi}$, which are referred to as QRcp sensors.

The combination of CROM with SSPOC is critical for making CROM applicable in realistic configurations that require in-time prediction, estimation, and control.
Our innovations facilitate the learning of a minimal number of optimized sensor locations tailored to a specific CROM to achieve maximal performance and, more generally, specifically targeted towards dynamical systems. 
Moreover, this generalization of SSPOC to unsupervised classifiers enables sensor placement for classification problems in an unsupervised manner.

\section{Results}
\label{Sec:Results}


We examine sensory placement for three examples from fluids that address different challenges (see Tab.~\ref{Tab:OverviewExamples}). 
\begin{table}[tb]
	\centering
	\begin{overpic}[width=\textwidth]{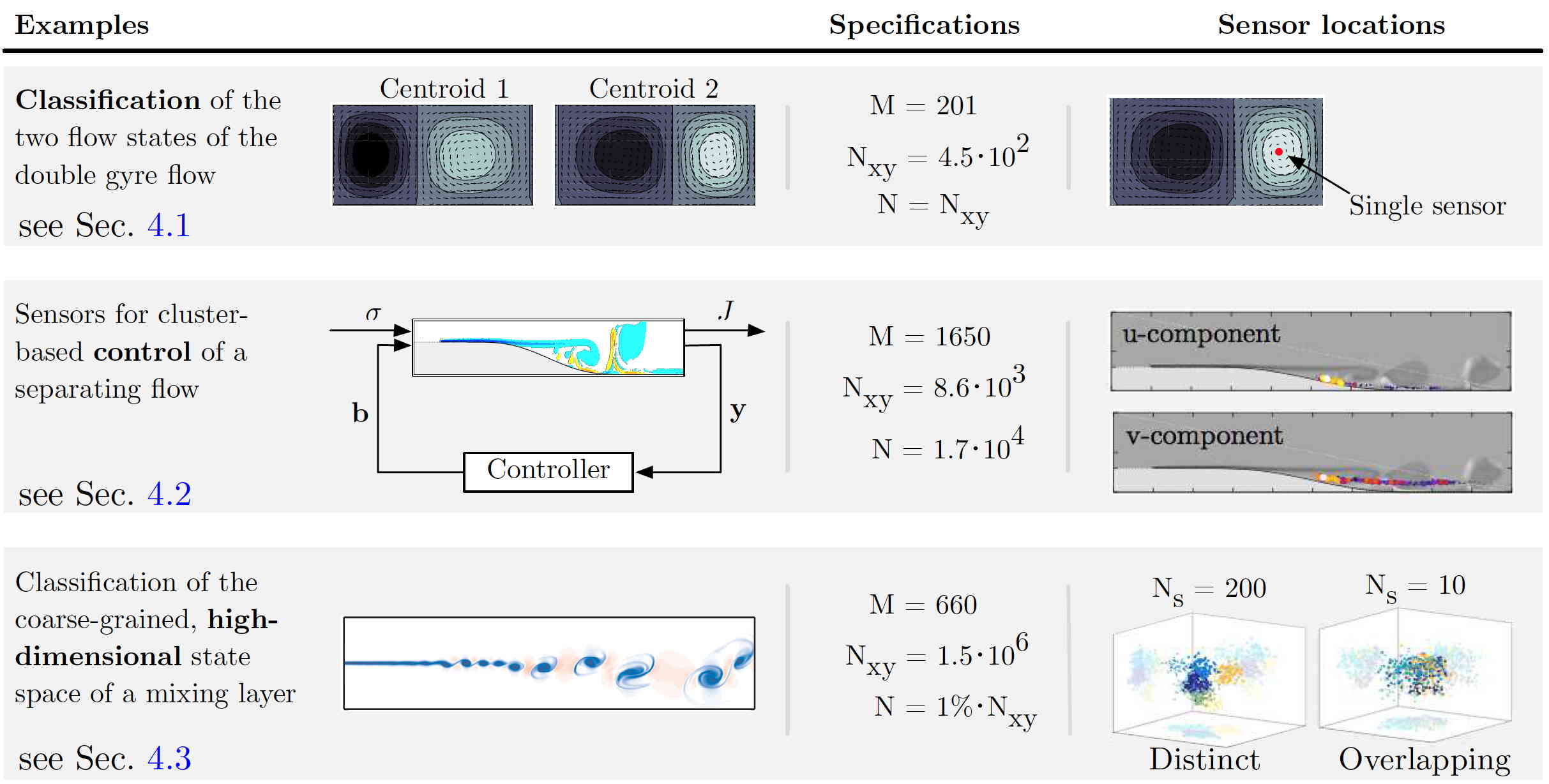}
		\put(1,35.5){\colorbox[rgb]{0.95,0.95,0.95}{see Sec.~\ref{Sec:DoubleGyreFlow}}}
		\put(1,18.5){\colorbox[rgb]{0.95,0.95,0.95}{see Sec.~\ref{Sec:SeparatingFlow}}}
		\put(1,1.25){\colorbox[rgb]{0.95,0.95,0.95}{see Sec.~\ref{Sec:MixingLayer}}}
	\end{overpic}	
	\caption{Overview of numerical examples.}
	\label{Tab:OverviewExamples}
\end{table}
The first example is the periodic double gyre flow of moderate dimension which serves as illustrative example. The state space is discretized into two clusters, each associated with the contraction and expansion of the two vortices.
The second flow system, a separating flow over a smoothly contoured ramp, has been previously employed as testbed for cluster-based control building on CROM \cite{Kaiser2016tcfd}.
Here, the control performance using optimized sensors is compared to full-state measurements.  
The third example is the spatially developing mixing layer, which exhibits high-dimensionality and strains computational resources.   
Thus, the optimization of sensors is facilitated by  using heavily subsampled data.

For cross-validation purposes, all datasets are first split into a training and test set; clustering and sensor placement is then learned on the training set and performance is assessed on the test set. 
The subsampling percentage of the data on which the sensors are learned is denoted $S\%$. 
For each example, the number of snapshots $M$, 
the number of spatial grid points $N_{xy}$ (all flow problems are two-dimensional), and the number of potential sensor locations $N$ are given in Tab.~\ref{Tab:OverviewExamples}.
Note that $N=2\,N_{xy}$ for the separating flow, as sensor placement distinguishes between the streamwise and transverse velocity component. 
In the double gyre and the mixing layer flows, vorticity snapshot data is considered.
The specifications and parameters for all cases are provided in Tab.~\ref{Tab:OverviewExamples_specs}.
\begin{table}[tb]
	\centering
	\begin{tabular}{L{0.2\textwidth}L{0.1\textwidth}L{0.2\textwidth}L{0.12\textwidth}L{0.11\textwidth}L{0.06\textwidth}}
		\textbf{Example} & \textbf{Clusters} & \textbf{SSPOC} & \textbf{QRcp} & \textbf{Random} & \textbf{Train/}\textbf{Test [\%]}\\\hline\\[0.5ex]
		\textbf{Double gyre flow}: Sec.~\ref{Sec:DoubleGyreFlow} 
		& $N_c = 2$ & $S\% = 100\%$ & $S\% = 100\%$ & & \\[1ex]
		& $N_f = 10$ &  $N_f \in [1,10]$ &  $N_f \in [1,10]$ & $N_s \in [1,10]$ & $80/20$\\[2ex]\arrayrulecolor[rgb]{0.9,0.9,0.9}\hline\\[1ex]
		\textbf{Separated flow}: Sec.~\ref{Sec:SeparatingFlow} 	  &  &  & & &\\[1ex]
		\hspace{0.5cm}Case 1: Sec.~\ref{Sec:Ramp2u:OSP-CI5} 		
		& $N_c = 10$ & $S\% = 100\%$ & $S\% = 100\%$ & & \\[1ex]
		& $N_f = 1480$ & $\lambda\in [0,10^6],\;N_f=20$   & $N_f \in [1,180]$ & $N_s \in [1,180]$ & $90/10$\\[1ex]
		&                                 & $N_f \in [1,180], \;\lambda=100$ & & & $90/10$\\[1ex]
		{\color{white}\rule[0mm]{3.5cm}{0.5pt}}{\color[rgb]{0.9,0.9,0.9}\rule[0mm]{11.7cm}{0.5pt}}\\[2ex]
		\hspace{0.5cm}Case 2: Sec.~\ref{Sec:Ramp2u:OSP-CROMc}		
		& $N_c = 10$ & $S\% = 100\%$ & $S\% = 100\%$ & &\\[1ex]
		& $N_f = 10$ & $\lambda\in [0,10^6],\;N_f=10$   & $N_f \in [1,10]$ & $N_s \in [1,90]$ & $90/10$\\[1ex]
		&                                 & $N_f \in [1,10], \;\lambda=10$ & & & $90/10$\\[1ex]\arrayrulecolor[rgb]{0.9,0.9,0.9}\hline\\[1ex]
		\textbf{Mixing layer}: Sec.~\ref{Sec:MixingLayer}
		& $N_c = 10$ & $S\% = 1\%$ & $S\% = 100\%$ & &\\[1ex]
		& $N_f = 600$ & $\lambda\in [0,10^6],\;N_f=40$ & $N_f \in [1,500]$& $N_s \in [1,500]$ & $90/10$\\[1ex]
		&                     & $N_f \in [1,40], \;\lambda=10$ & & & $90/10$\\[2ex]\arrayrulecolor[rgb]{0,0,0}\hline\\[2ex]
	\end{tabular}	
	\caption{Overview of specifications for CROM and SSPOC. For cross-validation the datasets have been split into training and test sets. Statistics are computed over $N_r=100$ random reshuffling of the training and test sets.}
	\label{Tab:OverviewExamples_specs}
\end{table}
The number of clusters is set to $N_c=2$ for the double gyre flow to identify the contraction and expansion behavior of the vortices.
In the remaining examples, $N_c=10$ clusters are used, motivated by the choice in previous work to which the results are compared.
The clusters are trained in the POD space where the number of employed features is denoted by $N_f$.

\subsection{Periodically driven double gyre flow as illustrative example}
\label{Sec:DoubleGyreFlow}
\begin{figure}[tb]
	\centering
	\includegraphics[width=\textwidth]{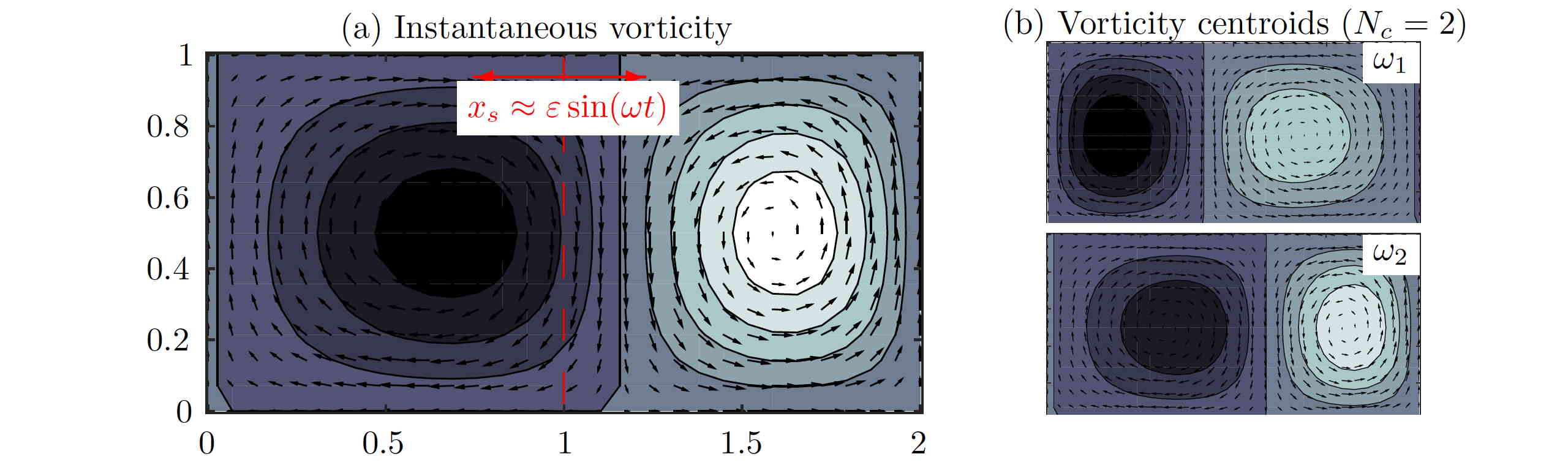}
	\caption{Periodic double gyre flow: (a) vorticity contours and velocity vectors of an instantaneous realization and
		(b) vorticity centroids for $N_c=2$ clusters.
	}
	\label{Fig:DoubleGyreFlow}
\end{figure}

The periodically driven double gyre flow models the transport  between convection rolls in the Rayleigh-B\'enard convection due to lateral oscillation, e.g.\ as a simple model for the gulf stream ocean front \cite{Solomon1988pra}. We employ here the same parameters as in Shadden's seminal work \cite{Shadden2005physica} on Lagrangian coherent structures.
Consider the stream function defined by
\begin{equation}
\psi(x,y,t) = A\sin(\pi f(x,t)) \sin(\pi y)
\end{equation}
with $f(x,t) = \varepsilon \sin(\omega t)\, x^2 + (1 - 2 \varepsilon \sin(\omega t))\, x$, 
where $A=0.25$, $\varepsilon=0.25$, and $\omega=2\pi/10$ are fixed parameters,
over the domain $\Omega = \{(x,y) \vert 0\leq x \leq 2,\, 0\leq y \leq 1  \}$,
discretized to obtain $N_x=30$ and $N_y=15$ grid nodes in the horizontal and vertical directions, respectively.
The parameter $\varepsilon$ 
represents the amplitude of the periodic oscillation, which yields a steady flow
for $\varepsilon = 0$ and oscillating flow for $\varepsilon > 0$.  
A visualization of the instantaneous vorticity is displayed in Fig.~\ref{Fig:DoubleGyreFlow}(a).
The separatrix between the two convection cells oscillates periodically with $\omega$, 
leading to a periodic expansion and contraction of the vortex cells.
These two dynamical regimes are identified in an unsupervised manner using the k-means clustering algorithm. 
The vorticity centroids of the two identified clusters,
denoted by $\boldsymbol{\omega}_1$ and $\boldsymbol{\omega}_2$,  
are shown in Fig.~\ref{Fig:DoubleGyreFlow}(b) and (c), respectively.
In the following, sparse sensor locations are learned to distinguish between these two states.

\begin{figure}[bt]
	\centering
	\includegraphics[width=\textwidth]{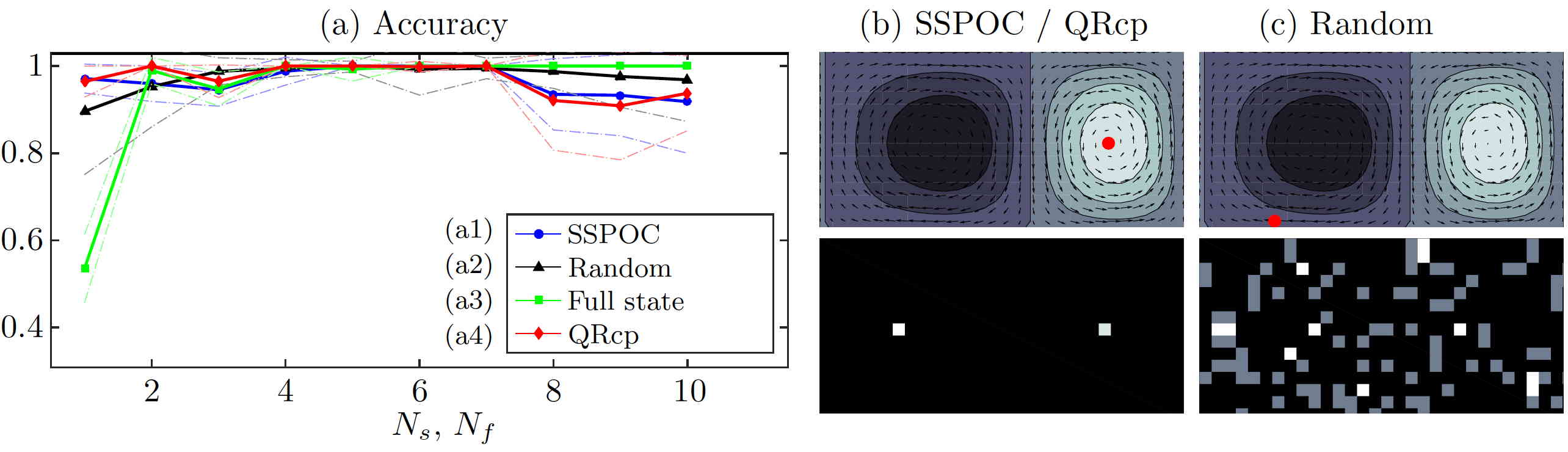}
	\caption{Classification results for the periodic double gyre flow:
		(a) accuracy for different numbers of sensors $N_s\in\{1,\ldots,10\}$
		for which the placement is determined using 
		(a1) SSPOC, 
		(a2) a random selection, or
		(a4) QR with column pivoting. 
		The results are compared with the accuracy for full-state sensors with increasing number of features $N_f$ (a3).
		The mean and standard deviation of the cross-validated accuracy are shown as solid and dashed lines, respectively.
		The optimal sensor location ($\color{red}{\bullet}$) for a single sensor (b)(top)
		lies slightly outwards of the vortex core when the vortices are symmetric.
		The symmetry of the problem results in two optimal locations with equal probability for a single sensor (see (b)bottom)
		which can be immediately found using SSPOC or QRcp in contrast to random sensor locations (c) or an exhaustive search.
	}
	\label{Fig:DG_results}
\end{figure}

The ability of vorticity sensors to classify the two dynamical regimes is shown in Fig.~\ref{Fig:DG_results}(a). 
In all cases, the accuracy improves with an increasing number of features $N_f$ or sensors $N_s$. 
The accuracy reaches a plateau at $N_s=N_f=5$, and decreases at $N_s=N_f=8$ due to overfitting.
SSPOC and QRcp achieve an average accuracy of $97.12\%$ and $96.44\%$, respectively, for a single point sensor. 
In contrast, the full-state projected onto a single POD mode achieves an average classification accuracy of $54\%$.
One example is shown in Fig.~\ref{Fig:DG_results}(b, top),
where the single point sensor (red circle) is located slightly off the center of one of the vortex cores. This sensor achieves $100\%$ accuracy.
In Fig.~\ref{Fig:DG_results}(b, bottom), the probability distribution of sensor locations found by SSPOC or QRcp (which are identical in this particular case) with a single feature ($N_f=1$), and hence a single sensor ($N_s=1$), is shown. Due to the symmetry of the problem, there are two optimal locations close to each of the vortex cores, which are found with equal probability. 
These sensor locations can be easily determined using SSPOC or QRcp in contrast to 
random sensors (see Fig.~\ref{Fig:DG_results}(c)) or brute-force search.

\begin{figure}[tb]
	\centering
	\includegraphics[width=\textwidth]{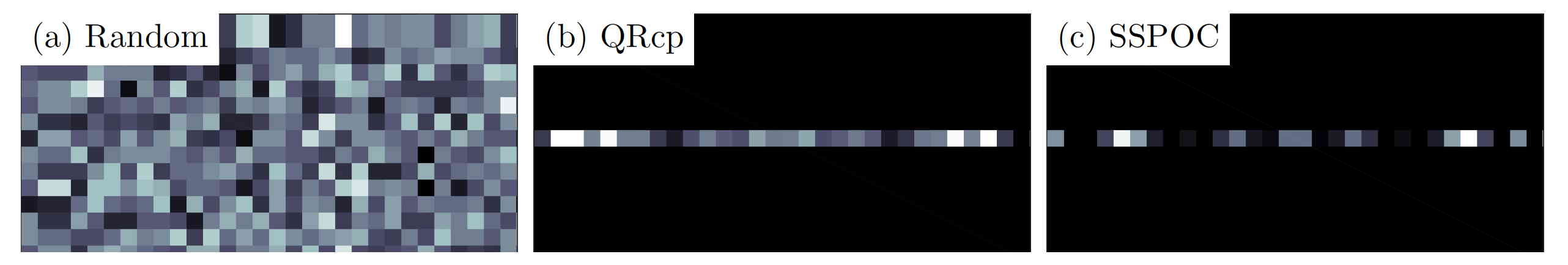}
	\caption{Probability of sensor locations for (a) random and optimal placment using (b) QR with column pivoting or (c) SSPOC.}
	\label{Fig:DG_ProbAll}
\end{figure}

The overall probability distributions of all sensor locations found for random, QRcp, or SSPOC sensor selection, are shown in Fig.~\ref{Fig:DG_ProbAll}.
The selection of sensor locations should be guided by the sensing or decision task. Here, sensors should be maximally informative observables with respect to the prevailing dynamical regime represented by the cluster.  
Both QRcp and SSPOC yield learned sensors along the horizontal center line, for which the double gyre flow exhibits a reflection symmetry. 
While the sensors found by QRcp are more equally distributed along that line, 
sensors found by SSPOC clearly favor the two distinct locations close to the vortex cores.
Overall, QRcp and SSPOC perform  equally well in this introductory example.

\subsection{Separating flow over a smooth ramp: Towards in-time control}
\label{Sec:SeparatingFlow}

\begin{figure}[tb]
	\centering
	\includegraphics[width=1\textwidth]{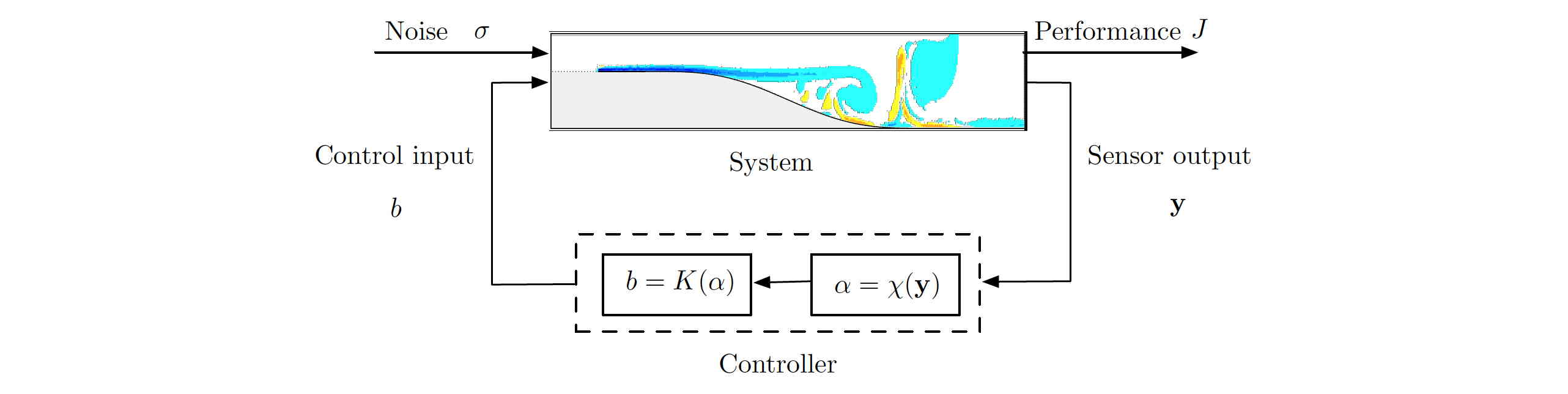}
	\caption{Schematic of cluster-based feedback control loop for the separating flow over a smoothly contoured ramp.}
	\label{Fig:FeedbackLoop}
\end{figure}

Sensor placement is studied for a controlled separating flow over a smooth ramp (see Fig.~\ref{Fig:FeedbackLoop}) governed by typical Kelvin-Helmholtz shedding with Reynolds number $Re = U_{\infty} L/\nu=7700$ based on the inflow velocity $U_{\infty}$, the length of the flat plate $L$ upstream of the curved wall, and the kinematic viscosity $\nu$. 
Learning optimized sparse sensor locations dramatically reduces the computational overhead in the online sensing and classification, reducing latency and improving bandwidth of in-time control.   
In the present study, we seek to identify few sensors that discriminate between different clusters (see Sec.~\ref{Sec:Ramp2u:OSP-CI5}) which is a key enabler for cluster-based control in experimental applications. 
Cluster-based control using a control-oriented CROM on full-state measurements has been previously studied to optimize an open-loop controller based on the optimal periodic excitation frequency for this configuration~\cite{Kaiser2016tcfd}.
In particular, a bang-bang controller is employed, which turns the periodic forcing on or off dependent on the cluster, exploiting the long relaxation times of the flow. 
In Sec.~\ref{Sec:Ramp2u:OSP-CROMc}, 
sensor locations are learned in a subspace specifically tailored towards this cluster-based control application and the performance of the optimal control laws using sparse sensor measurements and full-state measurements are compared.

\begin{figure}[tb]
	\centering
	\includegraphics[width=\textwidth]{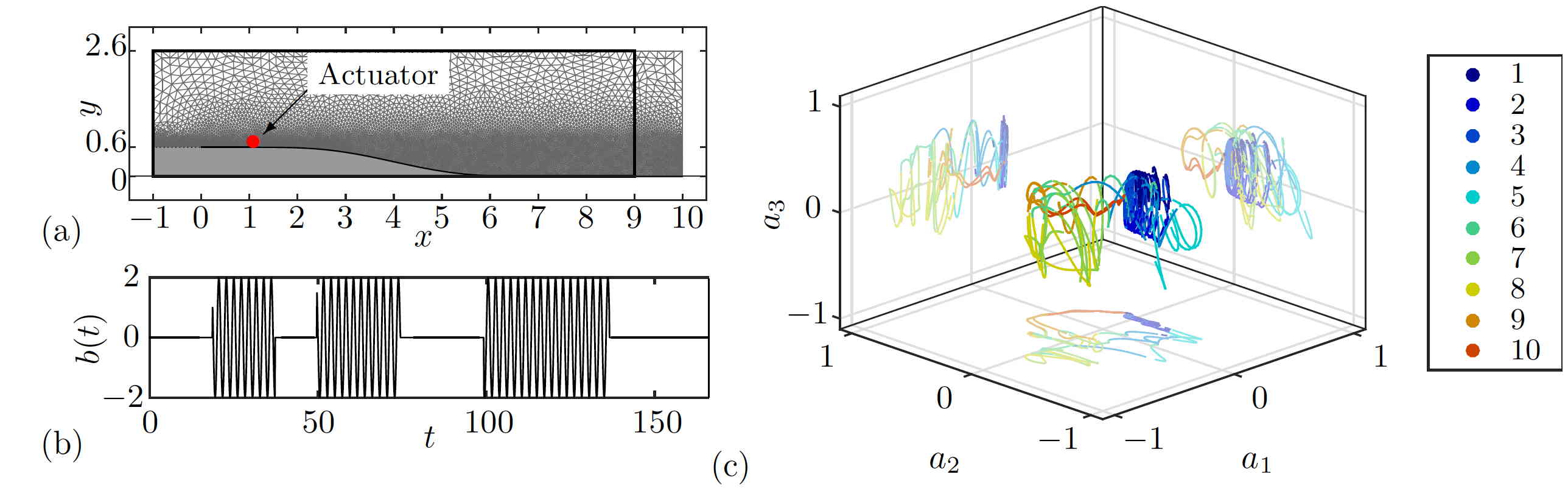}
	\caption{
		Separating flow over a smooth ramp:
		(a) Computational domain with increased resolution in boundary and shear layer region. 
		The location of the volume force is represented by $\color{red}{\bullet}$. 
		(b) Actuation signal applied to probe the natural and forced attractor. 
		(c) Phase plot of the first three POD coefficients ($a_1,a_2,a_3$) colored by cluster affiliation with $N_c=10$ clusters. Observations affiliated with different clusters are not well separated.}
	\label{Fig:Ramp2u:Clusters}
\end{figure}

In the following, we provide a brief description of the unsteady, two-dimensional, incompressible Navier-Stokes solver and data set, both previously described in~\cite{Kaiser2016tcfd}.  
The two-dimensional flow is defined by the velocity vector $\vec{u}(\vec{x}, t) := (u, v)^T$, where $u$ and $v$ are the streamwise and transverse velocity components, respectively.  
The computational domain $\Omega$, shown in Fig.~\ref{Fig:Ramp2u:Clusters}(a), is
discretized using mixed Taylor-Hood elements~\cite{Hood1974book} on an unstructured triangular mesh comprised of $8567$ nodes with increased resolution around the leading edge (located at $(x,y) = (0,0.6)$), 
in the boundary layer and in the shear layer region. 
A quadratic finite-element method formulation is used to discretize the evolution equations
with no-slip boundary on the ramp and stress-free outflow. 
A detailed description of the solver can be found in~\cite{Morzynski1987proc,Afanasiev2003phd}. 
A rectangular velocity profile $\boldsymbol{U}_{\infty} := \boldsymbol{u}(x=-1, y) = (1,0)^T$
is used for the inflow condition.
The numerical time step is $0.005$ and the sampling period of the snapshots is $20$ time steps, i.e. $\Delta t = 0.1$.
The function $b$ denotes the time-dependent control input amplitude, which has compact support in a circular region, centered at $x = 1$ and a $y$-position chosen such that the circular region is mostly inside the boundary layer (displayed as a red circle in Fig.~\ref{Fig:Ramp2u:Clusters}(a)).  

The curvature of the wall induces an adverse pressure gradient leading to flow separation. The developing free shear layer is convectively unstable giving rise to the Kelvin-Helmholtz instability~\cite{Ho1984arfm}. 
Behind the ramp a large recirculation area forms, the reduction of which benefits drag and lift forces.  
The recirculation area is here approximated by the area where the streamwise velocity component is negative. 
The corresponding time average of the recirculation area is defined by
\begin{equation}
\langle R(t) \rangle_T = \frac{1}{T_2-T_1} \int\limits_{T_1}^{T_2} \int\limits_{\Omega} H(-u({\bf x}))(t) \mathrm d{\bf x}\mathrm dt
\end{equation}
where $H$ denotes the Heaviside function. 
The recirculation area can be largely reduced by open-loop periodic forcing with excitation frequency close to the shedding frequency. 
In previous work~\cite{Kaiser2016tcfd}, a cluster-based feedback controller was developed to optimize this open-loop forcing by turning the actuation on or off depending on the prevailing cluster exploiting the fact that this flow exhibits long relaxation times. 

The particular dataset considered (see~\cite{Kaiser2016tcfd}) consists of instantaneous velocity fields, for which open-loop forcing is randomly turned on and off (see Fig.~\ref{Fig:Ramp2u:Clusters}(b)) with the optimal excitation frequency, $f_p=0.45$, known to achieve the smallest mean recirculation area. 
The data, comprised of $M=1650$ velocity snapshots, is stacked into a matrix and reduced using POD.
The data is clustered in the POD feature space into $N_c=10$ clusters. A representative clustering result, showing the phase plot of the first three POD coefficients with color-coded cluster affiliation, is displayed in Fig.~\ref{Fig:Ramp2u:Clusters}(c). 
The yellow/green-colored clusters represent flow realizations without forcing, while the dark blue clusters represent flow realizations with forcing and the corresponding lock-in between the flow and actuation.  
Transients between the natural and forced flows are colored in orange and light blue.
Despite abruptly switching the actuation on or off, the flow varies smoothly and hence the data points are not well separated into different clusters.  

In the cluster-based control loop, depicted in Fig.~\ref{Fig:FeedbackLoop},  
sensor measurements $\vec{y}$ are fed into the controller, which first determines the prevailing cluster $\alpha=\chi(\vec{y})$, where $\chi$ is a characteristic function affiliating a measurement with a particular cluster, and then enacts the next control input $b$. 
The control law $K$ is a piecewise constant function of the cluster index. 
The optimal control law with respect to a cost function can be determined using a control-oriented CROM.
The performance of each control law is evaluated with 
\begin{equation}
J = J_r + \gamma J_b = \langle R(t)\rangle_T + \langle b^2(t) \rangle_T
\end{equation}
where the penalization coefficient is $\gamma\approx 11$ for an equal weighting of the control objective and the input energy. 
We refer to~\cite{Kaiser2016tcfd} for details on the specific control approach and results.  

\subsubsection{Sensor placement for cluster classification}
\label{Sec:Ramp2u:OSP-CI5}
In this section, placement of sparse sensors for the purpose of classifying the full-state velocity fields into clusters is examined for the partitioned dataset described in the previous section.  
The flow switches smoothly between the unforced and controlled flow states. The data is clustered into a larger number of clusters compared with the previous example to resolve the probabilistic dynamics in the state space (we refer to~\cite{Kaiser2014jfm} and~\cite{Kaiser2016tcfd} for details). As flow states arising from the system with and without actuation may occupy the same cluster, classification from few measurements is considerably more difficult. All $8567$ spatial points are considered as potential sensor locations with discrimination between streamwise and transverse velocity components, thus there exist $17134$ potential sensor locations in total. 
The first $N_f\approx 20$ POD features are considered, representing about $90\%$ of the fluctuation energy.  

\begin{figure}[tb]
	\centering
	\includegraphics[width=\textwidth]{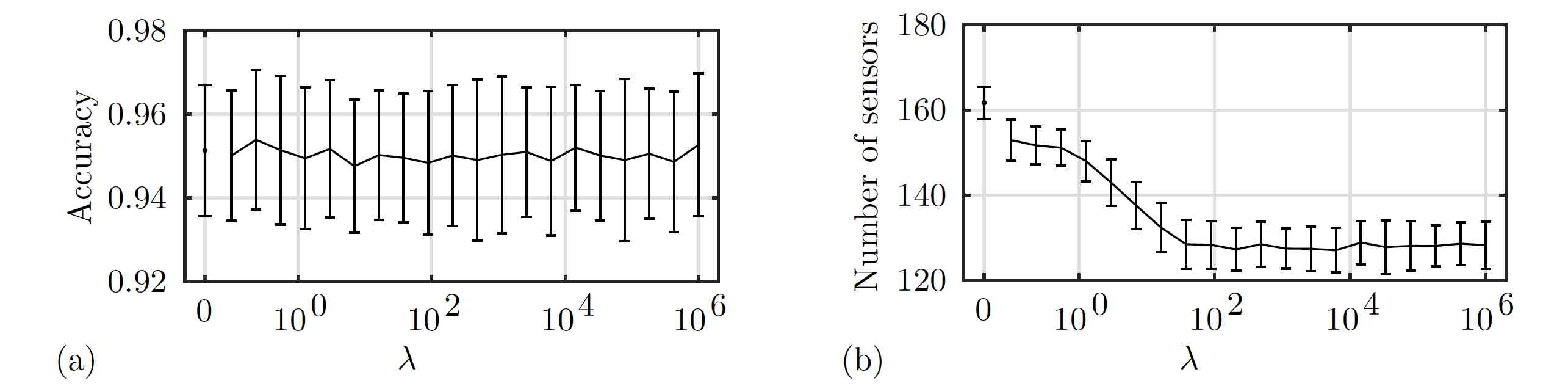}
	\caption{Classification results for $N_c=10$ clusters where $\lambda$ is varied between $0$ and $10^6$. 
		The accuracy in (a) appears to be independent of $\lambda$, even though the number of sensors (b) decreases. 
		The number of sensors saturates at about $\lambda=10^2$ with $N_s \approx 130$.}
	\label{Fig:Ramp2u:CI5:SSPOC}
\end{figure}

Cross-validated accuracy of SSPOC sensors (see approach in Sec.~\ref{Sec:SSPOC})
is shown in Fig.~\ref{Fig:Ramp2u:CI5:SSPOC}.
The average accuracy of $95\%$ does not change with increasing parameter $\lambda$,
while the number of sensors decreases saturating at about $\lambda=100$
with $N_s\approx 130$ learned sensors on average (corresponding to about $0.76\%$ of all potential sensor locations and $1.5\%$ of the grid points).  

\begin{figure}[tb]
	\centering
	\includegraphics[width=\textwidth]{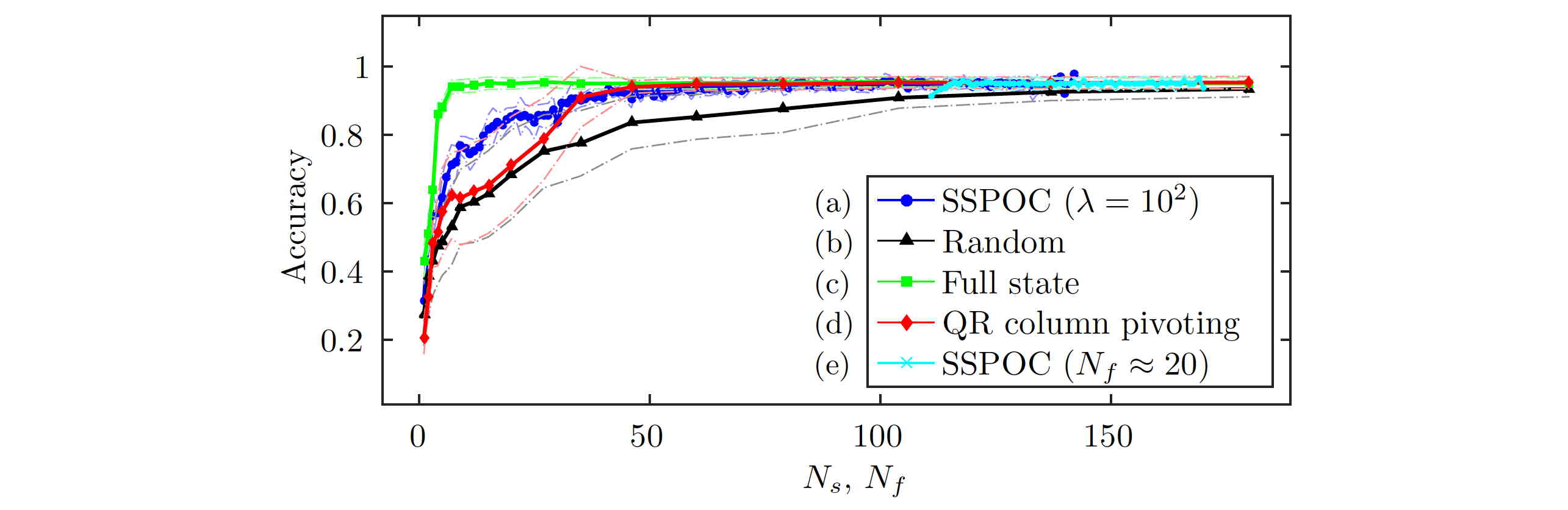}
	\caption{Classification accuracy for $N_c=10$ clusters using 
		(a) SSPOC sensors with varying $N_f$ and fixed $\lambda=10^2$, 
		(b) random sensors,
		(c) full-state sensors for varying $N_f$, 
		(d) QR with column pivoting for varying $N_f$ (which correspond to the number of sensors $N_s$), and
		(e) SSPOC sensors with fixed $N_f$ (results shown in Fig.~\ref{Fig:Ramp2u:CI5:SSPOC}) and varying $\lambda\in [0,10^6]$.
		Except for case (c), where the accuracy is plotted over the number of features $N_f$, 
		in all other cases the accuracy is shown as a function of the number of sensors $N_s$.
		Mean accuracy and its standard deviation are shown in solid and dashed lines, respectively.
		Both, SSPOC and QRcp yield better accuracy than random sensors and saturate at $N_s\approx 50$.
		In contrast to QRcp, SSPOC performs better if using fewer sensors and shows a smaller standard deviation in accuracy.
	}
	\label{Fig:Ramp2u:CI5:Accuracy_all}
\end{figure}

For a fixed $\lambda$, the number of sensors can be further tuned by adapting the number of features $N_f$. 
Cross-validated accuracy for SSPOC with fixed $\lambda=100$ is presented in Fig.~\ref{Fig:Ramp2u:CI5:Accuracy_all}. 
These results are compared to random sensors, full-state feature sensors, and sensors learned using QRcp. SSPOC results for varying $\lambda$ from Fig.~\ref{Fig:Ramp2u:CI5:SSPOC}  are also rearranged and shown with respect to the number of sensors (Fig.~\ref{Fig:Ramp2u:CI5:Accuracy_all}(e)).
Sensors learned using SSPOC or QRcp generally yield a better accuracy than random sensors.  Both reach a plateau of about $93\%$ for $N_s \approx 50$ sensors, corresponding to about $0.29\%$ of all potential sensor locations.
SSPOC sensors significantly outperform QRcp sensors for few sensors in the range of $10<N_s<40$.
Misclassification mainly occurs close to the cluster borders, which are defined by half of the distance between neighboring centroids. These clear cluster borders become fuzzy in the sensor space. Classification accuracy is increased by re-training the LDA classifier on the learned sensors, which generally performs better than classification based on re-trained centroids. 

\begin{figure}[tb] 
	\centering
	\includegraphics[width=\textwidth]{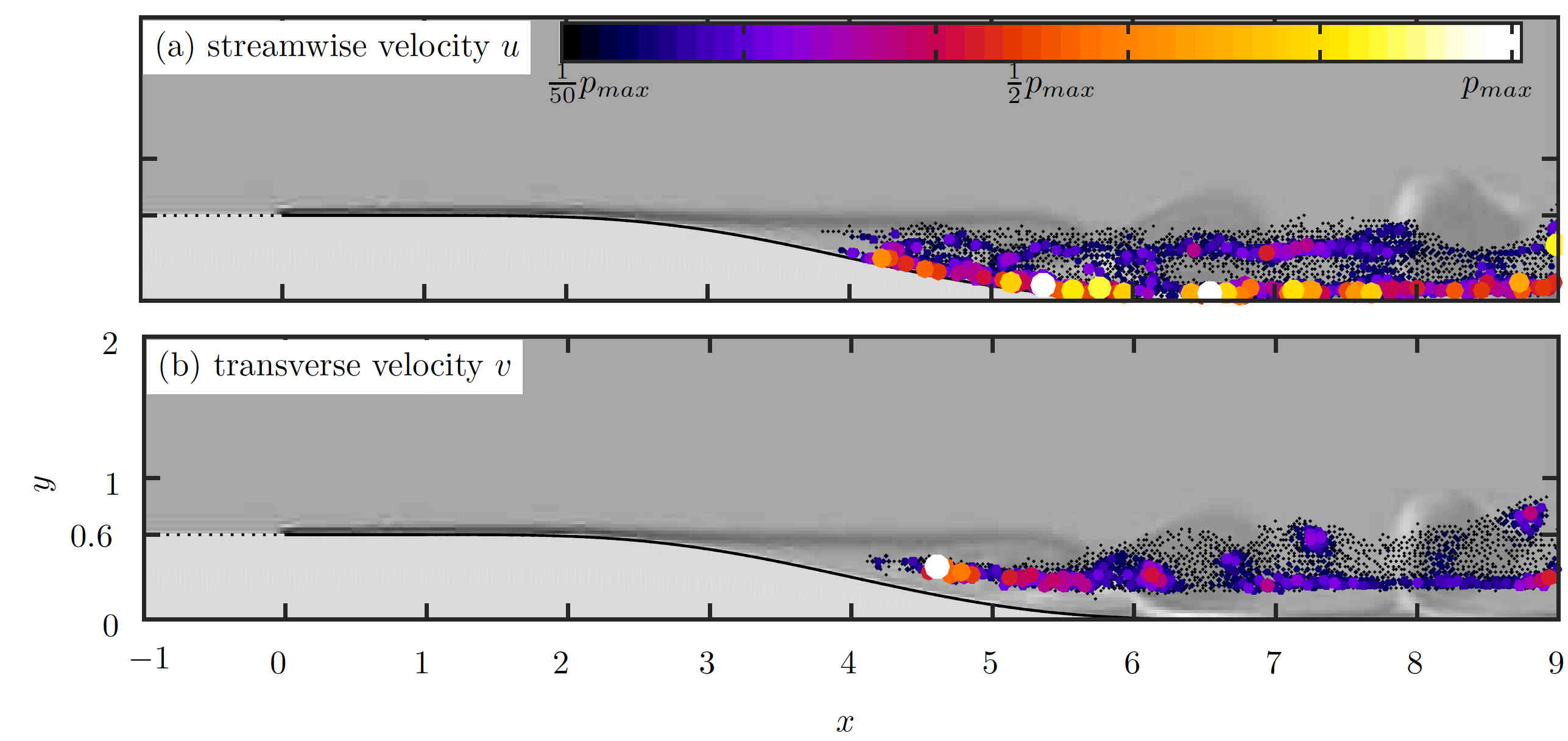}
	\caption{Distribution of sensor locations found using SSPOC (case (e) in Fig.~\ref{Fig:Ramp2u:CI5:Accuracy_all}).
		Bright color and large circle represent high probability that a particular sensor location is selected;
		probability is normalized with respect to the maximal probability $p_{max}$ any sensor location is selected.}
	\label{Fig:Ramp2u:CI5:SSPOC:Distribution}
\end{figure}

Sensors should be placed in sensitive regions capable of discriminating between different clusters.
The probability distribution that a SSPOC sensor is placed in a particular location is displayed in Fig.~\ref{Fig:Ramp2u:CI5:SSPOC:Distribution}, with the streamwise and transverse components plotted separately.   
Most sensors are placed in the recirculation region and further downstream; most of the upstream sensors are placed closely behind the separation point region (separation point is $x_{sp}^{nat} \approx 6$ for the unforced and $x_{sp}^{periodic} \approx 3.5$ for the periodically forced flow).
The favored sensor locations are different for the two velocity components:
The transverse velocity sensors are mainly placed along the lines associated with the convecting vortex cores, which can be close to the wall, when the flow locks in to the excitation frequency, or farther away for the unforced flow.
In contrast, the streamwise velocity sensors are placed close to the wall inside the boundary layer and (less frequently) along the convection lines of the vortex cores associated with the unforced flow.
The clusters contain kinematically similar snapshots, thus snapshots belonging to the same cluster exhibit a similar phase. However, a cluster may also contain snapshots from both the forced and unforced flows, as the transition occurs smoothly and the partitioning is coarse. 
Thus, sensors are placed where both the unforced and controlled flows exhibit distinct features that discriminate the clusters.
The aforementioned sensor locations are arguably the most sensitive regions, as these capture  
(1) whether the flow shows features from the forced or unforced flow,  
(2) to which phase bin the flow corresponds, and
(3) the extent of the instantaneous recirculation area.

\begin{figure}[tb] 
	\centering
	\includegraphics[width=\textwidth]{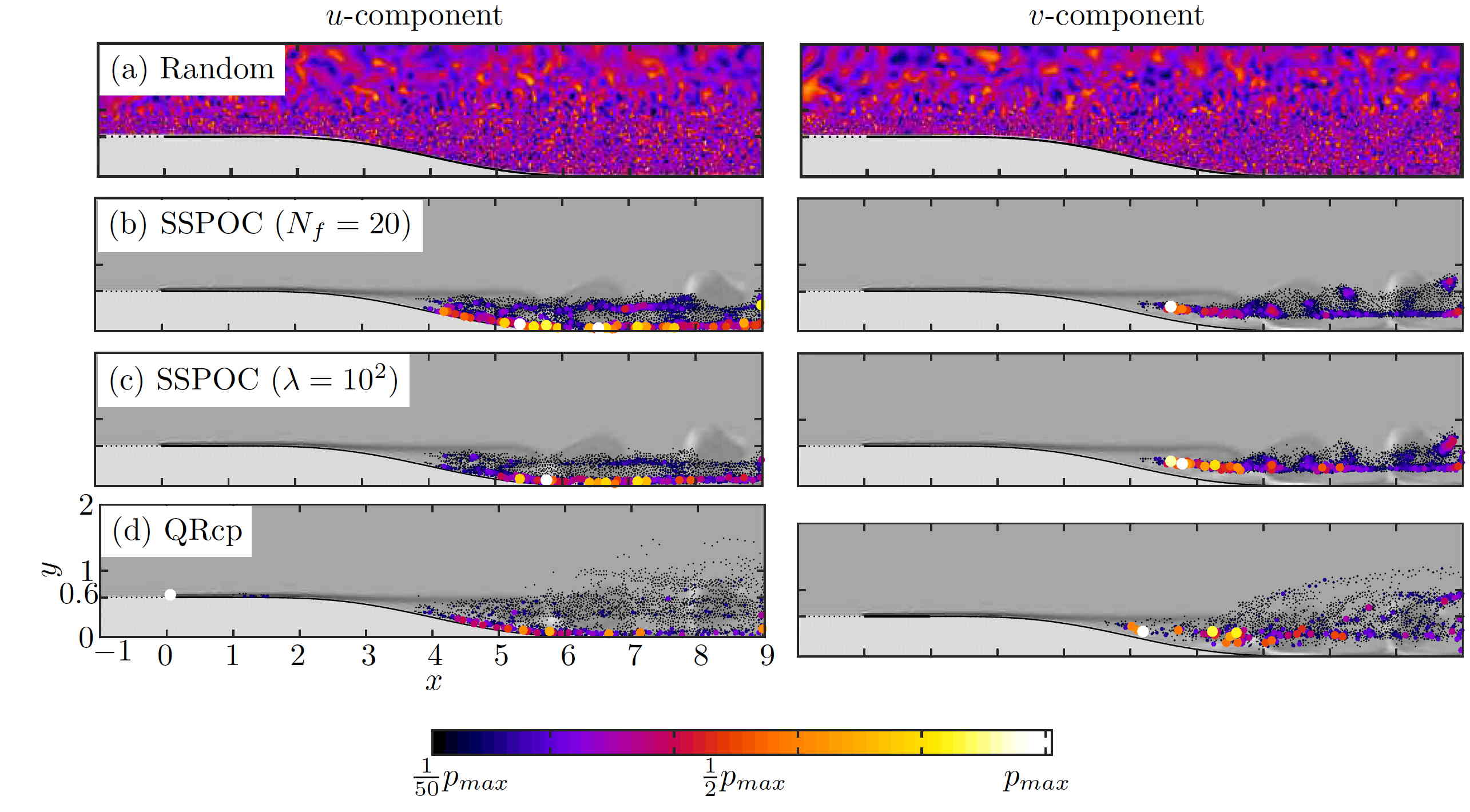}
	\caption{Distribution of sensor locations analog to Fig.~\ref{Fig:Ramp2u:CI5:SSPOC:Distribution} 
		and corresponding to the cases in Fig.~\ref{Fig:Ramp2u:CI5:Accuracy_all} comparing
		(a) a random selection of sensor locations, 
		(b) SSPOC sensors for varying $\lambda$, 
		(c) SSPOC sensors for varying $N_f$, and 
		(d) QRcp sensors, for the streamwise (left) and transverse (right) velocity component, respectively.
	}
	\label{Fig:Ramp2u:CI5:Distribution}
\end{figure}

In Fig.~\ref{Fig:Ramp2u:CI5:Distribution}, 
distributions are shown for random sensors, SSPOC sensors with varying $\lambda$ (the same as in Fig.~\ref{Fig:Ramp2u:CI5:Accuracy_all} to facilitate the comparison),  SSPOC sensors with varying features, and QRcp sensors. 
QRcp shows a similar preference as SSPOC for placing streamwise sensors in the boundary layer and 
transverse sensors along the line associated with the convected vortex cores.  
Despite the similarities between SSPOC and QRcp, there are also important differences. 
While SSPOC sensors are confined to a limited region downstream of the ramp, QRcp sensors are more distributed, showing a smaller preference in particular sensor locations. Moreover, QRcp places (few) streamwise velocity sensors at the leading edge of the plate, which do not contain information on the flow separation but instead measure non-physical behavior: The leading edge corner is approximated with only a few vertices, leading to numerical inaccuracies at that location. 
Nevertheless, QRcp also places streamwise velocity sensors around $1\leq x \leq 1.5$. These sensors capture disturbances introduced by the actuator located at $x=1$ that affect the flow behavior downstream.

\subsubsection{Comparing full-state and sparse sensors for control}
\label{Sec:Ramp2u:OSP-CROMc}
In this section, 
the performance of the best CROM-based control law using partial-information sensors is compared with full-state feature sensors employed in~\cite{Kaiser2016tcfd}. 
As the goal is optimization of the best open-loop periodic forcing, the considered feature space is the subspace spanned by the first $N_f=10$ POD modes ${\bf\Psi} = [ \boldsymbol{\psi}_1^{OL} \ldots \boldsymbol{\psi}_{N_f}^{OL}]$ associated with the best periodic forcing. 
Sparse sensors for classifying snapshots into $N_c=10$ clusters are then learned in that subspace. 
While in Sec.~\ref{Sec:Ramp2u:OSP-CI5}, clusters are learned repeatedly from the training set, in this section the cluster affiliation of each snapshot is fixed and corresponds to that used in~\cite{Kaiser2016tcfd} in order to compare results. 

\begin{figure}[tb]
	\centering
	\includegraphics[width=\textwidth]{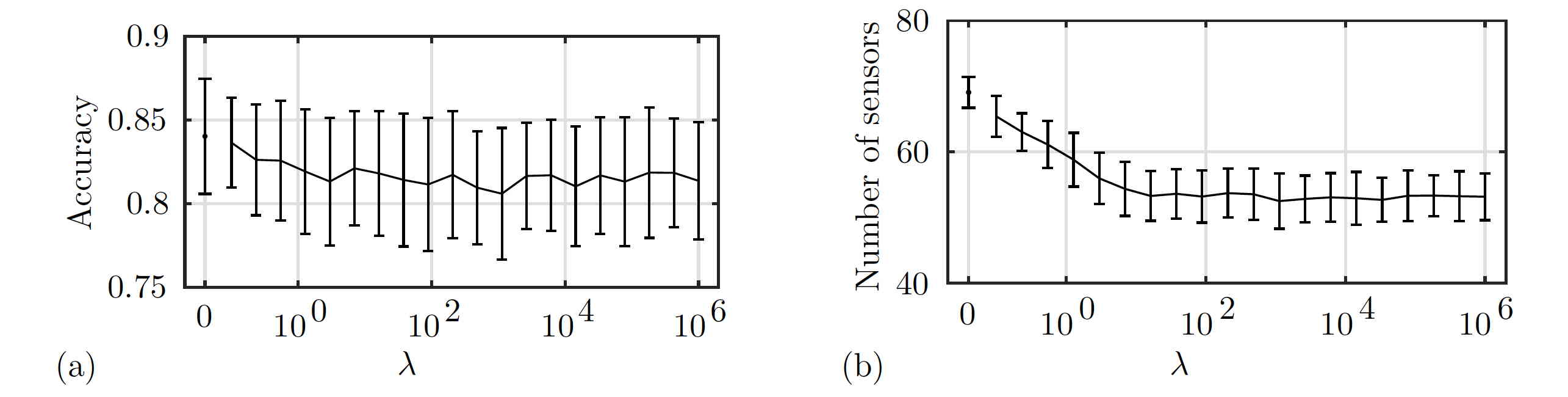}
	\caption{
		Cross-validated classification results for $N_c=10$ clusters. 
		The accuracy (a) decays up to $\lambda\approx 10^1$ and seems independent of $\lambda$ thereafter.
		Analogously, the number of sensors decreases to $N_s\approx 50$ for $\lambda\approx 10^1$ where it starts to saturate. 
	}
	\label{Fig:Ramp2u:CROMc:SSPOC}
\end{figure}

Cross-validated accuracy for SSPOC sensors is shown in Fig.~\ref{Fig:Ramp2u:CROMc:SSPOC}.
The number of learned sensors decays until $\lambda=10$ where it saturates with $N_s=50$ sensors and an average accuracy of $82\%$. The accuracy is lower than the results in the previous section. As the classification is performed in the subspace, a large amount of information from the snapshots is removed, which may be critical for discriminating the clusters.   
Although sensors are learned with respect to their sensitivity to the employed features, they provide unfiltered measurements, decreasing the accuracy. Considering this, the accuracy is still comparably high, which is partially achieved by re-training the classifier in the sensor space.

\begin{figure}[tb]
	\centering
	\includegraphics[width=\textwidth]{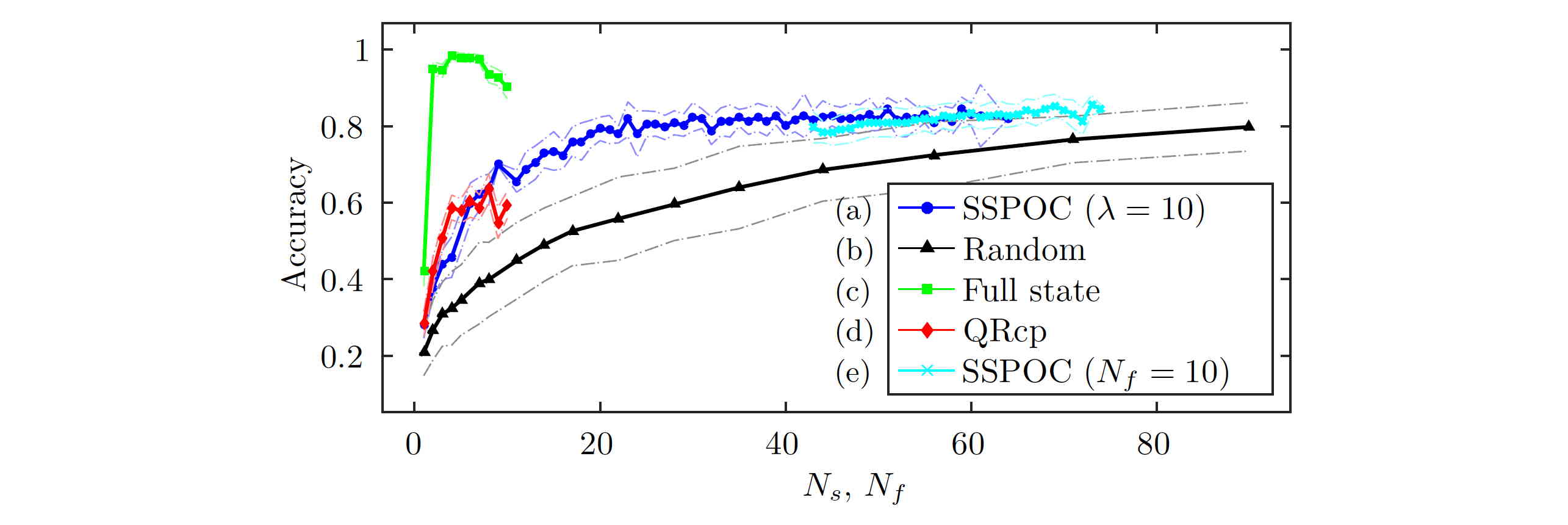}
	\caption{Classification accuracy (analog to Fig.~\ref{Fig:Ramp2u:CI5:Accuracy_all}) for $N_c=10$ clusters using 
		(a) SSPOC with varying $N_f$ and fixed $\lambda=10$, 
		(b) random sensors,
		(c) full-state sensors with varying $N_f$, 
		(d) QRcp with varying $N_f$, and
		(e) SSPOC with fixed $N_f=10$ and varying $\lambda\in [0,10^6]$.
		Both, SSPOC and QRcp yield better accuracy than random sensors.
	}
	\label{Fig:ramp2u:CROMCo:comparison_accuracy}
\end{figure}

Cross-validated accuracy for SSPOC sensors with varying $\lambda$ or varying $N_f$, respectively, random sensors, full-state feature sensors, and QRcp sensors are compared in Fig.~\ref{Fig:ramp2u:CROMCo:comparison_accuracy}.
Both SSPOC and QRcp yield better sensors than choosing random sensor locations. 
Since the maximum number of features is $N_f=10$, at most $N_s=10$ sensors can be determined using QRcp.
The accuracy using SSPOC saturates at about $80\%$ with $N_s \approx 20$ sensors, 
having the largest gain with respect to random measurements.
The full-state sensor accuracy decreases with increasing $N_f>4$ due to overfitting. 
The employed classifier relies on the discriminating directions found by LDA, while the true classification is based on the nearest cluster centroids. 

\begin{figure}[tb] 
	\centering
	\includegraphics[width=\textwidth]{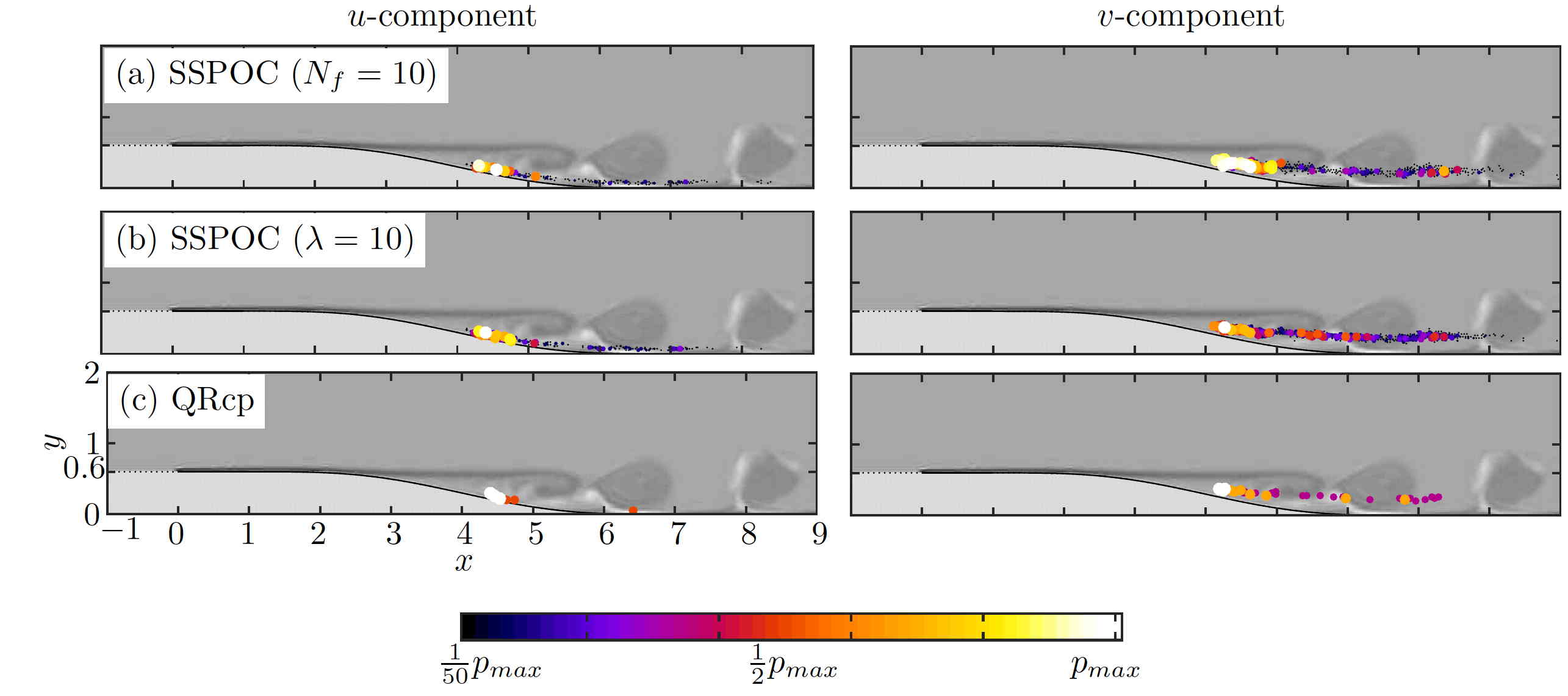}
	\caption{
		Distribution of sensor locations (analog to Fig.~\ref{Fig:Ramp2u:CI5:Distribution}) to discriminate $N_c=10$ clusters 
		based on 
		(a) SSPOC sensors for varying $\lambda$, 
		(b) SSPOC sensors for varying $N_f$, and 
		(c) QRcp sensors for varying $N_f$.
	}
	\label{Fig:Ramp2u:CROMc:Distribution}
\end{figure}

A comparison of the distribution of sensor locations is displayed in Fig.~\ref{Fig:Ramp2u:CROMc:Distribution}. 
Note that the distribution of random sensors is not included, as it is similar to Fig.~\ref{Fig:Ramp2u:CI5:Distribution}(a). 
Sensors learned using either SSPOC or QRcp have a clear location preference analogous to the results in
Fig.~\ref{Fig:Ramp2u:CI5:Distribution}, despite restricting the feature space to a subspace. 
Note that the reduction of the number of features also decreases the number of sensors placed, 
thus yielding fewer dominant sensor locations.

\begin{figure}[tb] 
	\centering
	\includegraphics[width=\textwidth]{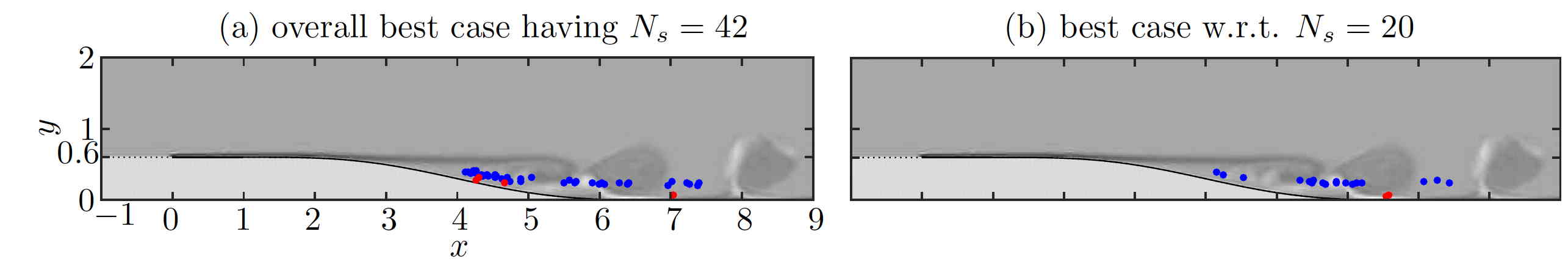}
	\caption{Sensor locations found using SSPOC for 
		(a) the overall best case with $N_s=42$ sensor locations (case \mysingleq{A}) and 
		(b) using $N_s=20$ sensors having the largest gain compared to random sensors (case \mysingleq{B}).
		The sensor locations are discriminated with respect to the streamwise ($\color{red}{\bullet}$)
		and transverse ($\color{blue}{\bullet}$) velocity component.
		Note that in both cases the majority of sensors measures only the transverse velocity component.
	}
	\label{Fig:CROMCo:SensorLocations}
\end{figure}

In the following, the performance of the optimal control law determined from the control-dependent CROM for this configuration is examined using only information from the learned sensors. 
Two particular cases are considered (see Fig.~\ref{Fig:ramp2u:CROMCo:comparison_accuracy}): 
(A) those SSPOC sensors that achieve the best accuracy among all cases for $\lambda=10$, and 
(B) using the best case for $N_s=20$ and $\lambda=10$, which yields the largest gain compared to random sensors, the latter achieving a similar accuracy with about $N_s=80$ sensors.
The distribution of sensors (red and blue circles) is displayed in Fig.~\ref{Fig:CROMCo:SensorLocations} with an instantaneous vorticity realization as the background.
For $N_s=42$, most sensors measure the transverse velocity component
and aggregate in the recirculation zone behind the backward-facing ramp or are distributed along the line the vortices associated with the forced flow are convected. 
The clustering of sensors in distinct regions suggests that fewer sensors could be sufficient to obtain similar information.
In comparison with $N_s=20$, the number of sensors is considerably reduced in $x\in [4,5]$ and $x\approx 7$.

\begin{figure}[tb]
	\centering
	\includegraphics[width=\textwidth]{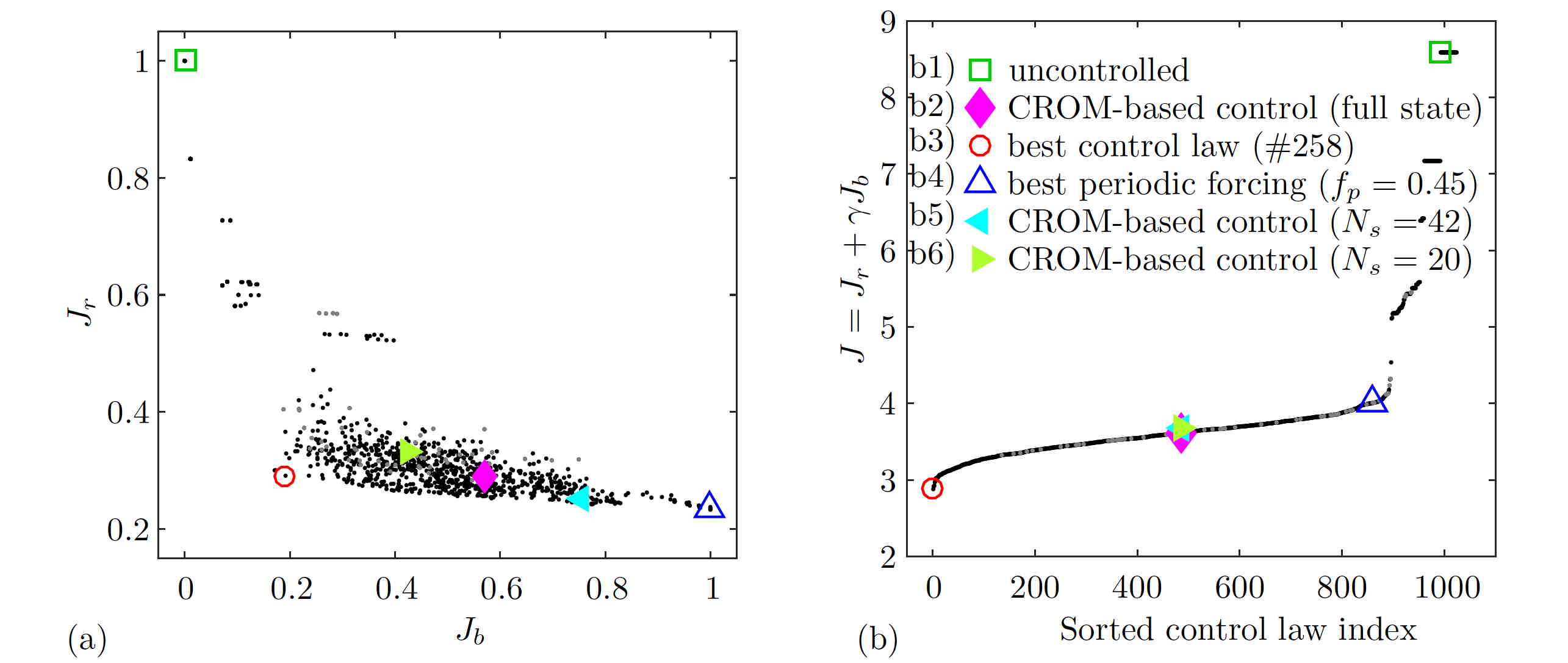}
	\caption{Performance results of cluster-based control laws.
		The best CROM-based control law (b2) performs similarly well using only (b5) $N_s=42$ or
		(b6) $N_s=20$ sensors.
		The difference in the performance (a) originates from misclassification due to the unfiltered sensor signal. 
		Both sensor-based control cases collapse with the full-state control when considering the overall performance (b).
	}
	\label{Fig:Ramp2u:CROMCo:ControlResults}
\end{figure}

Performance results of all evaluated control laws are displayed in Fig.~\ref{Fig:Ramp2u:CROMCo:ControlResults}.
All control laws are sorted with respect to their performance $J$. 
Particular cases are highlighted: 
(b1) the natural flow, 
(b2) the optimal control law determined with CROM using full-state POD feature sensors,
(b3) the overall best control law determined with a brute-force search, 
(b4) the best periodic forcing as reference, 
(b5) the best CROM-based control law from (b2) where classification is based on case \mysingleq{A} sensors ($N_s=42$), and
(b6) similar like (b5) but for case \mysingleq{B} sensors ($N_s=20$). 
The difference between the full-state controller and the sensor-based controller shown in Fig.~\ref{Fig:Ramp2u:CROMCo:ControlResults}(left) is due to the misclassification resulting from the unfiltered sensor signals. 
Nevertheless, all three cases (b2), (b5), and (b6) show a similar overall performance  (compare  Fig.~\ref{Fig:Ramp2u:CROMCo:ControlResults}(b)).

In conclusion, 
SSPOC has found few optimized sensor locations that perform equally well for control as full-state measurements.
More generally, 
SSPOC sensors outperform random sensors and perform equally well or better than QRcp sensors.
If enough random sensors are employed, these faithfully preserve the cluster geometry and can achieve a similar accuracy.

\subsection{Mixing layer with different dynamical regimes}
\label{Sec:MixingLayer}
\begin{figure}[tb]
	\centering
	\includegraphics[width=\textwidth]{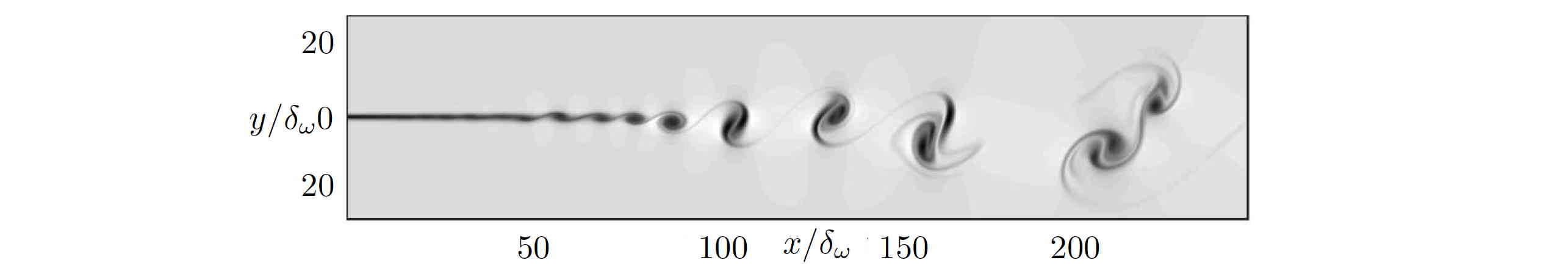}
	\caption{Instantaneous vorticity realization of the mixing layer.}
	\label{Fig:ML}
\end{figure}

In this section, sensor placement is optimized for a two-dimensional mixing layer flow undergoing vortex pairing. 
The flow exhibits the typical roll-up of vortices arising from the Kelvin-Helmholtz instability and vortex pairing further downstream.
This example is motivated by previous work \cite{Kaiser2014jfm}, 
in which CROM identifies two dynamical regimes associated with different wavenumbers and a particular cluster that acts as a switch between these regimes (depicted in Fig.~\ref{Fig:CROM_ML}).
The velocity ratio is $r = U_1/U_2=3$ where $U_1$ and $U_2$ denote the upper (fast) and lower (slow) stream velocities, respectively. The Reynolds number is $Re = \Delta U \delta_{\omega}\nu = 500$ based on the velocity difference $\Delta U = U_1-U_2$, the initial vorticity thickness $\delta_{\omega}$, and the kinematic viscosity $\nu$; the Mach number is $Ma=0.3$. We employ an ensemble of $M = 667$ snapshots with a sampling time of $3\Delta t$, non-dimensionalized with respect to $U_1$ and $\delta_{\omega}$.
The computational domain is $140 \delta_{\omega}$ long and $56 \delta_{\omega}$ high with increasing spatial resolution in the mixing region. 
Details of the finite-difference Navier-Stokes solver and the configuration
can be found in \cite{Daviller2010phd} and \cite{Cavalieri2011jsv}. 
An instantaneous vorticity realization of the flow is shown in Fig.~\ref{Fig:ML}.
There exist $N_{xy} \approx 1.5\cdot 10^6$ potential sensor locations. 
This high-dimensionality results in a computationally expensive optimization problem. 
Therefore, instead of using the full data, the data is randomly subsampled, and then POD is applied to this subset of measurements. 
Specifically, a random $1\%$ of the data is selected, reducing the number of potential sensors to $N \approx 1.5\cdot 10^4$.
Further, SSPOC sensors are only trained on the first $N_f=40$ POD features (see~\ref{Sec:CROM:DependencyOnFeatures} for an analysis of CROM's dependency on the number of features).

\begin{figure}[tb]
	\centering
	\includegraphics[width=\textwidth]{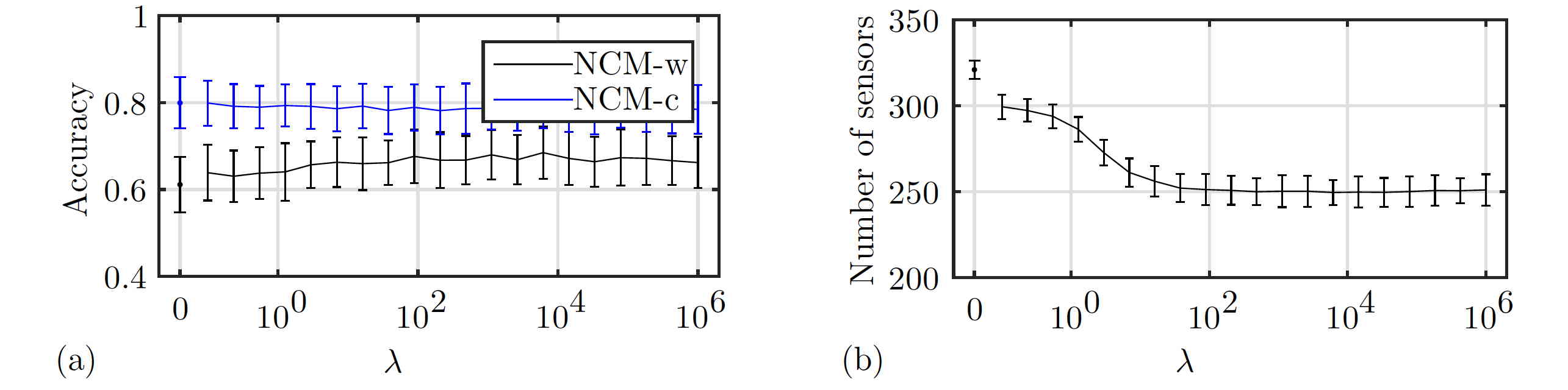}
	\caption{Classification results for $N_c=10$ clusters of the mixing layer.  
		Sensors are trained on a random selection of $1\%$ of the grid points.
		The accuracy in (a) appears to be independent of $\lambda$, even though the number of sensors (b) decreases. 
		The number of sensors saturates at about $\lambda=10^1$ with $N_s \approx 250$.}
	\label{Fig:ML:SSPOC}
\end{figure}

The mean and standard deviation of the cross-validated accuracy for SSPOC sensors are shown in Fig.~\ref{Fig:ML:SSPOC}. 
We compare two classifiers: (1) the nearest-centroid method applied in the subspace spanned by the LDA discriminating directions $\{\hat{\vec{w}}_i\}_{i=1}^{N_c-1}$, which will be denoted by \mysingleq{NCM-w}, and 
(2) the nearest-centroid method applied to the cluster centroids $\{\hat{\vec{c}}_k\}_{k=1}^{N_c}$ in the sensor space, which is denoted by \mysingleq{NCM-c}.
Although not shown, in the previous examples NCM-w generally outperformed NCM-c.
However, sensors learned on heavily subsampled data for the mixing layer perform better using the latter approach, on average by $10-20\%$.

\begin{figure}[tb]
	\centering
	\includegraphics[width=\textwidth]{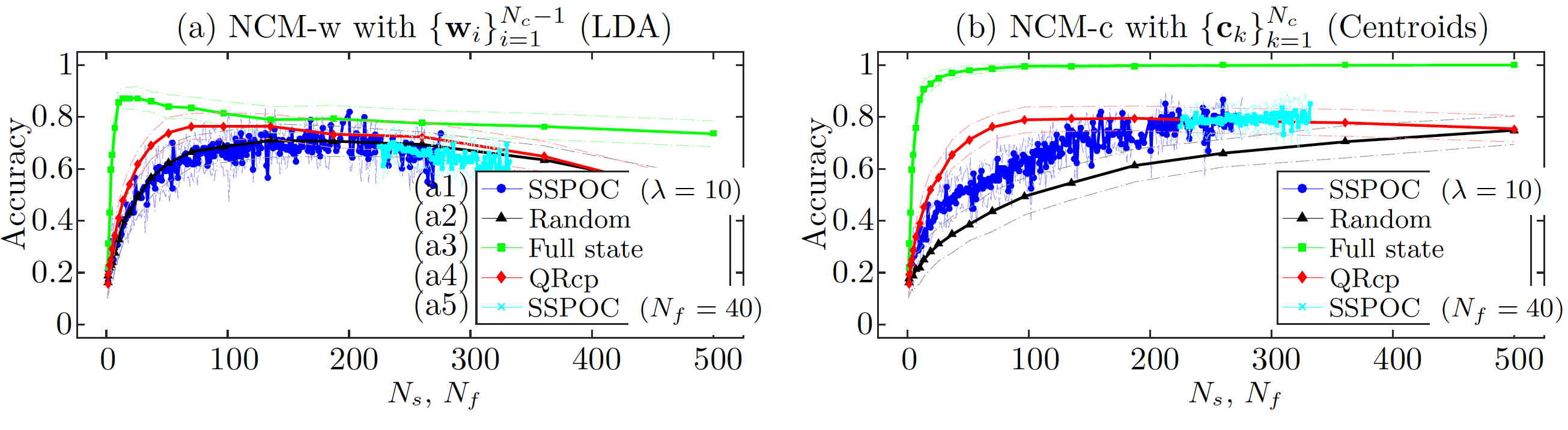}
	\caption{Comparison of cross-validated accuracy
		based on the nearest-centroid method using (a) LDA vectors or (b) centroids.
		Despite being trained on only a random $1\%$ of the data, SSPOC sensors yield a similar accuracy as QRcp sensors trained on $100\%$, if $N_s\geq 200$ and classification is based on centroids.}
	\label{Fig:ML:Accuracy}
\end{figure}

Classification performance is
compared using NCM-w (see Fig.~\ref{Fig:ML:Accuracy}(a)) and NCM-c (see Fig.~\ref{Fig:ML:Accuracy}(b))
for
sensors learned from SSPOC on $1\%$ subsampled data for a fixed $\lambda$, 
a random selection of sensors, 
full-state feature sensors using a varying number of features $N_f$, and 
sensors determined using QR with column pivoting (without subsampling). 
The SSPOC results from Fig.~\ref{Fig:ML:SSPOC} are also rearranged with respect to the number of sensors, and shown in Fig.~\ref{Fig:ML:Accuracy} (SSPOC $N_f=40$). 
A general observation is that the accuracy of random sensors can be increased by using NCM-w for classification. 
For fewer sensors, clusters tend to merge and overlap, which impedes their discrimination based on cluster centroids.
In contrast, LDA finds those features in sensor space that are most discriminating, increasing the performance.
However, this is not true for all cases examined, particularly because LDA suffers from overfitting in contrast to the cluster centroids.
Subsampled SSPOC sensors perform equally well compared with random sensors if the classification is based on NCM-w, 
but outperform random sensors if NCM-c is employed.
Further, if the number of sensors exceeds $N_s>130$, SSPOC sensors using NCM-c perform better than random sensors using NCM-w.
Despite being trained on only a random $1\%$ of the data, SSPOC sensors yield a similar accuracy as QRcp sensors trained on $100\%$, if $N_s\geq 200$ and classification is based on NCM-c. 
QRcp sensors achieve the largest gain for $N_s\approx 80$ sensors independent of the classification method.
The general decline of accuracy after $N_s>100$ in Fig.~\ref{Fig:ML:Accuracy}(a)
is associated with overfitting. 
The strong effect of overfitting can also be observed for full-state feature sensors using NCM-w, where the accuracy decays rapidly starting at $N_f\approx 20$.
In contrast, full-state features converge to $100\%$ accuracy based on the cluster centroids.

\begin{figure}[tb]
	\centering
	\includegraphics[width=\textwidth]{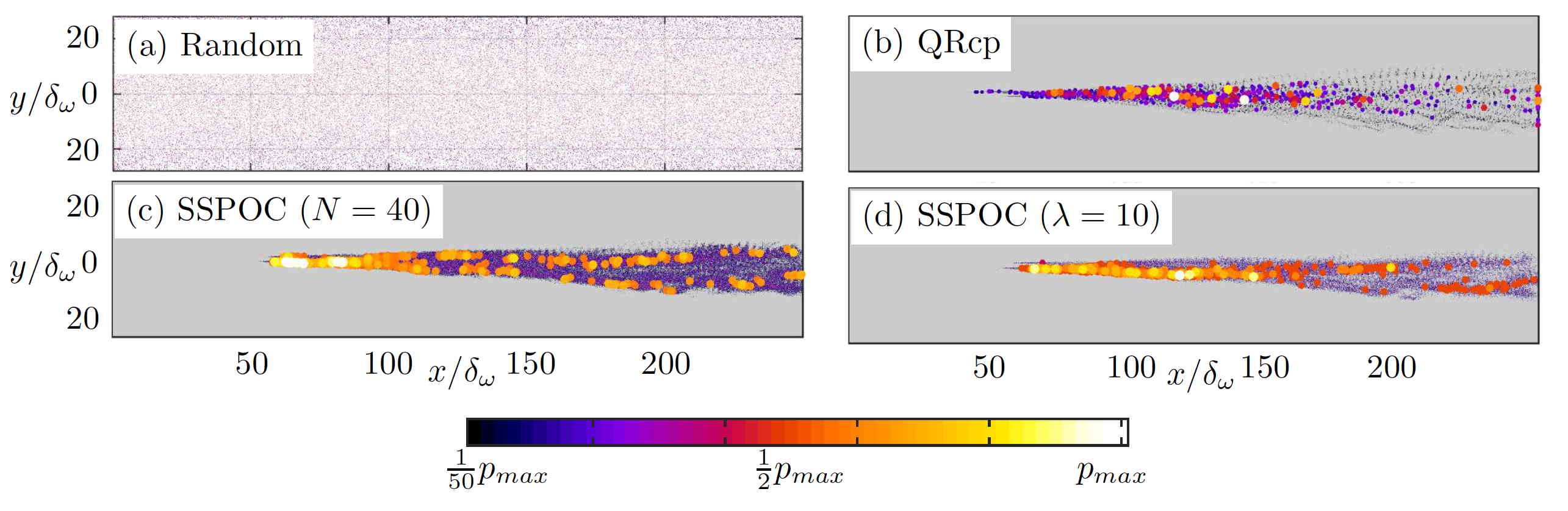}
	\caption{Probability distribution of sensor locations analogous to Fig.~\ref{Fig:Ramp2u:CI5:SSPOC:Distribution}. The background shows probability that a particular sensor location is selected, where gray refers to zero. In addition, the $1000$ most probable sensor locations are displayed as color-coded circles where color and size change with probability.}
	\label{Fig:ML:SensorDistribution}
\end{figure}

The distribution of selected sensor locations for each method is displayed in Fig.~\ref{Fig:ML:SensorDistribution}.
In both cases, SSPOC and QRcp sensors show a similar distribution with placement preference in the initial region where the shear layer instability develops. 
Clusters represent different phases but also discriminate the different dynamical regimes, where the flow is either governed by vortex shedding or dominated by vortex pairing.
The distributions suggest that the initial instability region is critical for the discrimination of the clusters.

\begin{figure}[tb]
	\centering
	\includegraphics[width=\textwidth]{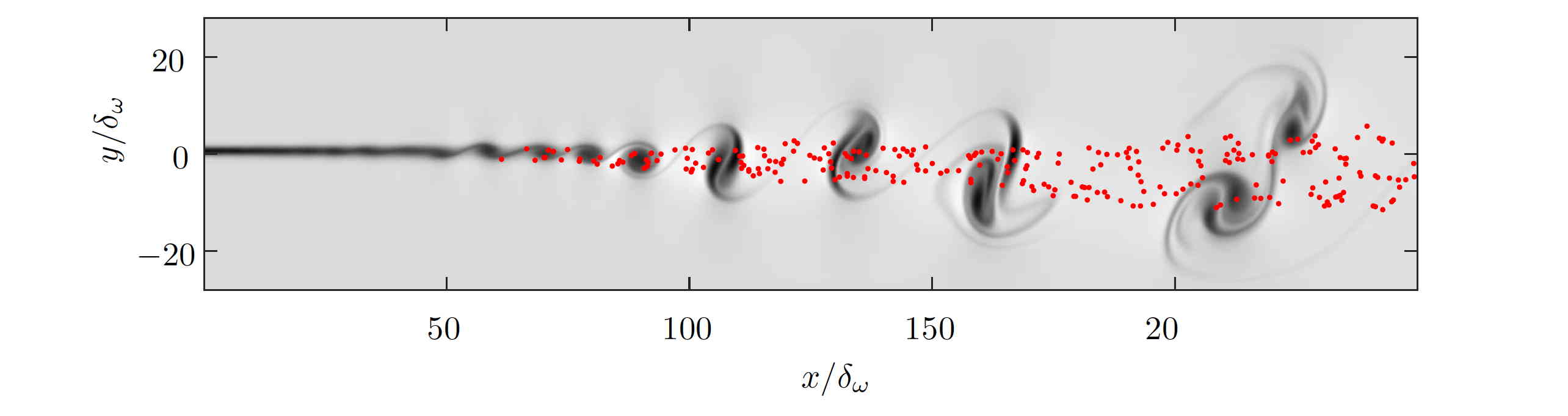}
	\caption{Set of $262$ sensor locations found using SSPOC, which achieves the highest accuracy of $91\%$ based on NCM-c.}
	\label{Fig:ML:SensorLocation_best}
\end{figure}

For a better assessment, we show the sensor locations (see Fig.~\ref{Fig:ML:SensorLocation_best}) found using SSPOC for the best case based on NCM-c. This set of $262$ sensors, which corresponds to about $0.017\%$ of all grid points, achieved the highest accuracy of $91\%$. 
Sensors are placed inside vortices and along the filaments, distributed along the direction of convection. 
Although the width of the shear layer is larger, sensors are restricted to a more confined region. 
Analogous to the growth of the mixing region, the spreading of the sensors in the transverse direction increases downstream with the streamwise direction.  
Note that the reason for the seemingly continuous distribution in the streamwise direction is that the flow is convective, similar to the separating flow and in contrast to the periodic double gyre.

\begin{figure}[tb]
	\centering
	\includegraphics[width=\textwidth]{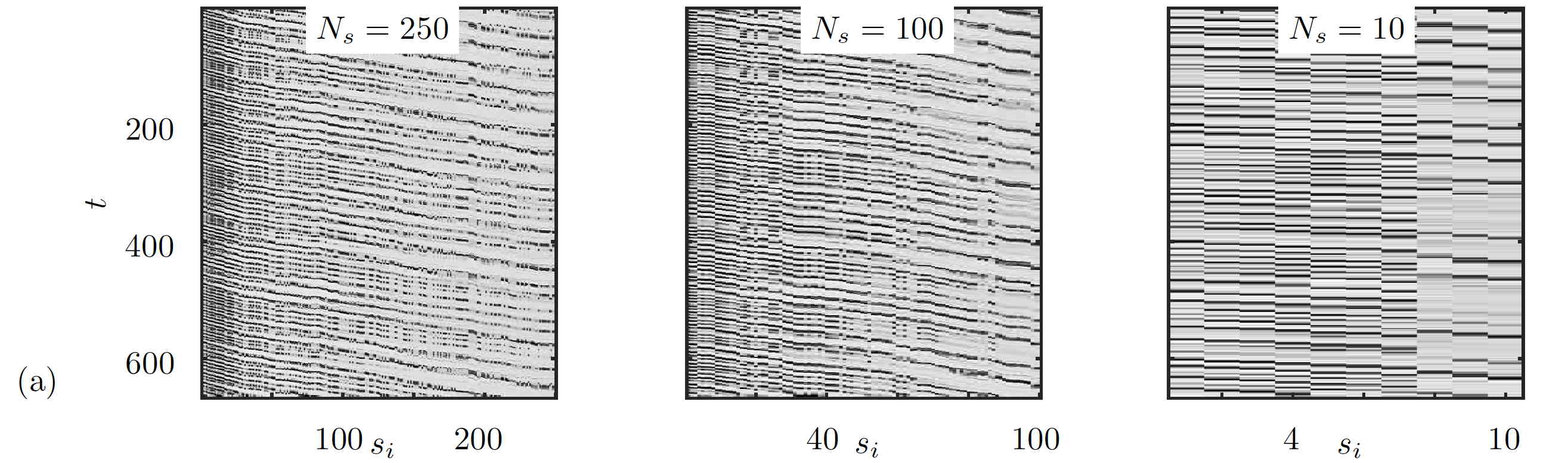}\\
	\includegraphics[width=\textwidth]{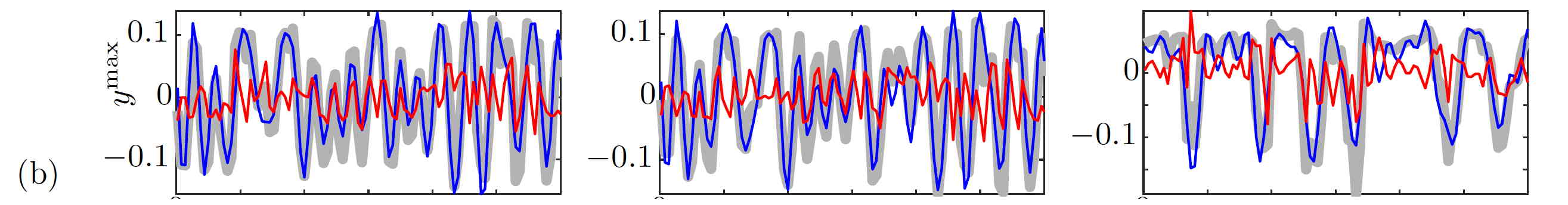}\\
	\includegraphics[width=\textwidth]{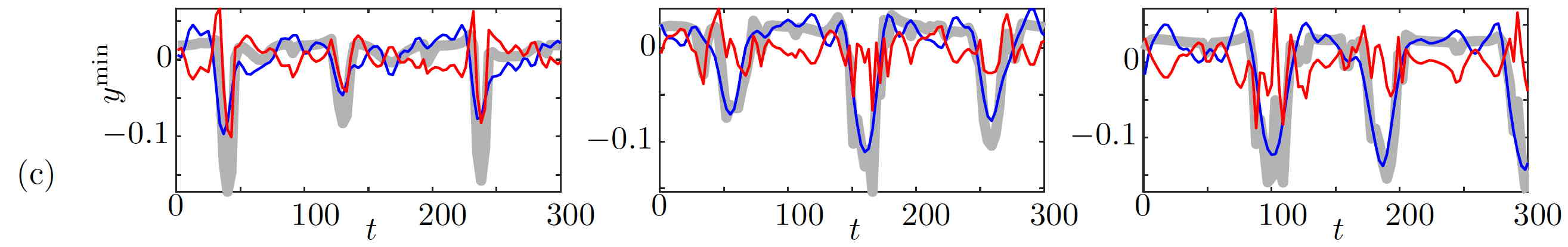}\\
	\includegraphics[width=\textwidth]{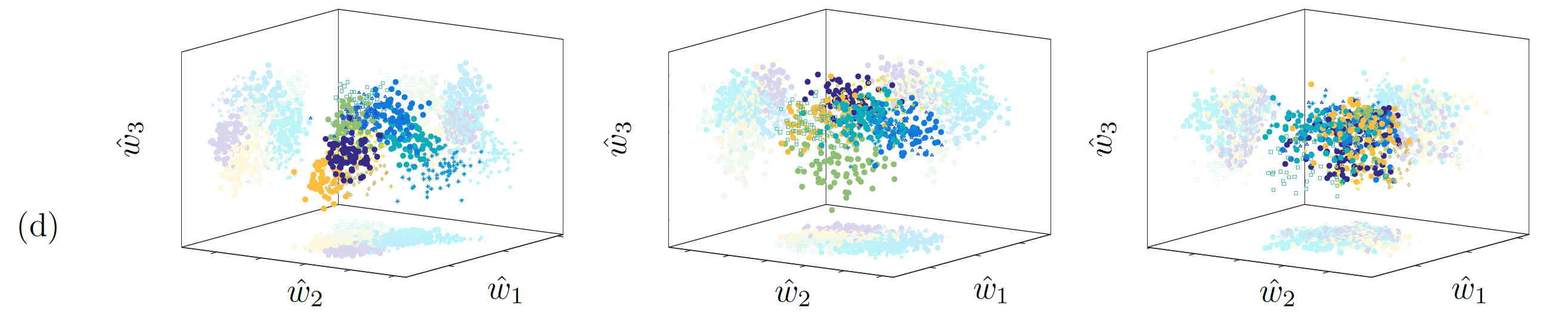}
	\caption{Dependency on the number of sensors:
		(a) Time history of sensor measurements (sorted with respect to streamwise location),
		(b) time series of sensor having maximum variance $y$ (gray thick line),  
		$y_{N\leq 40}$ (blue line), and  $y_{N> 40}$ (red line), respectively.
		(c) same as (b) but for sensor location showing minimum variance, and
		(d) cluster affiliation in the subspace given by $\hat{{\bf w}}_i$, $i=1,2,3$, of the sensors.
		Each point represents an observation color-coded by its cluster affiliation.}
	\label{Fig:ComparisonSensorNumber}
\end{figure}

To analyze the effect of the number of sensors, we present three cases with decreasing number of sensors in Fig.~\ref{Fig:ComparisonSensorNumber}.
Sensors are sorted with respect to their streamwise location.
Thus, the time history in Fig.~\ref{Fig:ComparisonSensorNumber}(a) 
depicts the convection of the vortices (dark lines corresponding to maximum measured vorticity value).
SSPOC sensors are only trained on the first $N_f=40$ POD features, 
thus we compare the following measurements: the fluctuating part of measurements denoted by $\vec{y}$, its reconstruction using only those modes $\vec{y}_{N\leq 40}$, where 
$\vec{y}_{N\leq 40} = \matr{\Phi}\vec{u}_{N\leq 40}$ with $\vec{u}_{N\leq 40} = \sum_{i=1}^{N_f=40} \vec{a}_i \boldsymbol{\psi}_i$,
and the remaining part $\vec{y}_{N> 40}$, which is computed analogously.
Two particular sensor locations, exhibiting the maximum and minimum variance in the considered set of sensors, 
are selected and the corresponding time history of $y$, $y_{N\leq 40}$, and $y_{N> 40}$ are displayed
in Fig.~\ref{Fig:ComparisonSensorNumber}(b) and (c), respectively.
The superscript \mysingleq{min} and \mysingleq{max} in Fig.~\ref{Fig:ComparisonSensorNumber}(b) and (c) refer to the two selected sensors. 
The accuracy can be increased by up to $12\%$ using NCM-c and up to $30\%$ using NCM-w for those cases shown in Fig.~\ref{Fig:ComparisonSensorNumber} and Fig.~\ref{Fig:ML:SensorLocation_best}, if 
filtered measurements $\vec{y}_{N\leq 40}$ are considered.
The influence of the number of sensors becomes evident when comparing the cluster affiliation of the observations in the subspace spanned by the LDA discriminating vectors, as shown in Fig.~\ref{Fig:ComparisonSensorNumber}(d). 
With decreasing number of sensors, clusters tend to merge and overlap, making the classification task more difficult.

Summarizing, for very high-dimensional systems it may be necessary to subsample the data on which sensors are trained.
SSPOC sensors outperform random sensors when using the nearest-centroid method based on the cluster centroids.
QR with column pivoting is computationally more efficient than solving the optimization problem, thus QRcp sensors can be trained on full-state data and using more features even for very high-dimensional systems. 
Generally, sensors are placed where they are most informative for the cluster discrimination, along the vortical structures in the direction of the convection.

\section{Conclusion}
\label{Sec:Conclusions}
Reduced-order models are of growing importance in a broad range of scientific applications as they enable simulations of large-scale engineering systems for design, optimization, and control thought impossible only a decade ago~\cite{benner2015survey}.  The success of ROMs centers on two key innovations:  (i) 
many complex systems exhibit low-dimensional dynamics~\cite{Cross:1993}
so that high-dimensional system can be projected to a low-dimensional subspace in a principled way, and (ii) sparse sampling of the state space for interpolating the nonlinear terms required for the subspace projection.  The low-rank embedding space for the ROM is typically computed via a POD reduction.
The efficient projection of the nonlinearity to the POD subspace 
can be accomplished with gappy POD methods~\cite{everson1995karhunen}, which include the modern principled approaches of discrete empirical interpolation method~\cite{eim,deim} and compressive sensing~\cite{bright:2013,Brunton2014siam,Sargsyan:2015}.
Although successful, the current POD-Galerkin method for producing a ROM has a number of important limitations, including that (i) the POD basis is expensive to compute and must be done in an {\em offline} manner, (ii) a nonlinear model is produced whose sensitivity to initial conditions make the ROM prediction only qualitative~\cite{carlberg2015arxiv}, and (iii)  the standard POD-Galerkin time-stepping algorithm is not robust and is prone to instability~\cite{carlberg2015arxiv}.   The nonlinear nature of standard ROMs limits the mathematical machinery available for the objective of prediction and control.  This suggests that alternatives to POD-Galerkin embeddings of the dynamics should be considered. 

There is a growing effort to represent nonlinear dynamics in a linear operator framework.  
This has motivated significant work on the infinite-dimensional Koopman and Perron-Frobenius operators.   
However, standard data-driven implementations, including dynamic mode decomposition for Koopman and Ulam-Galerkin methods for Perron-Frobenius, tend to result in high-dimensional models with their own associated challenges for computations and measurements.  
The recent cluster-based reduced-order model (CROM) framework provides an efficient low-dimensional representation of the Perron-Frobenius operator using a data-driven discretization of phase space into clusters, on which probabilistic dynamics evolve.  
Although the CROM is fundamentally low-dimensional, making it advantageous for real-time computations, uncertainty in the model grows with time so that data assimilation techniques must be incorporated.  
Because the clusters are typically defined in the ambient high-dimensional phase space, the data assimilation step is computationally expensive and relies on full-state data that may not be available in practical applications.  

In this work, we demonstrate the first algorithm that leverages sparse sensor selection for efficient operator-theoretic modeling of nonlinear systems, the so-called \emph{sparsity-enabled CROM}.  
We first show that a sufficient, but small number of random measurements of the state embed the cluster geometry and preserve the probabilistic dynamics, relying on compressed sensing and the restricted isometry property.  
Further, we demonstrate the ability to learn a minimal set of optimized sensors that are tailored to the specific CROM and provide performance on par with the full high-dimensional CROM.  
These sparsity enabled innovations are demonstrated on three high-dimensional nonlinear fluid systems of increasing complexity, and in all cases optimized sensors outperform randomly chosen sensors.  
We also show that the sparsity enabled CROM can be used for closed-loop control, resulting in control performance that is similar to that of full-state CROM.

The combination of sparsity promoting techniques with linear embeddings of nonlinear systems will become a key enabler for real-time estimation and control tasks because it overcomes many of the limitations of existing ROMs and/or linear operator models.  
A number of important future directions and extensions arise out of this work.  
First, it may be fruitful to explore not only selecting sparse sensor locations, but also which nonlinear measurements of the state are most informative for a Koopman or Perron-Frobenius embedding.  
The sparse sensor placement algorithm itself may also be modified to include more realistic cost functions that incorporate real costs associated with certain sensor locations and types; for example, sensors near the root of a wing may be less expensive than those at the tip, and sensors in the wake may be inadmissible.  
Finally, even though the sparse sensor optimization is an offline computation, it is currently prohibitively expensive for very high-dimensional state-spaces, such as that of the mixing layer, and further algorithmic developments are required to scale to larger problems.



\section*{Acknowledgments}
EK gratefully acknowledges funding by the Moore/Sloan foundation, the Washington Research Foundation and the
eScience Institute. 
JNK acknowledges support from the Air Force Office of Scientific Research (FA9550-15-1-0385).
SLB and JNK acknowledge support from the Defense Advanced Research Projects Agency (DARPA contract HR0011-16-C-0016). 
BWB, SLB, and EK acknowledge support from the Air Force Research Lab award (FA8651-16-1-0003).
We appreciate valuable
stimulating discussions with Bernd Noack and Joshua Proctor.

\appendix
\section{Model dependency on the number of features}
\label{Sec:CROM:DependencyOnFeatures}
\begin{figure}[tb]
	\centering
	\includegraphics[width=\textwidth]{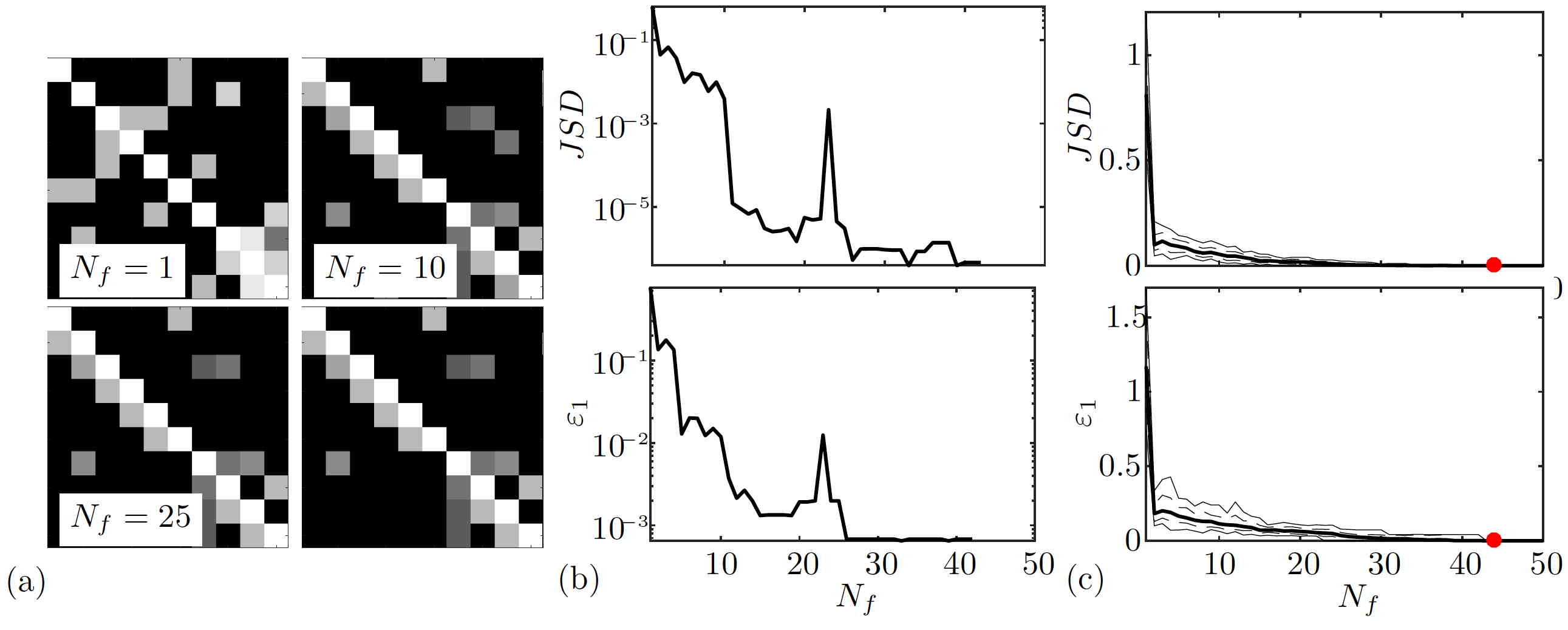}
	\caption{Dependency of estimated CROM on the number of features: 
		(a) Transition matrices for $N_f=1, 10, 25, M-1$ features based on the same clustering using $N_f=M-1$ features (best model chosen from 30 clustering repetitions), 
		(b) the Jensen-Shannon divergence and the $\varepsilon_1$ error decrease both with increasing number of features, both error measures vanish for $N_f > 43$, and
		(c) the mean (thick solid line), standard deviation (dashed line) and the minimum and maximum values (thin solid lines) for both error measures over 30 executions of the clustering algorithm.
		Transition probabilities are displayed in logarithmic scale ranging from zero probability ($\blacksquare$) to  probability of $1$ ($\square$).
		Both error measures vanish for $N_f > 43$ ($\color{red}{\bullet}$) irrespective of the clustering.}		
	\label{Fig:CROM:DependencyOnNumberFeatures}
\end{figure}

We examine the dependency of CROM on the number of features.
Specifically, the errors in the transition probabilities are assessed when less features are considered for computing the cluster affiliation. 
For this purpose, in addition to the Jensen-Shannon divergence (JSD) defined in \eqref{Eqn:JSD}, we give an estimate of the maximal error based on the $l_1$ norm:
\begin{equation}
\varepsilon_1 = \max{\boldsymbol{\varepsilon}} \quad \text{with} \quad \varepsilon_j =  \sum\limits_{i=1}^{N_c}\, \vert P_{i j}^{N_f} - Q_{i j}  \vert
\end{equation}
where $\matr{P}^{N_f}=(P_{ij}^{N_f})$ is the transition matrix recomputed using $N_f$ features, 
and $\matr{Q}=(Q_{ij})$ denotes the transition matrix based on all features.
The following steps are employed for calculating $\matr{P}^{N_f}$:
(1) CROM is computed from compressed data using POD as explained in Sec.~\ref{Sec:CROM}. This affiliates each $m$th observation with a label, $\{L_m\}_{m=1}^M$, and yields the transition matrix $\matr{Q}$.
For this step, all POD features are considered.
(2) The cluster affiliation is recomputed based on a reduced number of features. 
This relies on recomputed cluster centroids using the dominant $N_f$ POD coefficients.
(3) The transition matrix is recomputed based on the cluster affiliation in step (2). 
Example transition matrices and the error measures are displayed in Fig.~\ref{Fig:CROM:DependencyOnNumberFeatures}.
In Fig.~\ref{Fig:CROM:DependencyOnNumberFeatures}(a), 
it is observed that as the number of features increases, the transition matrix converges to that obtained using all features. Both JSD and $\varepsilon_1$ rapidly decay and vanish for $N_f>43$ (see Fig.~\ref{Fig:CROM:DependencyOnNumberFeatures}(b)).
The error measures, and particularly the minimum number of features to achieve zero error, do not depend significantly on the clustering (compare Fig.~\ref{Fig:CROM:DependencyOnNumberFeatures}(c)).

\bibliographystyle{model1-num-names} 
\bibliography{Literature.bib}

\begin{thebibliography}{90}
\expandafter\ifx\csname natexlab\endcsname\relax\def\natexlab#1{#1}\fi
\providecommand{\bibinfo}[2]{#2}
\ifx\xfnm\relax \def\xfnm[#1]{\unskip,\space#1}\fi
\bibitem[{Holmes et~al.(2012)Holmes, Lumley, Berkooz, and
  Rowley}]{Holmes2012book}
\bibinfo{author}{P.~Holmes}, \bibinfo{author}{J.~L. Lumley},
  \bibinfo{author}{G.~Berkooz}, \bibinfo{author}{C.~W. Rowley},
  \bibinfo{title}{Turbulence, {C}oherent {S}tructures, {D}ynamical {S}ystems
  and {S}ymmetry}, \bibinfo{publisher}{Cambridge University Press},
  \bibinfo{address}{Cambridge}, \bibinfo{edition}{2nd paperback} edition,
  \bibinfo{year}{2012}.
\bibitem[{Benner et~al.(2015)Benner, Gugercin, and Willcox}]{benner2015survey}
\bibinfo{author}{P.~Benner}, \bibinfo{author}{S.~Gugercin},
  \bibinfo{author}{K.~Willcox},
\newblock \bibinfo{title}{A survey of projection-based model reduction methods
  for parametric dynamical systems},
\newblock \bibinfo{journal}{SIAM review} \bibinfo{volume}{57}
  (\bibinfo{year}{2015}) \bibinfo{pages}{483--531}.
\bibitem[{Kaiser et~al.(2014)Kaiser, Noack, Cordier, Spohn, Segond, Abel,
  Daviller, \"Osth, Krajnovi\'c, and Niven}]{Kaiser2014jfm}
\bibinfo{author}{E.~Kaiser}, \bibinfo{author}{B.~R. Noack},
  \bibinfo{author}{L.~Cordier}, \bibinfo{author}{A.~Spohn},
  \bibinfo{author}{M.~Segond}, \bibinfo{author}{M.~Abel},
  \bibinfo{author}{G.~Daviller}, \bibinfo{author}{J.~\"Osth},
  \bibinfo{author}{S.~Krajnovi\'c}, \bibinfo{author}{R.~K. Niven},
\newblock \bibinfo{title}{Cluster-based reduced-order modelling of a mixing
  layer},
\newblock \bibinfo{journal}{J.\ Fluid Mech.} \bibinfo{volume}{754}
  (\bibinfo{year}{2014}) \bibinfo{pages}{365--414}.
\bibitem[{Brunton et~al.(2016)Brunton, Brunton, S.~L., and Kutz}]{Brunton2016}
\bibinfo{author}{B.~W. Brunton}, \bibinfo{author}{Brunton},
  \bibinfo{author}{J.~L. S.~L., Proctor}, \bibinfo{author}{J.~N. Kutz},
\newblock \bibinfo{title}{Sparse sensor placement optimization for
  classification},
\newblock \bibinfo{journal}{SIAM J. Appl. Math.} \bibinfo{volume}{76}
  (\bibinfo{year}{2016}) \bibinfo{pages}{2099--2122}.
\bibitem[{Everson and Sirovich(1995)}]{Everson1995josa}
\bibinfo{author}{R.~Everson}, \bibinfo{author}{L.~Sirovich},
\newblock \bibinfo{title}{Karhunen-lo\`eve procedure for gappy data},
\newblock \bibinfo{journal}{Journal of the Optical Society of America A}
  \bibinfo{volume}{12} (\bibinfo{year}{1995}) \bibinfo{pages}{1657--1664}.
\bibitem[{Chaturantabut and Sorensen(2010)}]{Chaturantabut2010siam-jsc}
\bibinfo{author}{S.~Chaturantabut}, \bibinfo{author}{D.~Sorensen},
\newblock \bibinfo{title}{Nonlinear model reduction via discrete empirical
  interpolation},
\newblock \bibinfo{journal}{SIAM Journal on Scientific Computing}
  \bibinfo{volume}{32} (\bibinfo{year}{2010}) \bibinfo{pages}{2737--2764}.
\bibitem[{Sargsyan et~al.(2015)Sargsyan, Brunton, and Kutz}]{Sargsyan:2015}
\bibinfo{author}{S.~Sargsyan}, \bibinfo{author}{S.~L. Brunton},
  \bibinfo{author}{J.~N. Kutz},
\newblock \bibinfo{title}{Nonlinear model reduction for dynamical systems using
  sparse sensor locations from learned libraries},
\newblock \bibinfo{journal}{Physical Review E} \bibinfo{volume}{92}
  (\bibinfo{year}{2015}) \bibinfo{pages}{033304}.
\bibitem[{Drma\u{c} and Gugercin(2016)}]{Drmac2016siam-jsc}
\bibinfo{author}{Z.~Drma\u{c}}, \bibinfo{author}{S.~Gugercin},
\newblock \bibinfo{title}{A new selection operator for the discrete empirical
  interpolation method---improved a priori error bound and extensions},
\newblock \bibinfo{journal}{SIAM J. Sci. Comput.} \bibinfo{volume}{38}
  (\bibinfo{year}{2016}) \bibinfo{pages}{A631--A648}.
\bibitem[{Brunton et~al.(2015)Brunton, Proctor, Tu, and Kutz}]{Brunton2015jcd}
\bibinfo{author}{S.~L. Brunton}, \bibinfo{author}{J.~L. Proctor},
  \bibinfo{author}{J.~H. Tu}, \bibinfo{author}{J.~N. Kutz},
\newblock \bibinfo{title}{Compressed sensing and dynamic mode decomposition},
\newblock \bibinfo{journal}{Journal of Computational Dynamics}
  \bibinfo{volume}{2} (\bibinfo{year}{2015}) \bibinfo{pages}{165--191}.
\bibitem[{Brunton et~al.(2016)Brunton, Proctor, and Kutz}]{Brunton2016pnas}
\bibinfo{author}{S.~L. Brunton}, \bibinfo{author}{J.~L. Proctor},
  \bibinfo{author}{J.~N. Kutz},
\newblock \bibinfo{title}{Discovering governing equations from data by sparse
  identification of nonlinear dynamical systems},
\newblock \bibinfo{journal}{Proceedings of the National Academy of Sciences}
  \bibinfo{volume}{113} (\bibinfo{year}{2016}) \bibinfo{pages}{3932--3937}.
\bibitem[{Jovanovi\'c et~al.(2014)Jovanovi\'c, Schmid, and
  Nichols}]{Jovanovic2014pof}
\bibinfo{author}{M.~R. Jovanovi\'c}, \bibinfo{author}{P.~J. Schmid},
  \bibinfo{author}{J.~W. Nichols},
\newblock \bibinfo{title}{Sparsity-promoting dynamic mode decomposition},
\newblock \bibinfo{journal}{Physics of Fluids} \bibinfo{volume}{26}
  (\bibinfo{year}{2014}) \bibinfo{pages}{024103}.
\bibitem[{Yildirim et~al.(2009)Yildirim, Chryssostomidis, and
  Karniadakis}]{Yildirim2009om}
\bibinfo{author}{B.~Yildirim}, \bibinfo{author}{C.~Chryssostomidis},
  \bibinfo{author}{G.~Karniadakis},
\newblock \bibinfo{title}{Efficient sensor placement for ocean measurements
  using low-dimensional concepts},
\newblock \bibinfo{journal}{Ocean Modeling} \bibinfo{volume}{273}
  (\bibinfo{year}{2009}) \bibinfo{pages}{160--173}.
\bibitem[{Bright et~al.(2013)Bright, Lin, and Kutz}]{Bright2013pof}
\bibinfo{author}{I.~Bright}, \bibinfo{author}{G.~Lin}, \bibinfo{author}{J.~N.
  Kutz},
\newblock \bibinfo{title}{Compressive sensing and machine learning strategies
  for characterizing the flow around a cylinder with limited pressure
  measurements},
\newblock \bibinfo{journal}{Physics of Fluids} \bibinfo{volume}{25}
  (\bibinfo{year}{2013}) \bibinfo{pages}{1--15}.
\bibitem[{Brunton et~al.(2014)Brunton, Tu, Bright, and Kutz}]{Brunton2014siam}
\bibinfo{author}{S.~L. Brunton}, \bibinfo{author}{J.~H. Tu},
  \bibinfo{author}{I.~Bright}, \bibinfo{author}{J.~N. Kutz},
\newblock \bibinfo{title}{Compressive sensing and low-rank libraries for
  classification of bifurcation regimes in nonlinear dynamical systems},
\newblock \bibinfo{journal}{SIAM Journal on Applied Dynamical Systems}
  \bibinfo{volume}{13} (\bibinfo{year}{2014}) \bibinfo{pages}{1716--1732}.
\bibitem[{Kim et~al.(2011)Kim, Park, Mohan, Gilbert, and
  Savarese}]{Kim2011proc}
\bibinfo{author}{B.~Kim}, \bibinfo{author}{J.~Y. Park},
  \bibinfo{author}{A.~Mohan}, \bibinfo{author}{A.~C. Gilbert},
  \bibinfo{author}{S.~Savarese},
\newblock \bibinfo{title}{Hierarchical classification of images by sparse
  approximation},
\newblock pp. \bibinfo{pages}{106.1--106.11}.
\bibitem[{Akhlaghi and Dogariu(2015)}]{Akhlaghi2015ol}
\bibinfo{author}{M.~I. Akhlaghi}, \bibinfo{author}{A.~Dogariu},
\newblock \bibinfo{title}{Compressive correlation imaging with random
  illumination},
\newblock \bibinfo{journal}{Optics Letters} \bibinfo{volume}{40}
  (\bibinfo{year}{2015}) \bibinfo{pages}{4464}.
\bibitem[{Bai et~al.(2017)Bai, Brunton, Brunton, Kutz, Kaiser, Spohn, and
  Noack}]{Bai2017}
\bibinfo{author}{Z.~Bai}, \bibinfo{author}{S.~L. Brunton},
  \bibinfo{author}{B.~W. Brunton}, \bibinfo{author}{J.~N. Kutz},
  \bibinfo{author}{E.~Kaiser}, \bibinfo{author}{A.~Spohn},
  \bibinfo{author}{B.~R. Noack}, \bibinfo{title}{Data-Driven Methods in Fluid
  Dynamics: Sparse Classification from Experimental Data},
  \bibinfo{publisher}{Springer International Publishing},
  \bibinfo{address}{Cham}, pp. \bibinfo{pages}{323--342}.
\bibitem[{Manohar et~al.(2016)Manohar, Brunton, and Kutz}]{Manohar2016arxiv}
\bibinfo{author}{K.~Manohar}, \bibinfo{author}{S.~L. Brunton},
  \bibinfo{author}{J.~N. Kutz},
\newblock \bibinfo{title}{Environment identification in flight using sparse
  approximation of wing strain},
\newblock \bibinfo{journal}{arXiv:1606.00034v1}  (\bibinfo{year}{2016}).
\bibitem[{Willcox(2006)}]{Willcox2006cf}
\bibinfo{author}{K.~Willcox},
\newblock \bibinfo{title}{Unsteady flow sensing and estimation via the gappy
  proper orthogonal decomposition},
\newblock \bibinfo{journal}{Computers \& Fluids} \bibinfo{volume}{35}
  (\bibinfo{year}{2006}) \bibinfo{pages}{208--226}.
\bibitem[{Carlberg et~al.(2013)Carlberg, Farhat, Cortial, and
  Amsallem}]{Carlberg2013jcp}
\bibinfo{author}{K.~Carlberg}, \bibinfo{author}{C.~Farhat},
  \bibinfo{author}{J.~Cortial}, \bibinfo{author}{D.~Amsallem},
\newblock \bibinfo{title}{The gnat method for nonlinear model reduction:
  Effective implementation and application to computational fluid dynamics and
  turbulent flows},
\newblock \bibinfo{journal}{Journal of Computational Physics}
  \bibinfo{volume}{242} (\bibinfo{year}{2013}) \bibinfo{pages}{623--647}.
\bibitem[{Koopman(1931)}]{Koopman1931pnas}
\bibinfo{author}{B.~O. Koopman},
\newblock \bibinfo{title}{Hamiltonian systems and transformation in hilbert
  space},
\newblock \bibinfo{journal}{Proceedings of the National Academy of Sciences}
  \bibinfo{volume}{17} (\bibinfo{year}{1931}) \bibinfo{pages}{315}.
\bibitem[{Mezi{\'c}(2005)}]{Mezic2005nd}
\bibinfo{author}{I.~Mezi{\'c}},
\newblock \bibinfo{title}{Spectral properties of dynamical systems, model
  reduction and decompositions},
\newblock \bibinfo{journal}{Nonlinear Dynamics} \bibinfo{volume}{41}
  (\bibinfo{year}{2005}) \bibinfo{pages}{309--325}.
\bibitem[{Perron(1907)}]{Perron1907ma}
\bibinfo{author}{O.~Perron},
\newblock \bibinfo{title}{Zur {T}heorie der {M}atrices},
\newblock \bibinfo{journal}{Math. Ann.} \bibinfo{volume}{64}
  (\bibinfo{year}{1907}) \bibinfo{pages}{248--263}.
\bibitem[{Ulam(1964)}]{Ulam1964book}
\bibinfo{author}{S.~Ulam}, \bibinfo{title}{Problems in {M}odern {M}athematics},
  \bibinfo{publisher}{Interscience}, \bibinfo{year}{1964}.
\bibitem[{Ryter(1987)}]{Ryter1987phys}
\bibinfo{author}{D.~Ryter},
\newblock \bibinfo{title}{On the eigenfunctions of the fokker-planck operator
  and of its adjoint},
\newblock \bibinfo{journal}{Physica A: Statistical Mechanics and its
  Applications} \bibinfo{volume}{142} (\bibinfo{year}{1987})
  \bibinfo{pages}{103--121}.
\bibitem[{Schmid(2010)}]{Schmid2010jfm}
\bibinfo{author}{P.~Schmid},
\newblock \bibinfo{title}{Dynamic mode decomposition of numerical and
  experimental data},
\newblock \bibinfo{journal}{J. Fluid Mech.} \bibinfo{volume}{65}
  (\bibinfo{year}{2010}) \bibinfo{pages}{5--28}.
\bibitem[{Rowley et~al.(2009)Rowley, Mezi\'c, Bagheri, Schlatter, and
  Henningson}]{Rowley2009jfm}
\bibinfo{author}{C.~W. Rowley}, \bibinfo{author}{I.~Mezi\'c},
  \bibinfo{author}{S.~Bagheri}, \bibinfo{author}{P.~Schlatter},
  \bibinfo{author}{D.~S. Henningson},
\newblock \bibinfo{title}{Spectral analysis of nonlinear flows},
\newblock \bibinfo{journal}{Journal of Fluid Mechanics} \bibinfo{volume}{641}
  (\bibinfo{year}{2009}) \bibinfo{pages}{115--127}.
\bibitem[{Kutz et~al.(2016)Kutz, Brunton, Brunton, and Proctor}]{Kutz2016book}
\bibinfo{author}{J.~N. Kutz}, \bibinfo{author}{S.~L. Brunton},
  \bibinfo{author}{B.~W. Brunton}, \bibinfo{author}{J.~L. Proctor},
  \bibinfo{title}{Dynamic Mode Decomposition: Data-Driven Modeling of Complex
  Systems}, \bibinfo{publisher}{SIAM}, \bibinfo{year}{2016}.
\bibitem[{Williams et~al.(2015)Williams, Kevrekidis, and
  Rowley}]{Williams2015jns}
\bibinfo{author}{M.~O. Williams}, \bibinfo{author}{I.~G. Kevrekidis},
  \bibinfo{author}{C.~W. Rowley},
\newblock \bibinfo{title}{A data--driven approximation of the koopman operator:
  Extending dynamic mode decomposition},
\newblock \bibinfo{journal}{Journal of Nonlinear Science} \bibinfo{volume}{25}
  (\bibinfo{year}{2015}) \bibinfo{pages}{1307--1346}.
\bibitem[{Liouville(1838)}]{Liouville1838}
\bibinfo{author}{J.~Liouville},
\newblock \bibinfo{title}{???},
\newblock \bibinfo{journal}{Journ. de Math.} \bibinfo{volume}{3}
  (\bibinfo{year}{1838}) \bibinfo{pages}{349}.
\bibitem[{Froyland(1998)}]{Froyland1998831}
\bibinfo{author}{G.~Froyland},
\newblock \bibinfo{title}{Approximating physical invariant measures of mixing
  dynamical systems in higher dimensions},
\newblock \bibinfo{journal}{Nonlinear Analysis: Theory, Methods \&
  Applications} \bibinfo{volume}{32} (\bibinfo{year}{1998}) \bibinfo{pages}{831
  -- 860}.
\bibitem[{Froyland and Dellnitz(2003)}]{Froyland2003jsc}
\bibinfo{author}{G.~Froyland}, \bibinfo{author}{M.~Dellnitz},
\newblock \bibinfo{title}{Detecting and locating near-optimal almost-invariant
  sets and cycles},
\newblock \bibinfo{journal}{SIAM J. Sci. Comput.} \bibinfo{volume}{24}
  (\bibinfo{year}{2003}) \bibinfo{pages}{1839--1863}.
\bibitem[{Dellnitz and Junge(2004)}]{Dellnitz2004}
\bibinfo{author}{M.~Dellnitz}, \bibinfo{author}{O.~Junge}, \bibinfo{title}{On
  the Approximation of Complicated Dynamical Behavior},
  \bibinfo{publisher}{Springer New York}, \bibinfo{address}{New York, NY}, pp.
  \bibinfo{pages}{400--424}.
\bibitem[{Froyland et~al.(2013)Froyland, Junge, and Koltai}]{Froyland2013jna}
\bibinfo{author}{G.~Froyland}, \bibinfo{author}{O.~Junge},
  \bibinfo{author}{P.~Koltai},
\newblock \bibinfo{title}{{E}stimating {L}ong-{T}erm {B}ehavior of {F}lows
  without {T}rajectory {I}ntegration: {T}he {I}nfinitesimal {G}enerator
  {A}pproach},
\newblock \bibinfo{journal}{SIAM J.\ Numer.\ Anal.} \bibinfo{volume}{51}
  (\bibinfo{year}{2013}) \bibinfo{pages}{223--247}.
\bibitem[{Bollt and Santitissadeekorn(2013)}]{Bollt2013book}
\bibinfo{author}{E.~M. Bollt}, \bibinfo{author}{N.~Santitissadeekorn},
  \bibinfo{title}{{A}pplied and {C}omputational {M}easurable {D}ynamics},
  \bibinfo{publisher}{SIAM}, \bibinfo{year}{2013}.
\bibitem[{Li(1976)}]{Li1976jat}
\bibinfo{author}{T.~Y. Li},
\newblock \bibinfo{title}{{F}inite approximation for the {F}robenius-{P}erron
  operator: {A} solution to {U}lam's conjecture},
\newblock \bibinfo{journal}{J.\ Approx.\ Theory} \bibinfo{volume}{17}
  (\bibinfo{year}{1976}) \bibinfo{pages}{177--186}.
\bibitem[{Bishop(2007)}]{Bishop2007book}
\bibinfo{author}{C.~M. Bishop}, \bibinfo{title}{Pattern Recognition and Machine
  Learning}, \bibinfo{publisher}{Springer}, \bibinfo{address}{???},
  \bibinfo{year}{2007}.
\bibitem[{Lasota and Mackey(1994)}]{Lasota1994book}
\bibinfo{author}{A.~Lasota}, \bibinfo{author}{M.~C. Mackey},
  \bibinfo{title}{{C}haos, {F}ractals, and {N}oise},
  \bibinfo{publisher}{Springer New York}, \bibinfo{edition}{2nd} edition,
  \bibinfo{year}{1994}.
\bibitem[{Birkhoff(1931)}]{Birkhoff1931pnas}
\bibinfo{author}{G.~D. Birkhoff},
\newblock \bibinfo{title}{Proof of the ergodic theorem},
\newblock \bibinfo{journal}{Proceedings of the National Academy of Sciences}
  \bibinfo{volume}{17} (\bibinfo{year}{1931}) \bibinfo{pages}{656--660}.
\bibitem[{Cvitanovi\'c et~al.(2012)Cvitanovi\'c, Artuso, Mainieri, Tanner, and
  Vattay}]{Cvitanovic2012book}
\bibinfo{author}{P.~Cvitanovi\'c}, \bibinfo{author}{R.~Artuso},
  \bibinfo{author}{R.~Mainieri}, \bibinfo{author}{G.~Tanner},
  \bibinfo{author}{G.~Vattay}, \bibinfo{title}{Chaos: Classical and Quantum},
  \bibinfo{publisher}{Niels Bohr Institute}, \bibinfo{address}{Copenhagen},
  \bibinfo{year}{2012}.
\bibitem[{Hopf(1952)}]{Hopf1952jrma}
\bibinfo{author}{E.~Hopf},
\newblock \bibinfo{title}{Statistical hydromechanics and functional analysis},
\newblock \bibinfo{journal}{J.\ Rat.\ Mech.\ Anal.} \bibinfo{volume}{1}
  (\bibinfo{year}{1952}) \bibinfo{pages}{87--123}.
\bibitem[{Venturi(2016)}]{Venturi2016arxiv}
\bibinfo{author}{D.~Venturi},
\newblock \bibinfo{title}{The numerical approximation of functional
  differential equations},
\newblock \bibinfo{journal}{arXiv preprint arXiv:1604.05250 [math.NA]}
  (\bibinfo{year}{2016}).
\bibitem[{H.(1965)}]{Mori1965ptp}
\bibinfo{author}{M.~H.},
\newblock \bibinfo{title}{Transport, collective motion and brownian motion},
\newblock \bibinfo{journal}{Prog. Theor. Phys.} \bibinfo{volume}{33}
  (\bibinfo{year}{1965}) \bibinfo{pages}{423450}.
\bibitem[{R.(1973)}]{Zwanzig1973jsp}
\bibinfo{author}{Z.~R.},
\newblock \bibinfo{title}{Nonlinear generalized langevin equations},
\newblock \bibinfo{journal}{J. Stat. Phys.} \bibinfo{volume}{9}
  (\bibinfo{year}{1973}) \bibinfo{pages}{215220}.
\bibitem[{Chorin and Hald(2009)}]{Chorin2009book}
\bibinfo{author}{A.~Chorin}, \bibinfo{author}{O.~Hald},
  \bibinfo{title}{Stochastic Tools for Mathematics and Science},
  \bibinfo{year}{2009}.
\bibitem[{Stinis(2015)}]{Stinis20140446}
\bibinfo{author}{P.~Stinis},
\newblock \bibinfo{title}{Renormalized mori{\textendash}zwanzig-reduced models
  for systems without scale separation},
\newblock \bibinfo{journal}{Proceedings of the Royal Society of London A:
  Mathematical, Physical and Engineering Sciences} \bibinfo{volume}{471}
  (\bibinfo{year}{2015}).
\bibitem[{Gouasmi et~al.(2016)Gouasmi, Parish, and
  Duraisamy}]{Gouasmi2016arxiv}
\bibinfo{author}{A.~Gouasmi}, \bibinfo{author}{E.~Parish},
  \bibinfo{author}{K.~Duraisamy},
\newblock \bibinfo{title}{Characterizing memory effects in coarse-grained
  nonlinear systems using the mori-zwanzig formalism},
\newblock \bibinfo{journal}{arXiv preprint arXiv:1611.06277}
  (\bibinfo{year}{2016}).
\bibitem[{Noack and Niven(2012)}]{Noack2012jfm}
\bibinfo{author}{B.~R. Noack}, \bibinfo{author}{R.~K. Niven},
\newblock \bibinfo{title}{Maximum-entropy closure for a {G}alerkin system of
  incompressible shear flow},
\newblock \bibinfo{journal}{J.\ Fluid Mech.} \bibinfo{volume}{700}
  (\bibinfo{year}{2012}) \bibinfo{pages}{187--213}.
\bibitem[{Steinhaus(1956)}]{Steinhaus1956}
\bibinfo{author}{H.~Steinhaus},
\newblock \bibinfo{title}{Sur la division des corps mat\'eriels en parties},
\newblock \bibinfo{journal}{Bull. Acad. Polon. Sci.} \bibinfo{volume}{4}
  (\bibinfo{year}{1956}) \bibinfo{pages}{801--804}.
\bibitem[{Du et~al.(1999)Du, Faber, and
  Gunzburger}]{Du_Faber_Gunzburger_SIAM_Review_1999}
\bibinfo{author}{Q.~Du}, \bibinfo{author}{V.~Faber},
  \bibinfo{author}{M.~Gunzburger},
\newblock \bibinfo{title}{Centroidal {V}oronoi {T}essellations: Applications
  and {A}lgorithms},
\newblock \bibinfo{journal}{SIAM review} \bibinfo{volume}{41}
  (\bibinfo{year}{1999}) \bibinfo{pages}{637--676}.
\bibitem[{Du and Gunzburger(2003)}]{Du2003}
\bibinfo{author}{Q.~Du}, \bibinfo{author}{M.~D. Gunzburger},
  \bibinfo{title}{Centroidal Voronoi Tessellation Based Proper Orthogonal
  Decomposition Analysis}, \bibinfo{publisher}{Birkh{\"a}user Basel},
  \bibinfo{address}{Basel}, pp. \bibinfo{pages}{137--150}.
\bibitem[{Amsallem et~al.(2009)Amsallem, Cortial, and
  Farhat}]{Amsallem2009aiaa}
\bibinfo{author}{D.~Amsallem}, \bibinfo{author}{J.~Cortial},
  \bibinfo{author}{C.~Farhat},
\newblock \bibinfo{title}{On-demand cfd-based aeroelastic predictions using a
  database of reduced-order bases and models},
\newblock in: \bibinfo{booktitle}{47th AIAA Aerospace Sciences Meeting
  Including The New Horizons Forum and Aerospace Exposition AIAA 2009-800 5 --
  8 January 2009, Orlando, Florida}.
\bibitem[{Amsallem et~al.(2012)Amsallem, Zahr, and Farhat}]{Amsallem2012nme}
\bibinfo{author}{D.~Amsallem}, \bibinfo{author}{M.~J. Zahr},
  \bibinfo{author}{C.~Farhat},
\newblock \bibinfo{title}{Nonlinear model order reduction based on local
  reduced-order bases},
\newblock \bibinfo{journal}{International Journal for Numerical Methods in
  Engineering} \bibinfo{volume}{92} (\bibinfo{year}{2012})
  \bibinfo{pages}{891--916}.
\bibitem[{Giannakis and Majda(2011)}]{Giannakis2011joc}
\bibinfo{author}{D.~Giannakis}, \bibinfo{author}{A.~J. Majda},
\newblock \bibinfo{title}{Quantifying the predictive skill in long-range
  forecasting. part i: Coarse-grained predictions in a simple ocean model},
\newblock \bibinfo{journal}{J. of Climate} \bibinfo{volume}{???}
  (\bibinfo{year}{2011}) \bibinfo{pages}{???}
\bibitem[{Budi{\v{s}}i{\'c} et~al.(2012)Budi{\v{s}}i{\'c}, Mohr, and
  Mezi{\'c}}]{Budivsic2012chaos}
\bibinfo{author}{M.~Budi{\v{s}}i{\'c}}, \bibinfo{author}{R.~Mohr},
  \bibinfo{author}{I.~Mezi{\'c}},
\newblock \bibinfo{title}{Applied {K}oopmanism a)},
\newblock \bibinfo{journal}{Chaos: An Interdisciplinary Journal of Nonlinear
  Science} \bibinfo{volume}{22} (\bibinfo{year}{2012}) \bibinfo{pages}{047510}.
\bibitem[{Mezic(2013)}]{Mezic2013arfm}
\bibinfo{author}{I.~Mezic},
\newblock \bibinfo{title}{Analysis of fluid flows via spectral properties of
  the {K}oopman operator},
\newblock \bibinfo{journal}{Annual Review of Fluid Mechanics}
  \bibinfo{volume}{45} (\bibinfo{year}{2013}) \bibinfo{pages}{357--378}.
\bibitem[{Brunton et~al.(2016)Brunton, Brunton, Proctor, and
  Kutz}]{Brunton2016plosone}
\bibinfo{author}{S.~L. Brunton}, \bibinfo{author}{B.~W. Brunton},
  \bibinfo{author}{J.~L. Proctor}, \bibinfo{author}{J.~N. Kutz},
\newblock \bibinfo{title}{Koopman observable subspaces and finite linear
  representations of nonlinear dynamical systems for control},
\newblock \bibinfo{journal}{PLoS ONE} \bibinfo{volume}{11}
  (\bibinfo{year}{2016}) \bibinfo{pages}{e0150171}.
\bibitem[{Cand\`es et~al.(2006)Cand\`es, Romberg, and Tao}]{Candes2006ieee}
\bibinfo{author}{E.~J. Cand\`es}, \bibinfo{author}{J.~Romberg},
  \bibinfo{author}{T.~Tao},
\newblock \bibinfo{title}{Robust uncertainty principles: exact signal
  reconstruction from highly incomplete frequency information},
\newblock \bibinfo{journal}{IEEE Transactions on Information Theory}
  \bibinfo{volume}{52} (\bibinfo{year}{2006}) \bibinfo{pages}{489--509}.
\bibitem[{Cand\`es et~al.(59)Cand\`es, Romberg, and Tao}]{Candes2006c}
\bibinfo{author}{E.~J. Cand\`es}, \bibinfo{author}{J.~Romberg},
  \bibinfo{author}{T.~Tao},
\newblock \bibinfo{title}{Stable signal recovery from incomplete and inaccurate
  measurements},
\newblock \bibinfo{journal}{Communications in Pure and Applied Mathematics}
  \bibinfo{volume}{8} (\bibinfo{year}{59}).
\bibitem[{Donoho(2006{\natexlab{a}})}]{Donoho2006ieee}
\bibinfo{author}{D.~L. Donoho},
\newblock \bibinfo{title}{Compressed sensing},
\newblock \bibinfo{journal}{IEEE Transactions on Information Theory}
  \bibinfo{volume}{52} (\bibinfo{year}{2006}{\natexlab{a}})
  \bibinfo{pages}{1289--1306}.
\bibitem[{Donoho(2006{\natexlab{b}})}]{Donoho2006book}
\bibinfo{author}{D.~L. Donoho},
\newblock \bibinfo{title}{For most large underdetermined systems of linear
  equations, the minimal l1-norm solution is also the sparsest solution},
\newblock \bibinfo{journal}{Communications in Pure and Applied mathematics}
  \bibinfo{volume}{59} (\bibinfo{year}{2006}{\natexlab{b}})
  \bibinfo{pages}{797--829}.
\bibitem[{Grant and Boyd(2014)}]{cvx}
\bibinfo{author}{M.~Grant}, \bibinfo{author}{S.~Boyd}, \bibinfo{title}{{CVX}:
  Matlab software for disciplined convex programming, version 2.1},
  \bibinfo{howpublished}{http://cvxr.com/cvx}, \bibinfo{year}{2014}.
\bibitem[{Grant and Boyd(2008)}]{gb08}
\bibinfo{author}{M.~Grant}, \bibinfo{author}{S.~Boyd},
\newblock \bibinfo{title}{Graph implementations for nonsmooth convex programs},
\newblock in: \bibinfo{editor}{V.~Blondel}, \bibinfo{editor}{S.~Boyd},
  \bibinfo{editor}{H.~Kimura} (Eds.), \bibinfo{booktitle}{Recent Advances in
  Learning and Control}, Lecture Notes in Control and Information Sciences,
  \bibinfo{publisher}{Springer-Verlag Limited}, \bibinfo{year}{2008}, pp.
  \bibinfo{pages}{95--110}.
\bibitem[{Tropp and Gilbert(2007)}]{Tropp2007ieee}
\bibinfo{author}{J.~A. Tropp}, \bibinfo{author}{A.~C. Gilbert},
\newblock \bibinfo{title}{Signal recovery from random measurements via
  orthogonal matching pursuit},
\newblock \bibinfo{journal}{IEEE Transactions on Information Theory}
  \bibinfo{volume}{53} (\bibinfo{year}{2007}) \bibinfo{pages}{4655--4666}.
\bibitem[{Tropp(2006)}]{Tropp2006sp}
\bibinfo{author}{J.~A. Tropp},
\newblock \bibinfo{title}{Algorithms for simultaneous sparse approximation.
  part ii: Convex relaxation},
\newblock \bibinfo{journal}{Signal Processing} \bibinfo{volume}{86}
  (\bibinfo{year}{2006}) \bibinfo{pages}{589--602}.
\bibitem[{Ozoli{\c{n}}{\v{s}} et~al.(2013)Ozoli{\c{n}}{\v{s}}, Lai, Caflisch,
  and Osher}]{Ozolicnvs2013pnas}
\bibinfo{author}{V.~Ozoli{\c{n}}{\v{s}}}, \bibinfo{author}{R.~Lai},
  \bibinfo{author}{R.~Caflisch}, \bibinfo{author}{S.~Osher},
\newblock \bibinfo{title}{Compressed modes for variational problems in
  mathematics and physics},
\newblock \bibinfo{journal}{Proceedings of the National Academy of Sciences}
  \bibinfo{volume}{110} (\bibinfo{year}{2013}) \bibinfo{pages}{18368--18373}.
\bibitem[{Schaeffer et~al.(2013)Schaeffer, Caflisch, Hauck, and
  Osher}]{Schaeffer2013pnas}
\bibinfo{author}{H.~Schaeffer}, \bibinfo{author}{R.~Caflisch},
  \bibinfo{author}{C.~D. Hauck}, \bibinfo{author}{S.~Osher},
\newblock \bibinfo{title}{Sparse dynamics for partial differential equations},
\newblock \bibinfo{journal}{Proceedings of the National Academy of Sciences
  USA} \bibinfo{volume}{110} (\bibinfo{year}{2013})
  \bibinfo{pages}{6634--6639}.
\bibitem[{Mackey et~al.(2014)Mackey, Schaeffer, and
  Osher}]{mackey2014compressive}
\bibinfo{author}{A.~Mackey}, \bibinfo{author}{H.~Schaeffer},
  \bibinfo{author}{S.~Osher},
\newblock \bibinfo{title}{On the compressive spectral method},
\newblock \bibinfo{journal}{Multiscale Modeling \&amp; Simulation}
  \bibinfo{volume}{12} (\bibinfo{year}{2014}) \bibinfo{pages}{1800--1827}.
\bibitem[{Tu et~al.(2014)Tu, Rowley, Kutz, and Shang}]{Tu2014ef}
\bibinfo{author}{J.~H. Tu}, \bibinfo{author}{C.~W. Rowley},
  \bibinfo{author}{J.~N. Kutz}, \bibinfo{author}{J.~K. Shang},
\newblock \bibinfo{title}{Spectral analysis of fluid flows using
  sub-nyquist-rate piv data},
\newblock \bibinfo{journal}{Experiments in Fluids} \bibinfo{volume}{55}
  (\bibinfo{year}{2014}) \bibinfo{pages}{1--13}.
\bibitem[{Gueniat et~al.(2015)Gueniat, Mathelin, and Pastur}]{Gueniat2015pof}
\bibinfo{author}{F.~Gueniat}, \bibinfo{author}{L.~Mathelin},
  \bibinfo{author}{L.~Pastur},
\newblock \bibinfo{title}{A dynamic mode decomposition approach for large and
  arbitrarily sampled systems},
\newblock \bibinfo{journal}{Physics of Fluids} \bibinfo{volume}{27}
  (\bibinfo{year}{2015}) \bibinfo{pages}{025113}.
\bibitem[{Kramer et~al.(2015)Kramer, Grover, Boufounos, Benosman, and
  Nabi}]{Kramer2015arxiv}
\bibinfo{author}{B.~Kramer}, \bibinfo{author}{P.~Grover},
  \bibinfo{author}{P.~Boufounos}, \bibinfo{author}{M.~Benosman},
  \bibinfo{author}{S.~Nabi},
\newblock \bibinfo{title}{Sparse sensing and dmd based identification of flow
  regimes and bifurcations in complex flows},
\newblock \bibinfo{journal}{arXiv preprint arXiv:1510.02831}
  (\bibinfo{year}{2015}).
\bibitem[{Davenport et~al.(2012)Davenport, Duarte, Eldar, and
  Kutyniok}]{Davenport2012book}
\bibinfo{author}{M.~A. Davenport}, \bibinfo{author}{M.~F. Duarte},
  \bibinfo{author}{Y.~C. Eldar}, \bibinfo{author}{G.~Kutyniok},
  \bibinfo{title}{Introduction to Compressed Sensing},
  \bibinfo{publisher}{Cambridge University Press}.
\bibitem[{Lin(1991)}]{Lin1991ieee}
\bibinfo{author}{J.~Lin},
\newblock \bibinfo{title}{Divergence measures based on the shannon entropy},
\newblock \bibinfo{journal}{IEEE Transactions on information theory}
  \bibinfo{volume}{37} (\bibinfo{year}{1991}) \bibinfo{pages}{145--151}.
\bibitem[{Kullback and Leibler(1951)}]{Kullback1951ams}
\bibinfo{author}{S.~Kullback}, \bibinfo{author}{R.~A. Leibler},
\newblock \bibinfo{title}{On information and sufficiency},
\newblock \bibinfo{journal}{Annals Math. Stat.} \bibinfo{volume}{22}
  (\bibinfo{year}{1951}) \bibinfo{pages}{79--86}.
\bibitem[{Kullback(1959)}]{Kullback1959book}
\bibinfo{author}{S.~Kullback}, \bibinfo{title}{Information {T}heory and
  {S}tatistics}, \bibinfo{publisher}{John Wiley}, \bibinfo{address}{New York},
  \bibinfo{edition}{1st} edition, \bibinfo{year}{1959}.
\bibitem[{Kaiser et~al.(2016)Kaiser, Noack, Spohn, Cattafesta, and
  Morzy\'nski}]{Kaiser2016tcfd}
\bibinfo{author}{E.~Kaiser}, \bibinfo{author}{B.~R. Noack},
  \bibinfo{author}{A.~Spohn}, \bibinfo{author}{L.~N. Cattafesta},
  \bibinfo{author}{M.~Morzy\'nski},
\newblock \bibinfo{title}{Cluster-based control of nonlinear dynamics},
\newblock \bibinfo{journal}{under review in Theoret. and Comp. Fluid Dynamics}
  \bibinfo{volume}{arXiv:1602.05416} (\bibinfo{year}{2016}).
\bibitem[{Solomon and Gollub(1988)}]{Solomon1988pra}
\bibinfo{author}{T.~H. Solomon}, \bibinfo{author}{J.~P. Gollub},
\newblock \bibinfo{title}{Chaotic particle transport in time-dependent
  rayleigh-b\'enard convection},
\newblock \bibinfo{journal}{Physical Review A} \bibinfo{volume}{38}
  (\bibinfo{year}{1988}) \bibinfo{pages}{6280--6286}.
\bibitem[{Shadden et~al.(2005)Shadden, Lekien, and
  Marsden}]{Shadden2005physica}
\bibinfo{author}{S.~Shadden}, \bibinfo{author}{F.~Lekien},
  \bibinfo{author}{J.~Marsden},
\newblock \bibinfo{title}{Definition and properties of lagrangian coherent
  structures from finite-time lyapunov exponents in two-dimensional aperiodic
  flows.},
\newblock \bibinfo{journal}{Physica D: Nonlinear Phenomena}
  \bibinfo{volume}{212} (\bibinfo{year}{2005}) \bibinfo{pages}{271--304}.
\bibitem[{Hood and Taylor(1974)}]{Hood1974book}
\bibinfo{author}{P.~Hood}, \bibinfo{author}{C.~Taylor}, \bibinfo{title}{Finite
  Element Methods in Flow Problems}, \bibinfo{publisher}{University of Alabama
  in Huntsville Press}, pp. \bibinfo{pages}{121--132}.
\bibitem[{Morzy\'nski(1987)}]{Morzynski1987proc}
\bibinfo{author}{M.~Morzy\'nski},
\newblock \bibinfo{title}{Numerical solution of navier-stokes equations by the
  finite element method},
\newblock in: \bibinfo{booktitle}{Proceedings of SYMKOM 87, Compressor and
  Turbine Stage Flow Path -- Theory and Experiment}, \bibinfo{year}{1987}, pp.
  \bibinfo{pages}{119--128}.
\bibitem[{Afanasiev(2003)}]{Afanasiev2003phd}
\bibinfo{author}{K.~Afanasiev}, \bibinfo{title}{Stabilit\"atsanalyse,
  niedrigdimensionale Modellierung und optimale Kontrolle der
  Kreiszylinderumstr\"omung (trans.: Stability analysis, low-dimensional
  modeling, and optimal control of the flow around a circular cylinder)}, Ph.D.
  thesis, Fakult\"at Maschinenwesen, Technische Universit\"at Dresden,
  \bibinfo{year}{2003}.
\bibitem[{Ho and Huerre(1984)}]{Ho1984arfm}
\bibinfo{author}{C.-M. Ho}, \bibinfo{author}{P.~Huerre},
\newblock \bibinfo{title}{Perturbed free shear layers},
\newblock \bibinfo{journal}{Ann.\ Rev.\ Fluid Mech.} \bibinfo{volume}{16}
  (\bibinfo{year}{1984}) \bibinfo{pages}{365--424}.
\bibitem[{Daviller(2010)}]{Daviller2010phd}
\bibinfo{author}{G.~Daviller}, \bibinfo{title}{\'{E}tude num\'{e}rique des
  effets de temp\'{e}rature dans les jets simples et coaxiaux}, Ph.D. thesis,
  \'{E}cole {N}ationale {S}up\'{e}rieure de {M}\'{e}canique et
  d'{A}\'{e}rotechnique, \bibinfo{year}{2010}.
\bibitem[{Cavalieri et~al.(2011)Cavalieri, Daviller, Comte, Jordan, Tadmor, and
  Gervais}]{Cavalieri2011jsv}
\bibinfo{author}{A.~Cavalieri}, \bibinfo{author}{G.~Daviller},
  \bibinfo{author}{P.~Comte}, \bibinfo{author}{P.~Jordan},
  \bibinfo{author}{G.~Tadmor}, \bibinfo{author}{Y.~Gervais},
\newblock \bibinfo{title}{Using large eddy simulation to explore sound-source
  mechanisms in jets},
\newblock \bibinfo{journal}{J.\ Sound Vib.} \bibinfo{volume}{330}
  (\bibinfo{year}{2011}) \bibinfo{pages}{4098--4113}.
\bibitem[{Cross and Hohenberg(1993)}]{Cross:1993}
\bibinfo{author}{M.~C. Cross}, \bibinfo{author}{P.~C. Hohenberg},
\newblock \bibinfo{title}{Pattern formation outside of equilibrium},
\newblock \bibinfo{journal}{Reviews of modern physics} \bibinfo{volume}{65}
  (\bibinfo{year}{1993}) \bibinfo{pages}{851}.
\bibitem[{Everson and Sirovich(1995)}]{everson1995karhunen}
\bibinfo{author}{R.~Everson}, \bibinfo{author}{L.~Sirovich},
\newblock \bibinfo{title}{Karhunen--lo\`eve procedure for gappy data},
\newblock \bibinfo{journal}{JOSA A} \bibinfo{volume}{12} (\bibinfo{year}{1995})
  \bibinfo{pages}{1657--1664}.
\bibitem[{Barrault et~al.(2004)Barrault, Maday, Nguyen, and Patera}]{eim}
\bibinfo{author}{M.~Barrault}, \bibinfo{author}{Y.~Maday},
  \bibinfo{author}{N.~C. Nguyen}, \bibinfo{author}{A.~T. Patera},
\newblock \bibinfo{title}{An empirical interpolation method: application to
  efficient reduced-basis discretization of partial differential equations},
\newblock \bibinfo{journal}{Comptes Rendus Mathematique} \bibinfo{volume}{339}
  (\bibinfo{year}{2004}) \bibinfo{pages}{667--672}.
\bibitem[{Chaturantabut and Sorensen(2010)}]{deim}
\bibinfo{author}{S.~Chaturantabut}, \bibinfo{author}{D.~C. Sorensen},
\newblock \bibinfo{title}{Nonlinear model reduction via discrete empirical
  interpolation},
\newblock \bibinfo{journal}{SIAM Journal on Scientific Computing}
  \bibinfo{volume}{32} (\bibinfo{year}{2010}) \bibinfo{pages}{2737--2764}.
\bibitem[{Bright et~al.(2013)Bright, Lin, and Kutz}]{bright:2013}
\bibinfo{author}{I.~Bright}, \bibinfo{author}{G.~Lin}, \bibinfo{author}{J.~N.
  Kutz},
\newblock \bibinfo{title}{Compressive sensing based machine learning strategy
  for characterizing the flow around a cylinder with limited pressure
  measurements},
\newblock \bibinfo{journal}{Physics of Fluids (1994-present)}
  \bibinfo{volume}{25} (\bibinfo{year}{2013}) \bibinfo{pages}{127102}.
\bibitem[{Carlberg et~al.(2015)Carlberg, Barone, and Antil}]{carlberg2015arxiv}
\bibinfo{author}{K.~Carlberg}, \bibinfo{author}{M.~Barone},
  \bibinfo{author}{H.~Antil},
\newblock \bibinfo{title}{Galerkin v. discrete-optimal projection in nonlinear
  model reduction},
\newblock \bibinfo{journal}{arXiv preprint arXiv:1504.03749}
  (\bibinfo{year}{2015}).

\end{thebibliography}

\end{document}